\documentclass[12pt]{article}
\usepackage{epsf}
\usepackage{amsmath,amssymb}

\usepackage{graphicx}
\usepackage{comment}

\begin{document}

\renewcommand{\theequation}{\thesection.\arabic{equation}}

\renewcommand{\thefootnote}{\fnsymbol{footnote}}
\setcounter{footnote}{0}

\baselineskip 0.5cm

\begin{titlepage}

\def\thefootnote{\fnsymbol{footnote}}

\begin{center}

\hfill IPMU17-0117\\
\hfill UT-17-29\\
\hfill September, 2017\\

\vskip .3in

{\Large \bf 

  Revisiting Big-Bang Nucleosynthesis Constraints \\
  on Long-Lived Decaying Particles

}

\vskip .5in

{\large 
  Masahiro Kawasaki$^{(a,b)}$, Kazunori Kohri$^{(c,d)}$, 
  Takeo Moroi$^{(e,b)}$,\\ and Yoshitaro Takaesu$^{(e,f)}$
}

\vskip 0.5in

$^{(a)}$
{\em Institute for Cosmic Ray Research, The University of Tokyo, Kashiwa
277-8582, Japan}

\vspace{2mm}

$^{(b)}$
{\em Kavli IPMU (WPI), UTIAS, The University of Tokyo, Kashiwa 277-8583, Japan}

\vspace{2mm}

$^{(c)}$
{\em Theory Center, IPNS, KEK, Tsukuba 305-0801, Japan}

\vspace{2mm}

$^{(d)}$
{\em 
The Graduate University of Advanced Studies,
Tsukuba 305-0801, Japan}

\vspace{2mm}


\vspace{2mm}

$^{(e)}$
{\em Department of Physics, The University of Tokyo,
Tokyo 113-0033, Japan}

\vspace{2mm}

$^{(f)}$
{\em Research Institute for Interdisciplinary Science, Okayama University,\\
Okayama 700-8530, Japan}

\end{center}
\vskip .5in

\begin{abstract}
  We study effects of long-lived massive particles, which decay during
  the big-bang nucleosynthesis (BBN) epoch, on the primordial
  abundances of light elements.  Compared to the previous studies, (i)
  the reaction rates of the standard BBN reactions are updated, (ii)
  the most recent observational data of the light element abundances
  and cosmological parameters are used, (iii) the effects of the
  interconversion of energetic nucleons at the time of inelastic
  scatterings with background nuclei are considered, and (iv) the
  effects of the hadronic shower induced by energetic high energy
  anti-nucleons are included.  We compare the theoretical predictions
  on the primordial abundances of light elements with the latest
  observational constraints, and derive upper bounds on the relic
  abundance of the decaying particle as a function of its lifetime.
  We also apply our analysis to unstable gravitino, the superpartner
  of graviton in supersymmetric theories, and obtain constraints on
  the reheating temperature after inflation.
\end{abstract}

\end{titlepage}

\setcounter{page}{1}
\renewcommand{\thefootnote}{\#\arabic{footnote}}
\setcounter{footnote}{0}

\section{Introduction}
\label{sec:intro}
\setcounter{equation}{0}

The big-bang nucleosynthesis (BBN) is one of the most important
predictions of big-bang cosmology.  At the cosmic temperature around
$0.1\ {\rm MeV}$, the typical energy of the cosmic microwave
background (CMB) photon becomes sufficiently lower than the binding
energies of the light elements (deuterium D, $^3$He, $^4$He, and so
on) so that the light elements can be synthesized with avoiding the
dissociation due to the scattering with background photons.  The cross
sections for nuclear reactions governing the BBN are well understood
so that a precise theoretical calculation of the primordial light
element abundances is possible with the help of numerical calculation.
In addition, the primordial abundances of the light elements are well
extracted from astrophysical observations.  Comparing theoretical
predictions with observational constraints, a detailed test of the BBN
is now possible; currently the predicted values of D and $^4$He in the
standard BBN (SBBN) reasonably agree with observations.

It has been well recognized that, if there exists new physics beyond
the standard model which may induce non-standard BBN reactions, the
predictions of the SBBN change.\footnote
{For earlier studies, see Refs.\ \cite{Lindley:1984bg, Khlopov:1984pf,
    Balestra:1984xxx, Ellis:1984eq, Juszkiewicz:1985gg, Ellis:1984er,
    Audouze:1985be, Lindley:1986wt, Kawasaki:1986my, Scherrer:1988xxx,
    Ellis:1990nb, Khlopov:1993ye}.}
In particular, with long-lived unstable particles decaying into
electromagnetic \cite{Kawasaki:1994af, Kawasaki:1994sc,
  Holtmann:1998gd, Jedamzik:1999di, Kawasaki:2000qr, Cyburt:2002uv,
  Kusakabe:2008kf, Poulin:2015opa} or hadronic \cite{Reno:1987qw,
  Dimopoulos:1987fz, Dimopoulos:1988zz, Dimopoulos:1988ue,
  Kohri:2001jx, Jedamzik:2004er, Kawasaki:2004yh, Kawasaki:2004qu,
  Kohri:2005wn, Jedamzik:2006xz, Kawasaki:2008qe, Cyburt:2009pg}
particles, the light element abundances are affected by
photodissociation, hadrodissociation, and $p\leftrightarrow n$
conversion processes.  In order not to spoil the agreement between
theoretical predictions and observational constraints, upper bounds on
the primordial abundances of the unstable particles are obtained.
Such constraints have been intensively studied in the past.
Remarkably, the BBN constraints may shed light on
beyond-the-standard-model particles to which collider studies cannot
impose constraints.  One important example is gravitino, which is the
superpartner of graviton in supersymmetric (SUSY) models
\cite{Weinberg:1982zq}.  Gravitino is very weakly interacting, and is
produced by the scattering processes of particles in the thermal bath
after inflation; the primordial abundance of gravitino is
approximately proportional to the reheating temperature after
inflation.  Consequently, with the reheating temperature being fixed,
we may acquire a constraint on the gravitino mass assuming that
gravitino is unstable.  Such a constraint can be converted to the
upper bound on the reheating temperature after inflation, which
provides important information in studying cosmology based on SUSY
models.

The purpose of this paper is to revisit the BBN constraints on
long-lived particles, which we call $X$, taking into account recent
progresses in theoretical and observational studies of the primordial
abundances of the light elements.  Theoretically, the understandings
of the cross sections of the SBBN reactions have been improved, which
results in smaller uncertainties in the theoretical calculations of
the light element abundances.  In addition, the observational
constraints on the primordial abundances of the light elements have
been updated.  These affect the BBN constraints on the primordial
abundances of long-lived particles.

In this paper, we study the BBN constraints on the primordial
abundance of long-lived exotic particles, which we parameterize by
using the so-called yield variable:
\begin{align}
  Y_X \equiv \left[ \frac{n_X}{s} \right]_{t\ll\tau_X},
\end{align}
where $n_X$ is the number density of $X$, $s$ is the entropy density,
and the quantity is evaluated at the cosmic time much earlier than the
lifetime of $X$ (denoted as $\tau_X$).  We take into account
theoretical and observational progresses.  In particular, compared to
the previous studies:
\begin{itemize}
\item The reaction rates of the SBBN reactions are updated.
\item The most recent observational constraints on the primordial
  abundances of the light elements are adopted.
\item The calculation of the evolution of the hadronic showers induced
  by energetic nucleons from the decay is improved.
\item We include the effect of hadronic shower induced by energetic
  anti-nucleons from the decay.\footnote{For annihilating massive
    particles, we had partly considered the effects induced by
    anti-nucleons emitted from the
    annihilation~\cite{Kawasaki:2015yya}.}
\end{itemize}
We consider various decay modes of long-lived particles and derive
upper bounds on their abundances.  We also apply our analysis to the
study of the effects of unstable gravitino on the light element
abundances.  In the study, we adopt several patterns of mass
spectra of superparticles (i.e., squarks, sleptons, gauginos, and
Higgsinos) suggested by viable SUSY models, based on which the partial
decay rates of gravitino are calculated.

The organization of this paper is as follows.  In Section
\ref{sec:obs}, we summarize the observational constraints on the light
element abundances we adopt in our analysis.  In Section
\ref{sec:decay}, we explain how the theoretical calculation of the
light element abundances is performed with taking into account the
effects of the decay of long-lived particles.  Upper bounds on the
primordial abundances of generic decaying particles are given in
Section \ref{sec:const_generic}.  Then, our analysis is applied to the
case of unstable gravitino in Section \ref{sec:gravitino}.  Section
\ref{sec:conclusions} is devoted for conclusions and discussion.

\section{Observed Abundances of the Light Elements}
\label{sec:obs}
\setcounter{equation}{0}

We first summarize the current observational constraints on the
primordial abundances of the light elements D, $^4$He, $^3$He and
$^7$Li.  In the following (A/B) denotes the ratio of number densities
of light elements A and B, and the subscript ``$p$'' indicates the
primordial value.

\begin{itemize}
\item \textbf{D}

  The primordial abundance of D is inferred from D absorption in
  damped Ly$\alpha$ systems (DLAs).  Most recently Cook {\it et
    al.}~\cite{Cooke:2013cba} measured D/H by observing a DLA toward
  QSO SDSS J1358+6522.  Moreover they reanalyzed four previously
  observed DLAs and from the total five DLA data they obtained the
  primordial D abundance as
\begin{equation} 
  ({\rm D}/{\rm H})_p = (2.53 \pm 0.04) \times 10^{-5}.
  \label{eq:obs_const_D}
\end{equation}
The quoted error is much smaller (by a factor of $\sim 5$) than those
obtained in the previous study.  The improvement of the D measurement
is a main reason why we obtain more stringent BBN constraints than
those in the previous work~\cite{Kawasaki:2004qu} as seen in later
sections.

\item \textbf{$^3$He}

  The $^3$He abundance is measured in protosolar objects.  As
  described in the previous work~\cite{Kawasaki:2004qu} we use the
  ratio $^3$He/D as observational constraint instead of using
  $^3$He/H.  This is because chemical evolution can increase or
  decreases the $^3$He abundance and it is difficult to infer the
  primordial value for $^3$He.  On the other hand the D abundance
  always decrease in chemical evolution and D is more fragile than
  $^3$He.  Consequently, the ratio $^3$He/D increases monotonically
  with time, which allows us to use the measured $^3$He/D as an upper
  bound on the primordial value~\cite{Sigl:1995kk}.  From $^3$He
  abundances observed in protosolar clouds~\cite{GG03}, we adopt
\begin{equation}
  ({\rm ^3He}/{\rm D})_p < 0.83 + 0.27.
  \label{eq:obs_const_He3D}
\end{equation}

\item \textbf{$^4$He}

  The primordial mass fraction of $^4$He, $Y_p$, is inferred from
  measurement of recombination lines of HeII (and HII) emitted from
  extra-galactic HII regions.  Izotov, Thuan and
  Guseva~\cite{Izotov:2014fga} reported a new determination of $Y_p$
  with the use of the infrared as well as visible $^4$He emission
  lines in 45 extragalactic HII regions.  Their result is
\begin{equation}
   Y_p = 0.2551 \pm 0.0022.
   \label{eq:obs_const_He4_ITG}
\end{equation}
After Ref.~\cite{Izotov:2014fga} was reported, Aver, Olive and
Skillman~\cite{Aver:2015iza} reanalyzed the data of Ref.\
\cite{Izotov:2014fga}.  They estimated the $^4$He abundance and its
error by using Markov chain Monte Carlo (MC) analysis and obtained
\begin{equation}
  Y_p = 0.2449 \pm 0.0040.
  \label{eq:obs_const_He4_AOS}
\end{equation}
Thus, the two values are inconsistent, and the discrepancy is more
than a $2\sigma$ level.  If we adopt the baryon-to-photon ratio $\eta$
determined by Planck, the BBN prediction for $Y_p$ is well consistent
with Eq.~(\ref{eq:obs_const_He4_AOS}) but not with
Eq.~(\ref{eq:obs_const_He4_ITG}).  For this reason we adopt the value
given in Eq.~(\ref{eq:obs_const_He4_AOS}) as a constraint on $Y_p$.
We will also show how constraints change if we adopt Eq.\
(\ref{eq:obs_const_He4_ITG}).

\item \textbf{$^7$Li (and $^6$Li)}

  The primordial abundance of $^7$Li was determined by measurement
  of $^7$Li in atmospheres of old metal-poor stars.  The observed
  $^7$Li abundances in stars with $[\text{Fe/H}] =-(2.5-3)$ showed
  almost a constant value ($\log_{10}(^7{\rm Li}/{\rm H}) \simeq
  -9.8$) called Spite plateau which was considered as
  primordial.\footnote
  { $[\text{Fe/H}] \equiv \log_{10}(\text{Fe/H}) -
    \log_{10}(\text{Fe/H})_\odot$, where $\odot$ indicates the solar
  abundance.  }
However, the plateau value turns out to be smaller than the standard
BBN prediction by a factor of nearly 3.  In fact,
Ref.~\cite{Sbordone:2010zi} reported the plateau value
$\log_{10}(^7{\rm Li}/{\rm H}) = -9.801\pm 0.086$ while the BBN
prediction is $\log_{10}(^7{\rm Li}/{\rm H}) = -9.35\pm 0.06$ for the
central value of $\eta$ suggested from the CMB data~\cite{Ade:2015xua}
(see, Section \ref{sec:decay}). This discrepancy is called the lithium
problem.  Moreover, the recent observation shows much smaller $^7$Li
abundances ($\log_{10}(^7{\rm Li}/{\rm H}) <-10$) for metal-poor stars
with metalicity below $[\text{Fe/H}] \sim -3$ \cite{Sbordone:2010zi}.
Thus, the situation of the $^7$Li observation is now controversial.
Since we do not know any mechanism to make $^7$Li abundances small in
such metal poor stars, we do not use $^7$Li as a constraint in this
paper.  We do not use $^6$Li either because $^6$Li abundance is
observed as the ratio to the number density of $^7$Li.

\end{itemize}

\section{BBN with Decaying Particles}
\label{sec:decay}
\setcounter{equation}{0}

In this section, we explain how we calculate the light element
abundances taking into account the effects of decaying massive
particles.  Our procedure is based on that developed in Refs.\
\cite{Kawasaki:2004yh, Kawasaki:2004qu} with several modifications
which will be explained in the following subsections.

\subsection{Overview}

The BBN constraints strongly depend on how $X$ decays.  In order to
make our discussion simple, we first concentrate on the case where $X$
decays only into a particle and its anti-particle. The subsequent
decays of the daughter particles from $X$ decay as well as the
hadronization processes of colored particles (if emitted) are studied
by using the PYTHIA 8.2 package \cite{Sjostrand:2014zea}.  We note
here that, even if $X$ primarily decays into a pair of
electromagnetically interacting particles (like $e^+e^-$ or
$\gamma\gamma$ pair), colored particles are also emitted by the
final-state radiation; such effects are included in our study.

\begin{figure}[t]
  \begin{center}
    \includegraphics[width=0.65\textwidth]{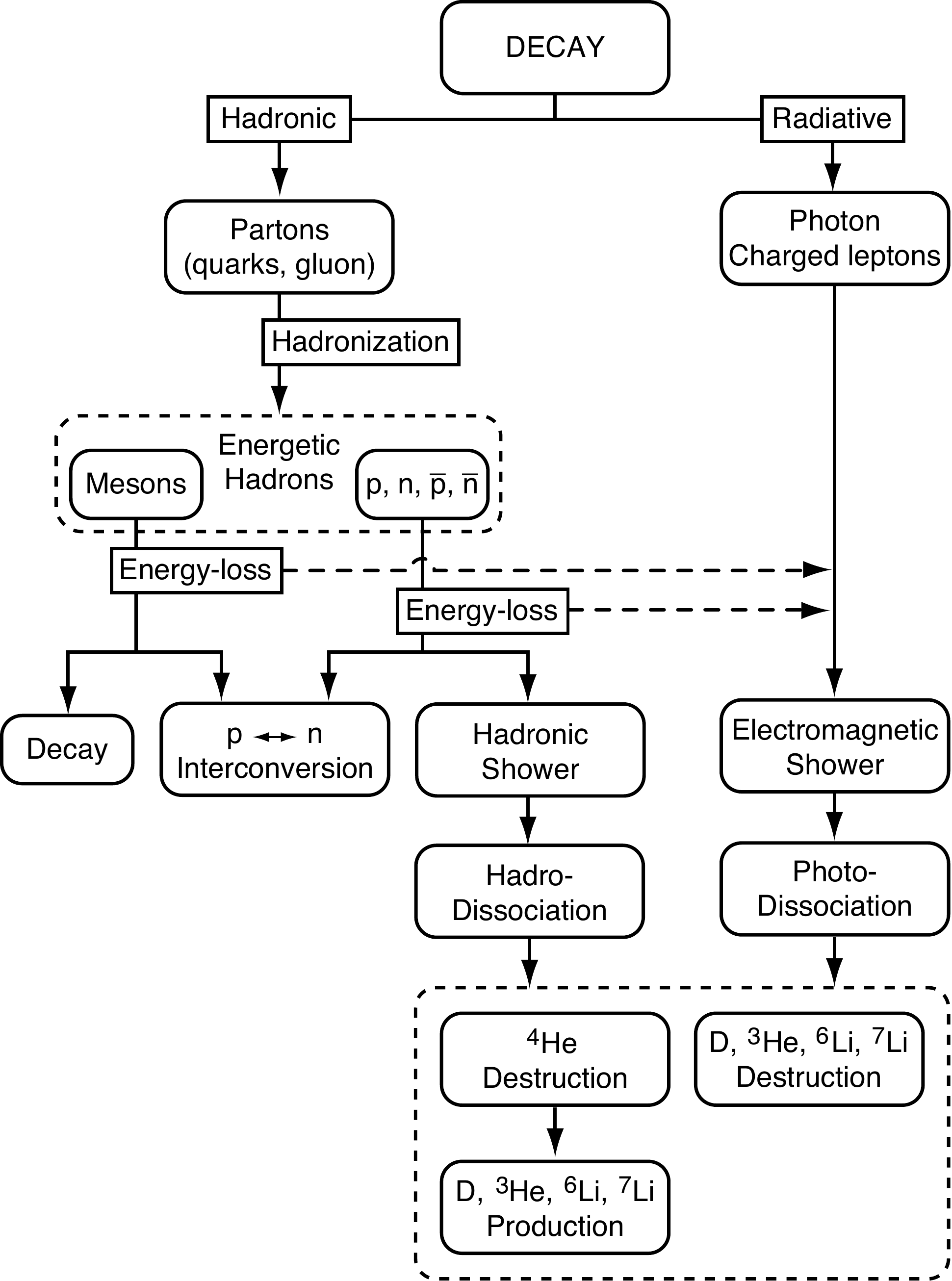}
  \end{center}
  \caption{ Flow-chart of the effects of the decay of massive
    particles. New effects induced by ``energetic'' anti-nucleons
    ($\bar{n}$ and $\bar{p}$) are included in this study.  }
\label{fig:scheme2017}
\end{figure}

There are two categories of the decay processes in studying the
effects of long-lived particles on the BBN:
\begin{itemize}
\item \textbf{Radiative decay}

  With the decay of the long-lived particles, high-energy photons and
  electromagnetically-charged particles are emitted.
\item \textbf{Hadronic decay}

  High-energy colored particles (i.e., quarks or gluons) are emitted
  in the decay.
\end{itemize}
We first overview them in the following.  (A flow-chart of the effects
induced by the decay of $X$ is given in Fig.\ \ref{fig:scheme2017}.)

\subsubsection{Radiative decay modes}

Energy injections by energetic electromagnetic particles induce
electromagnetic showers through their scatterings off the background
photons and electrons \cite{Ellis:1990nb, Kawasaki:1994af,
  Kawasaki:1994sc, Holtmann:1998gd, Jedamzik:1999di, Kawasaki:2000qr,
  Cyburt:2002uv, Jedamzik:2004er,Jedamzik:2006xz,Cyburt:2009pg}.  The
electromagnetic particles include $\gamma$ and $e^\pm$ as well as
charged hadrons.  Energetic photons in the shower can destroy the
light elements, in particular, D and $^4$He, produced by the SBBN
reactions.  (Such processes are called photodissociations).  The
photon spectrum in the electromagnetic shower is determined by the
total amount of the visible energy injected by the decay and the
temperature of the background thermal bath \cite{Ellis:1990nb,
  Kawasaki:1994af, Kawasaki:1994sc}; in particular, at a high
precision, the normalization of the photon spectrum is proportional to
the total amount of the energy injection.  In our numerical
calculation, the total visible energy, which is the sum of the
energies of photon and charged particles after the hadronization, is
calculated, based on which the normalization of the photon spectrum is
determined.
 
The photodissociation processes become effective when the threshold
energy for the $e^+e^-$ pair creation by the scattering of high energy
photons and background photons, which is approximately given by
$E^{(\gamma)}_{\rm th}\sim m_e^2/22T$ (with $m_e$ and $T$ being the
electron mass and the cosmic temperature, respectively)
\cite{Kawasaki:1994af}, is larger than the threshold energy for the
dissociation processes of the light elements.  For the
photodissociation of $^4$He, this is the case at the cosmic
temperature lower than $\sim 1\ {\rm keV}$ (which corresponds to the
cosmic time $t \gtrsim 10^6$~sec).  With the photodissociation of
$^4$He, both $^3$He and D are copiously produced. Thus, for the
long-lived particle with the lifetime longer than $\sim 10^6$~sec, the
D and $^3$He abundances give stringent constraints on the
radiatively-decaying modes.  The photodissociation of D becomes
effective at higher temperature because of the smallness of its
binding energy; the photodissociation of D may be important for
long-lived particle with lifetime longer than $\sim 10^4\ {\rm sec}$,
for which significant destruction of D takes
place~\cite{Holtmann:1998gd, Kawasaki:2000qr}. Another effect of the
electromagnetic shower on BBN is non-thermal production of $^6$Li.
The energetic T and $^3$He are produced through photodissociation of
$^4$He and they scatter off the background $^4$He and synthesize
$^6$Li. The productions of $^7$Li and $^7$Be due to energetic $^4$He
scattered by energetic photons through inelastic $\gamma$ + $^4$He are
negligible.

\subsubsection{Hadronic decay modes}

\begin{figure}[t]
  \begin{center}
    \includegraphics[width=0.95\textwidth]{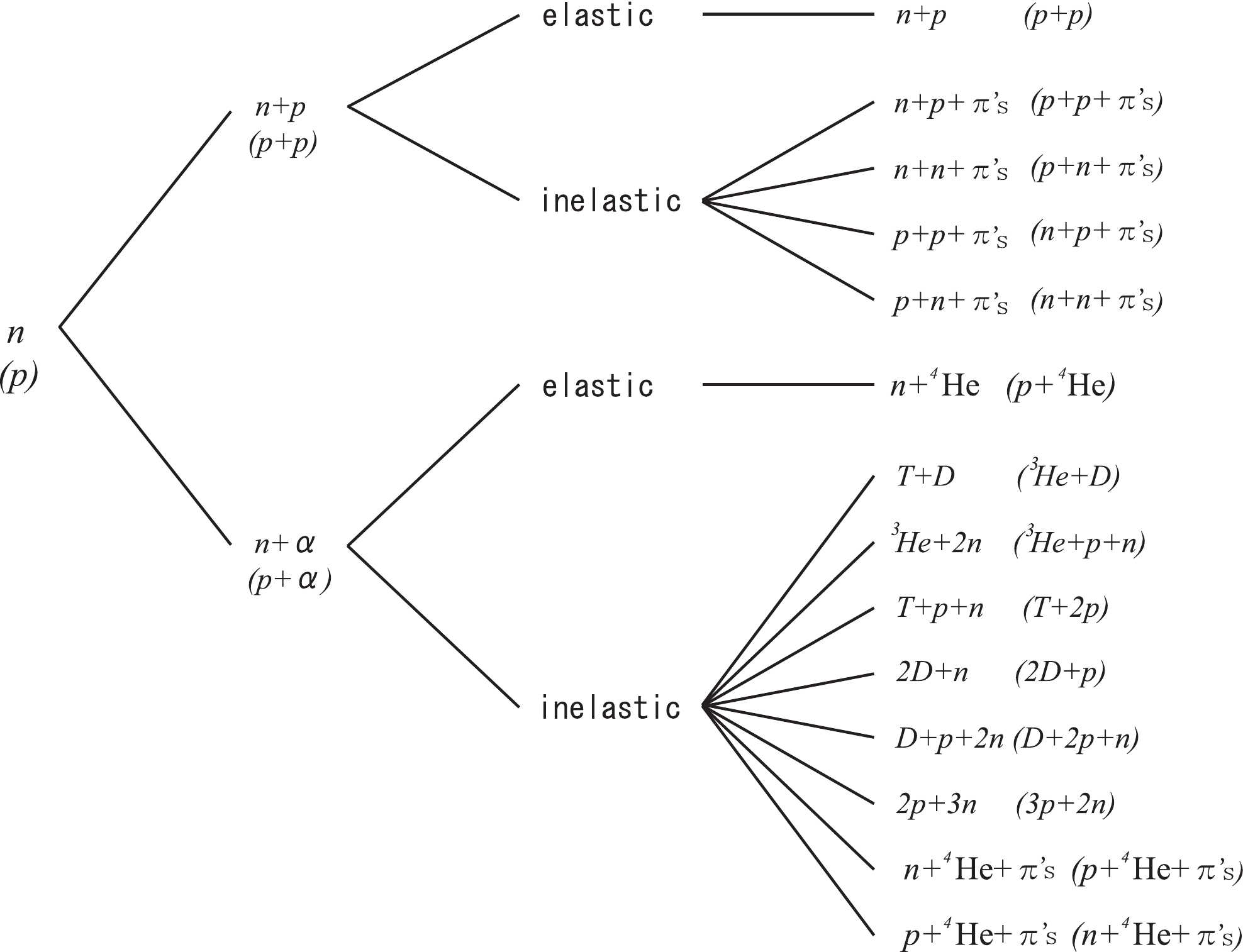}
  \end{center}
  \caption{Schematic picture of hadronic shower induced by a high
    energy projectile $n$ (or $p$) which scatters off the background
    proton or the background $^4$He. (Here $T$ denotes tritium.)
}
\label{fig:hadron-showers2017}
\end{figure}

\begin{figure}[t]
  \begin{center}
    \includegraphics[width=0.95\textwidth]{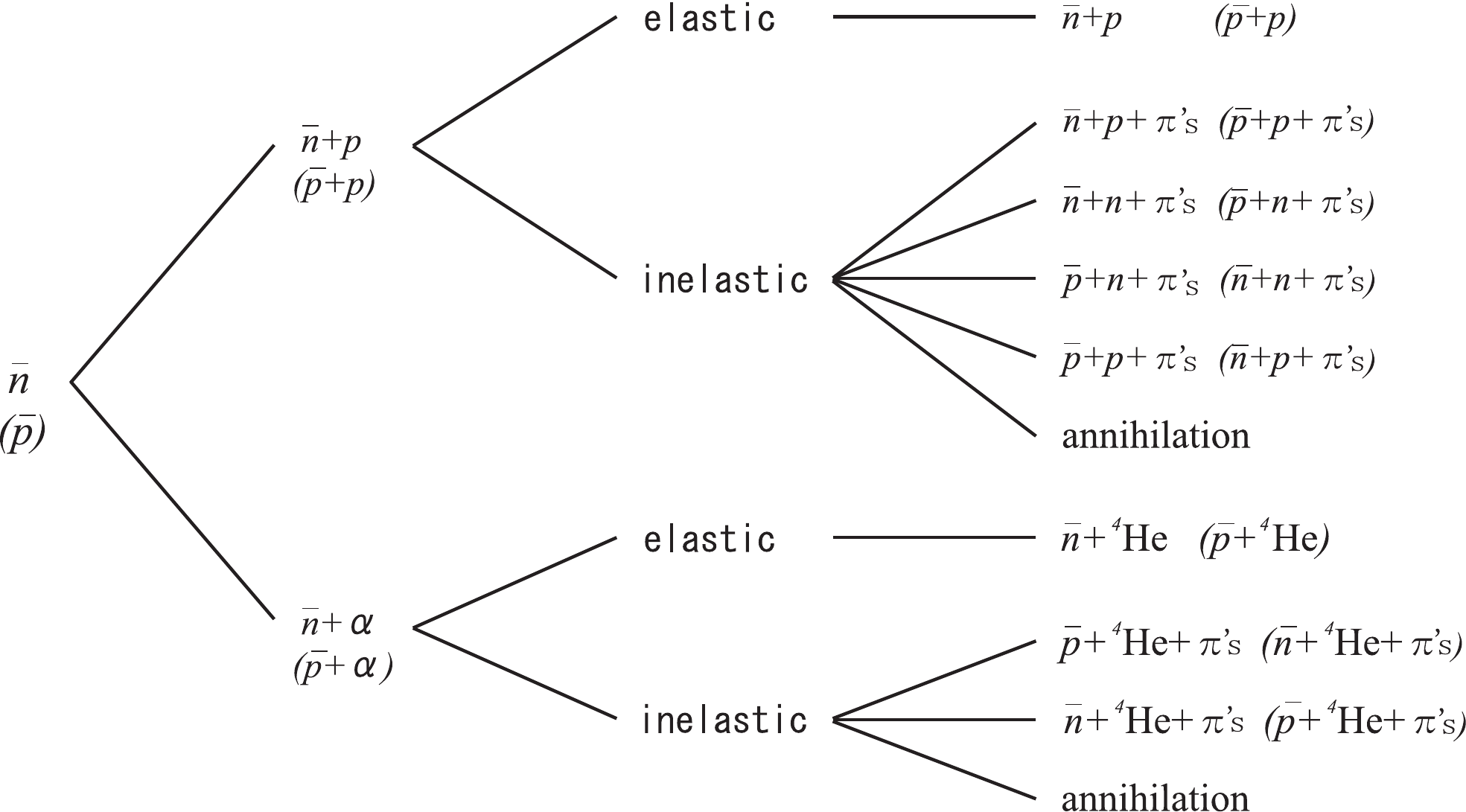}
  \end{center}
  \caption{ Same as Fig.~\ref{fig:hadron-showers2017}, but for a high
    energy projectile $\bar{n}$ (or $\bar{p}$).  }
\label{fig:hadron-showers2017part2}
\end{figure}

In hadronic decays, the emitted colored particles fragment into
hadrons such as pions, kaons, nucleons (i.e., neutron $n$ and proton
$p$), and anti-nucleons (i.e., anti-neutron $\bar{n}$ and anti-proton
$\bar{p}$).  Hereafter, $N=p$ or $n$ ($\bar{N}=\bar{p}$ or $\bar{n}$)
denotes the nucleon (anti-nucleon).  Energetic hadrons, in particular,
the nucleons, induce hadronic showers and hadrodissociation processes.
(The effect of the Lorentz suppression of the decay rate is included
for energetic unstable hadrons.)  In addition, even after being
stopped, some of the hadrons (in particular, charged pions and
nucleons) change the neutron-to-proton ratio in the background plasma,
which affects the $^4{\rm He}$ and D abundances.  The most important
effects of the hadronic decay modes are summarized as follows:
\begin{itemize}
\item When a massive particle decays into hadrons at $T \gtrsim {\cal
    O}(0.1)\ {\rm MeV}$ (i.e., when the lifetime is shorter than $\sim
  {\cal O}(10^2)\ {\rm sec}$), the high-energy hadrons are stopped in the
  thermal plasma \cite{Kawasaki:2004qu}.  The extra pions and nucleons
  affect the neutron-to-proton ratio after the neutron freeze-out by
  interchanging background $p$ and $n$ through the
  strongly-interacting interconversion processes like
  \begin{align}
     \pi^+ + n ~&\longrightarrow ~p + \pi^0 ,\\
     \pi^- + p ~&\longrightarrow ~n + \pi^0 .
  \end{align}
  There also exist similar interconversion processes caused by the
  injected nucleons.  The neutron-to-proton ratio $n/p$ increases due
  to the strongly-interacting conversions, resulting in the increase
  of the abundances of $^4$He and D~\cite{Reno:1987qw,Kohri:2001jx}.
\item If the background temperature at the time of the decay is low
  enough, the emitted hadrons cannot be stopped in the plasma; this is
  the case for energetic $n$ and $\bar{n}$ ($p$ and $\bar{p}$) if $T
  \lesssim 0.1\ {\rm MeV}$ ($T \lesssim 30\ {\rm keV}$)
  \cite{Kawasaki:2004qu}.  The energetic nucleons scatter off and
  destroy the background nuclei \cite{Dimopoulos:1987fz,
    Dimopoulos:1988zz, Dimopoulos:1988ue}.  The processes considered
  in this study are summarized in Figs.\ \ref{fig:hadron-showers2017}
  and \ref{fig:hadron-showers2017part2}.  In particular, through the
  destruction of $^4$He by the high-energy neutrons, an
  overproduction of D may occur, which leads to a stringent constraint on the
  primordial abundance of the hadronically decaying long-lived
  particles \cite{Kawasaki:2004yh, Kawasaki:2004qu, Kawasaki:2008qe,
    Kawasaki:2015yya}.  In addition, energetic $^3$He, $^4$He and T
  produce $^7$Li, $^7$Be and $^6$Li non-thermally through scattering
  off the background $^4$He.
\end{itemize}

\subsection{New implementations in the numerical calculation}

As mentioned before, our analysis is based on
Refs.~\cite{Kawasaki:2004yh, Kawasaki:2004qu}.
Here, we summarize the new implementations in the numerical
calculation added after Refs.~\cite{Kawasaki:2004yh, Kawasaki:2004qu,
  Kawasaki:2008qe}.
\begin{itemize}
\item We update the SBBN reaction rates and their uncertainties.
\item We revise the algorithm to calculate the evolution of the
  hadronic shower induced by the injections of energetic $p$ and $n$.
  In particular, we include the $p\leftrightarrow n$ interconversion
  of the energetic nucleons via inelastic scatterings.\footnote
  {The interconversions of energetic $p$ and $n$ should not be
    confused with the interconversions of background $p$ and $n$.  The
    former affects the stopping rate of the energetic nucleons which
    induces hadronic showers as well as hadrodissociation processes.
    On the contrary, the latter affects the neutron-to-proton ratio
    after the neutron freeze-out to which $Y_p$ is sensitive; the
    effects of the latter has been already taken into account in the
    previous analysis \cite{Kohri:2001jx, Kawasaki:2004yh,
      Kawasaki:2004qu, Kawasaki:2008qe}.}
  In the inelastic scattering processes of energetic nucleon (i.e.,
  $p$, $n$, $\bar{p}$, or $\bar{n}$) with the background nuclei, it is
  likely that the projectile nucleon carries away most of the energy
  while other final state particles are less energetic; the kinetic
  energies of the final-state particles other than the most energetic
  one are typically $K_T\sim 140\ {\rm MeV}$.  The conversion of the
  projectile nucleon via the inelastic scatterings affects the
  (effective) energy loss rate of the nucleon as we discuss below.

\item We newly include the effects of energetic $\bar{p}$ and
  $\bar{n}$ injections by the decay.
\end{itemize}

\subsubsection{SBBN reactions}

Compared to the previous studies
\cite{Kawasaki:2004qu,Kawasaki:2008qe}, we update the SBBN reaction
rates adopting those given in Ref.~\cite{Serpico:2004gx,
  Cyburt:2008up}, which are obtained by fitting relatively-new
experimental data.  In order to take into account uncertainties in the
reaction rates, we perform MC simulations to estimate error
propagations to the light element abundances.  For given values of the
lifetime and the primordial abundance of $X$, we perform $1,000$ runs
of the calculation of the light element abundances with assuming that
the reaction rates (as well as other parameters in the calculation)
are random Gaussian variables.  In each run, the reaction rate $R_i$
for $i$-th reaction is determined as
\begin{equation}
  R_i(T) = \bar{R}_{i}(T) + \sigma_i(T)\, y_i,
  \label{eq:rate_uncertainty}
\end{equation}
where $\bar{R}_{i}$ and $\sigma_i$ are the central value and the
standard deviation of the reaction $i$, and $y_i$ is the temperature
independent Gaussian variable with probability distribution
\begin{equation}
   P(y) =\frac{1}{\sqrt{2\pi}}\exp\left(-\frac{1}{2}y^2\right) .
\end{equation}
Notice that, as explicitly shown in Eq.~(\ref{eq:rate_uncertainty}),
some of uncertainties of reaction rates depend on the cosmic
temperature.  Throughout each run, such a reaction rate is determined
by adding the temperature-dependent uncertainty multiplied by the
temperature-independent random Gaussian variable to the central value.
If the upper and lower uncertainties of a reaction rate are
asymmetric, we take upward and downward fluctuations with equal
probability, assuming that they obey one-sided Gaussian distributions.

\begin{figure}
  \begin{center}
    \includegraphics[width=0.75\textwidth]{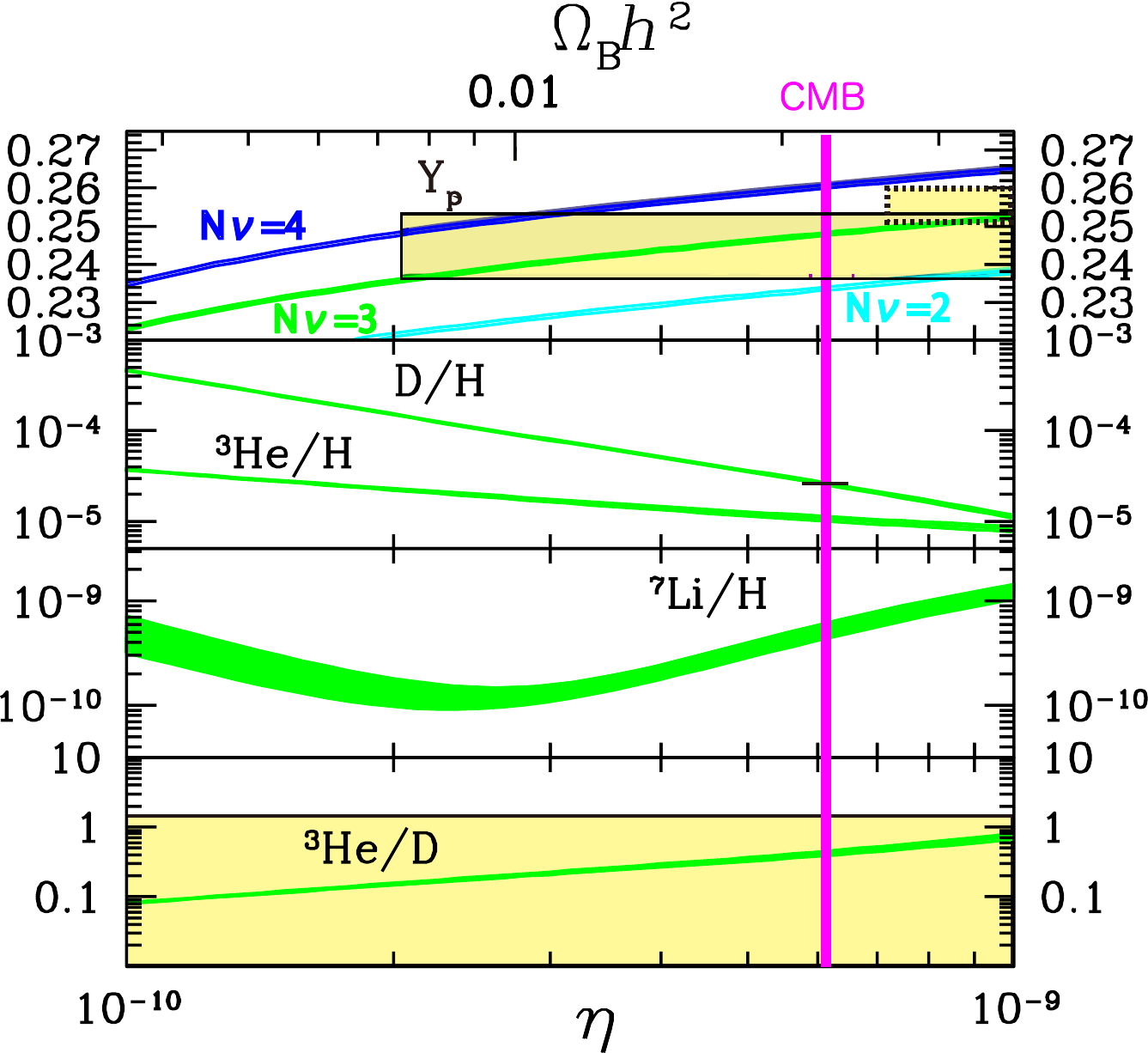}
  \end{center}
  \caption{Theoretical predictions of the light element abundances as
    functions of the baryon-to-photon ratio $\eta={n_{\rm
        B}}/{n_{\gamma}}$.  (The upper horizontal axis shows
    $\Omega_{\rm B}h^2$, where $\Omega_{\rm B}$ is the density
    parameter of baryon and $h$ is the Hubble constant in units of
    $100\ {\rm km/sec/Mpc}$.)  The vertical band shows $\eta = (6.11
    \pm 0.08) \times 10^{-10}$, which is the $2\sigma$ band of
    baryon-to-photon ratio suggested by the CMB observations
    \cite{Ade:2015xua}.  For $Y_p$, we plot results for effective
    number of neutrino species $N_{\nu} = 2$, $3$, and $4$ (from the
    bottom to the top).  The boxes indicate the observational values
    with $2\sigma$ uncertainties.  (For $Y_p$, the box surrounded by
    the solid and dashed lines correspond to Eq.\
    \eqref{eq:obs_const_He4_AOS} and Eq.\
    \eqref{eq:obs_const_He4_ITG}, respectively.)}
\label{fig:yield2017}
\end{figure}

In order to check the consistency between the observational
constraints summarized in Section \ref{sec:obs} and the SBBN
predictions, we compare them in Fig.~\ref{fig:yield2017}; we plot the
SBBN values of the light element abundances as functions of the baryon
to photon ratio $\eta={n_{\rm B}}/{n_{\gamma}}$.  From the top to the
bottom, we plot (i) the mass fraction of $^4$He ($Y_p$), (ii) the
deuterium to hydrogen ratio (D/H), (iii) the $^3$He to hydrogen ratio
($^3$He/H), (iv) the $^7$Li to hydrogen ratio ($^7$Li/H), and (v) the
helium 3 to deuterium ratio ($^3$He/D).  The theoretical predictions
show $2\sigma$ bands due to the uncertainties in experimental data of
cross sections and lifetimes of nuclei.  The boxes indicate the
$2\sigma$ observational constraints (see Section \ref{sec:obs}).  The
CMB observations provide independent information about the
baryon-to-photon ratio; we adopt
\begin{align}
  \eta = (6.11 \pm 0.04) \times 10^{-10}~~~(\mbox{68$\%$ C.L.}),
  \label{eta}
\end{align}
which is based on the TT,TE,EE+lowP+BAO analysis of the Planck
collaboration, $\Omega_{\rm B} h^2 = 0.02229^{+0.00029}_{-0.00027}$
(95$\%$ C.L.), where $\Omega_{\rm B}$ is the density parameter of
baryon and $h$ is the Hubble constant in units of $100\ {\rm
  km/sec/Mpc}$ \cite{Ade:2015xua}. In Fig.~\ref{fig:yield2017}, we
also show the CMB constraint on $\eta$ at $2\sigma$ (the vertical
band).  We can see that the SBBN predictions based on the
baryon-to-photon ratio suggested from the CMB observations are in
reasonable agreements with observations.

\subsubsection{High energy $n$ and $p$ injections}

The effects of the high energy $n$ and $p$ injected into the thermal
bath have been studied in Refs.\ \cite{Dimopoulos:1987fz,
  Dimopoulos:1988zz, Dimopoulos:1988ue,
  Kawasaki:2004yh,Kawasaki:2004qu}.  Compared to the previous studies,
we newly include the interconversion reactions between energetic $p$
and $n$ at the time of the inelastic scatterings accompanied by pion emissions
(see Fig.~\ref{fig:hadron-showers2017}).  Once an energetic $p$ or $n$
is injected into thermal plasma, it can be converted to $n$ or $p$
through inelastic scatterings off the background nuclei, i.e., $p$ or
$^4$He.  (Hereafter, the background $^4$He is denoted as $\alpha_{\rm
  BG}$.)  As we discuss below, such an interconversion affects the
(effective) energy loss rates of nucleons during their propagation in
thermal plasma, resulting in a change of the hadrodissociation rates.

The effects of the interconversions of energetic $p$ and $n$ are
important when the cosmic temperature becomes lower than $\sim 0.1\
{\rm MeV}$.  This is because the energetic nucleons are stopped in the
thermal plasma if the background temperature is high enough; as we
have mentioned, the neutron is likely to be stopped when $T\gtrsim
0.1\ {\rm MeV}$ while the proton is stopped when $T\gtrsim 30\ {\rm
  keV}$.  Once stopped, the nucleons do not induce hadrodissociation
processes.  An energetic neutron, which would have a sufficient time to
induce hadronic showers without the interconversion, does not cause hadronic showers when it is converted to proton.  Thus, the
inclusion of the energetic $p\leftrightarrow n$ conversion makes the
BBN constraints milder in particular when 30~keV~$\lesssim T \lesssim
$~0.1 MeV (i.e., $10^2$~sec~$\lesssim t \lesssim 10^3$~sec).  The
inelastic scatterings occur with the targets of $p$ and $\alpha_{\rm
  BG}$.  The interconversion rates are estimated as follows:
\begin{itemize}
\item Because of the lack of the experimental data of a cross section
  of each mode for the $n$ (projectile)+ $p$ (target) and $p$
  (projectile)+ $p$ (target) processes shown in
  Fig.~\ref{fig:hadron-showers2017}, we assume that all of the cross
  sections for those 8 modes ($n+p,\, p+p\rightarrow
  p+p+\pi\text{'s},\, p+n+\pi\text{'s},\, n+p+\pi\text{'s},\,
  n+n+\pi\text{'s}$) are equal.  (In the case of $p+p\rightarrow
  n+n+\pi\text{'s}$, at least two pions should be emitted and hence we
  consider the process only if it is kinematically allowed.)  In
  addition, we neglect the effects of the interconversions induced by
  the emitted pions from the inelastic scatterings. That is because
  those pions decay before they scatters off the background particles.
\item With inelastic $n+\alpha_{\rm BG}$ scatterings, the energetic
  $n$ can be again converted to $p$.  Because of the lack of
  experimental data, we assume that, when the inelastic processes with
  pion emissions are concerned, the cross section for $n+^4{\rm He}$
  scattering is equal to that for $p+^4{\rm He}$ scattering.  Since
  the rates of the inelastic scatterings $n+^4\text{He}\rightarrow
  n+^4\text{He}$ and $p+^4\text{He}$ are relatively
  small~\cite{Meyer:1972}, effects of the interconversion due to the
  inelastic $n+\alpha_{\rm BG}$ scatterings are unimportant and does
  not change the hadronic shower evolution.
\end{itemize}

We also comment on the non-thermal production processes of $^6$Li,
$^7$Li, and $^7$Be.  With the inelastic scatterings $N+\alpha_{\rm
  BG}\rightarrow\cdots$, the final state may contain energetic T,
$^3$He, and $^4$He.  They may scatter off $\alpha_{\rm BG}$ to produce
heavier elements, i.e., $^6$Li, $^7$Li, and $^7$Be
\cite{Dimopoulos:1987fz, Dimopoulos:1988zz,
  Dimopoulos:1988ue,Kawasaki:2004qu}.  Although we do not use $^6$Li
and $^7$Li to constrain the primordial abundance of $X$, these
non-thermal production processes are included in our numerical
calculation.  In particular, we include the processes $^4{\rm
  He}+\alpha_{\rm BG}\rightarrow ^6{\rm Li}+\cdots$, $^7{\rm
  Li}+\cdots$, and $^7{\rm Be}+\cdots$, whose effects were not taken
into account in Ref.~\cite{Kawasaki:2004qu}.  For the study of these
processes, energy distribution of $^4$He produced by the inelastic
hadronic scatterings is necessary.  We determine energy distribution
of the final-state $^4$He using the prescription given in Appendix C
of Ref.~\cite{Kawasaki:2004qu}.  Notice that, for the non-thermal
production processes of $^6$Li induced by T and $^3$He, the energy
distributions of T and $^3$He produced by the hadrodissociations of
$\alpha_{\rm BG}$ are obtained by fitting experimental data (see
Ref.~\cite{Kawasaki:2004qu}).

\subsubsection{Injections of energetic $\bar{n}$ or $\bar{p}$}

In the present study, we newly include the effects of the injections
of energetic anti-nucleons, $\bar{n}$ or $\bar{p}$.  We only consider
the scatterings of energetic anti-nucleons off the background $p$ and
$\alpha_{\rm BG}$, and the anti-nucleons are treated as sources of
hadronic showers.\footnote{On the other hand, for anti-nucleons
  stopped in the plasma, we have already explained their effects on
  the inter-conversion reactions between background $n$ and $p$ in
  previous subsections.} The high-energy nucleons in the hadronic
showers induced by $\bar{n}$ and $\bar{p}$ can destroy
$\alpha_{\rm BG}$ and further produce copious high-energy daughter
nuclei.  In Fig.~\ref{fig:hadron-showers2017part2}, we show a
schematic picture for the reactions induced by energetic
anti-nucleons.  Because of the lack of experimental data for the
processes including the anti-nucleons, we adopt several approximations
and assumptions on the reactions induced by the energetic
anti-nucleon.  We note here that, in adopting approximations or
assumptions, we require that the constraints become conservative in
order not to over-constrain the properties of long-lived particles.

First, let us consider the scattering of energetic anti-nucleons with
the background $p$.  We use the experimental data of the differential
cross sections for the $\bar{p}$-$p$ scatterings by referring
Ref.~\cite{Olive:2016xmw} for the total and elastic cross sections and
Ref.~\cite{Bracci:1973cg} for the annihilation cross sections.
Because of the lack of experimental data, we use the data for the
$\bar{p}$-$p$ scatterings to estimate the cross sections for other
processes.  (i) We assume that the differential cross sections for
$\bar{n}$-$p$ scatterings are the same as those of corresponding
$\bar{p}$-$p$ scatterings paired in Fig.\
\ref{fig:hadron-showers2017part2}.  We expect that this assumption is
reasonable because the Coulomb corrections to the cross sections are
estimated to be less than a few percent for the energy of our interest~\cite{Kawasaki:2004qu}.  (ii)
Through inelastic scatterings off the background $p$, the strongly-interacting
$p\leftrightarrow n$ interconversion reactions occur with emitting
pions.  Then, there are four possible combinations of final-state
anti-nucleon and scattered nucleon; the final-state anti-nucleon may
be $\bar{p}$ or $\bar{n}$, while the nucleon may be $p$ or $n$ (see
Fig.\ \ref{fig:hadron-showers2017part2}).  We assume that the
differential cross sections for these processes are equal and use the
data for the $\bar{p}$-$p$ scatterings for all of these processes
because the cross sections for some of the final states are unknown.  Here
one remark is that the process $\bar{n} \text{(projectile)}+ p
\text{(target)}\rightarrow \bar{p}+n$ needs at least two pion emission
so we take into account it if kinematically allowed.

Next, we discuss scatterings of energetic anti-nucleons off
$\alpha_{\rm BG}$.  Unfortunately, experimental data of the cross
sections for the scattering processes of anti-nucleon with $^4$He are
insufficient to perform a detailed study.  Therefore, we only consider
the energy-loss and interconversions of anti-nucleon induced by the
inelastic scatterings with $\alpha_{\rm BG}$.  Approximating that the
energetic anti-nucleon scatters off individual nucleons in $^4$He, the
cross sections for the scattering processes with $^4$He are taken to
be four times larger than those with the background nucleons.  Notice that
this assumption is reasonable when the energy of $\bar{N}$ is larger
than the binding energy of $^4$He.\footnote
{For example, see the plots shown in Ref.\ \cite{Olive:2016xmw} in
  which the cross section of the process $\bar{p} + {\rm D}$ is
  approximately twice larger than that of $\bar{p} + n$ for a
  high-energy $\bar{p}$. Similarly, it is also known that the cross
  sections of $\bar{p} +^4$He are approximately four times larger than
  that of $\bar{p} + n$ at high energies.}
In addition, for the scatterings induced by the anti-nucleon, we
neglect the subsequent reactions induced by the cascade products from
$\alpha_{\rm BG}$; in other words, we neglect the destruction of
$\alpha_{\rm BG}$ and the recoil energy of $\alpha_{\rm BG}$ (if it is
not destroyed).  These would give the most conservative bounds.

For annihilation reactions of anti-nucleon with background $p$ or
$\alpha_{\rm BG}$, we expect emissions of energetic hadrons, by which
the light element abundances should be affected. However, because we
do not have sufficient data of differential cross sections, we
conservatively neglect any effects after the annihilation of
anti-nucleon.  Thus, in our calculation, only the effect of the pair
annihilations of anti-nucleon is to reduce the number of energetic
anti-nucleon.

\subsubsection{${\xi}$ parameters}

In the present study, the effects of the hadrodissociation are
parameterized by the functions called $\xi_{A_i}$ (with ${A_i}$ being
$n$, D, T, $^3$He, $\alpha$, $^6$Li, $^7$Li and $^7$Be)
\cite{Dimopoulos:1987fz, Dimopoulos:1988zz, Dimopoulos:1988ue,
  Kawasaki:2004yh, Kawasaki:2004qu}.  For $A_i\neq\alpha$, $\xi_{A_i}$
is the total number of $A_i$ produced by the hadronic decay of one
parent particle $X$, while $\xi_\alpha$ is the total number of
$\alpha_{\rm BG}$ destroyed.  With the properties of the long-lived
particle $X$ being given, $\xi_{A_i}$ depend on three quantities:
the cosmic temperature $T$, the mass fraction of $^4$He, and the baryon-to-photon ratio $\eta$.

It is notable that $\xi_{A_i}$ is defined as the value just after the
shower evolutions. Notice that the timescale of the shower evolution
is much shorter than those of any nuclear reactions in
SBBN. Therefore, the face value of $\xi_{A_i}$ does not represent a
net increase or decrease of the nuclei $A_i$. For example, $^6$Li
produced by the non-thermal reactions may be further processed in the
SBBN reaction as $^6$Li + $p \to$ $^3$He + $^4$He.

In the case of our interest, the hadronic showers are triggered by the
injections of high-energy nucleons (i.e., $N=p$ and $n$) and
anti-nucleons (i.e., $\bar{N}=\bar{p}$ and $\bar{n}$) which originate
from the hadronizations of colored particles emitted by the decay of
long-lived particles.  $\xi_{A_i}$ are given by the convolutions of
two functions as
\begin{eqnarray}
  \label{eq:x}
   \xi_{A_i} (T,Y_p,\eta) = \sum_{N= n, \bar{n}, p, \bar{p}} 
  \int 
  \tilde{\xi}_{A_i,N}(E_N, T,Y_p,\eta)
  \frac{dn_N}{dE_N} dE_N,
\end{eqnarray}
where $(dn_N/dE_N)$ is the energy spectrum of the injected nucleon $N$
with energy $E_N$ per a single massive particle decay, while
$\tilde{\xi}_{A_i,N}$ is the number of the light element $A_i$
produced (or destroyed for $A_i=\alpha$) by the products of hadronic
shower induced by the injection of single $N$ with energy $E_N$.  We
calculate the functions $\tilde{\xi}_{A_i,N}$ based on the procedure
explained in Ref.\ \cite{Kawasaki:2004qu}.\footnote
{In the calculation of $\xi_{^6{\rm Li}}$ in
  Ref.~\cite{Kawasaki:2004qu}, there was an error in numerical
  calculation of the Coulomb stopping powers of energetic T and
  $^3$He, with which $\xi_{^6{\rm Li}}$ was maximized at the cosmic
  temperature of $\sim$ 40 -- 50 keV.  The error was corrected in
  subsequent studies, e.g., in Ref.~\cite{Hisano:2009rc}, and
  $\xi_{^6{\rm Li}}$ is peaked at the cosmic temperature of $T\sim 20\
  {\rm keV}$.  However, the error did not affect the resultant value
  of relic $^6{\rm Li}$ abundance because $^6{\rm Li}$, if it exists,
  is completely destructed by the process of $^6$Li + $p \to$ $^3$He +
  $^4$He.  We thank K. Jedamzik for his comments on this issue.}

\begin{figure}
  \begin{center}
    \includegraphics[height=0.3\textheight]{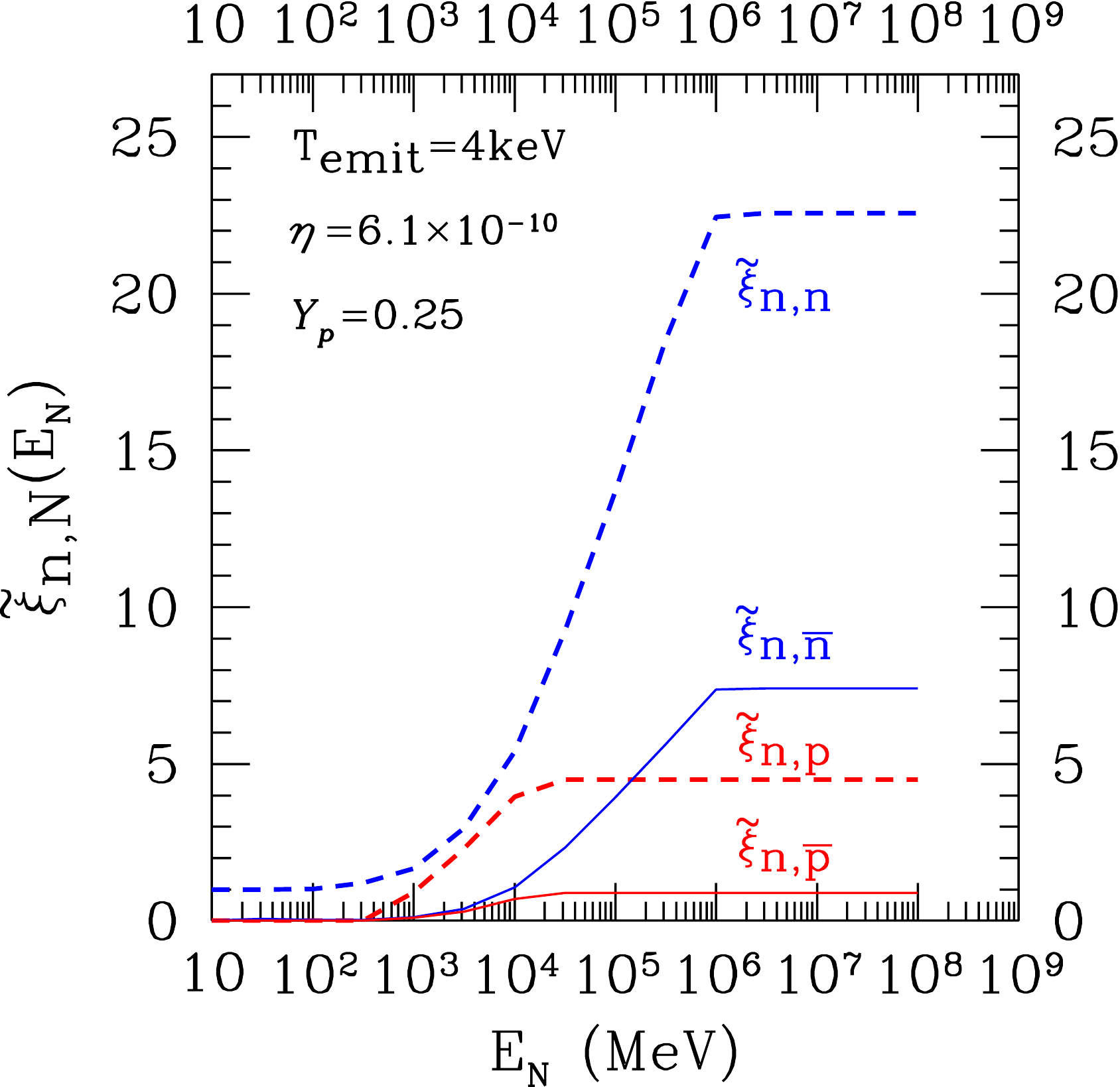}~~~
    \includegraphics[height=0.3\textheight]{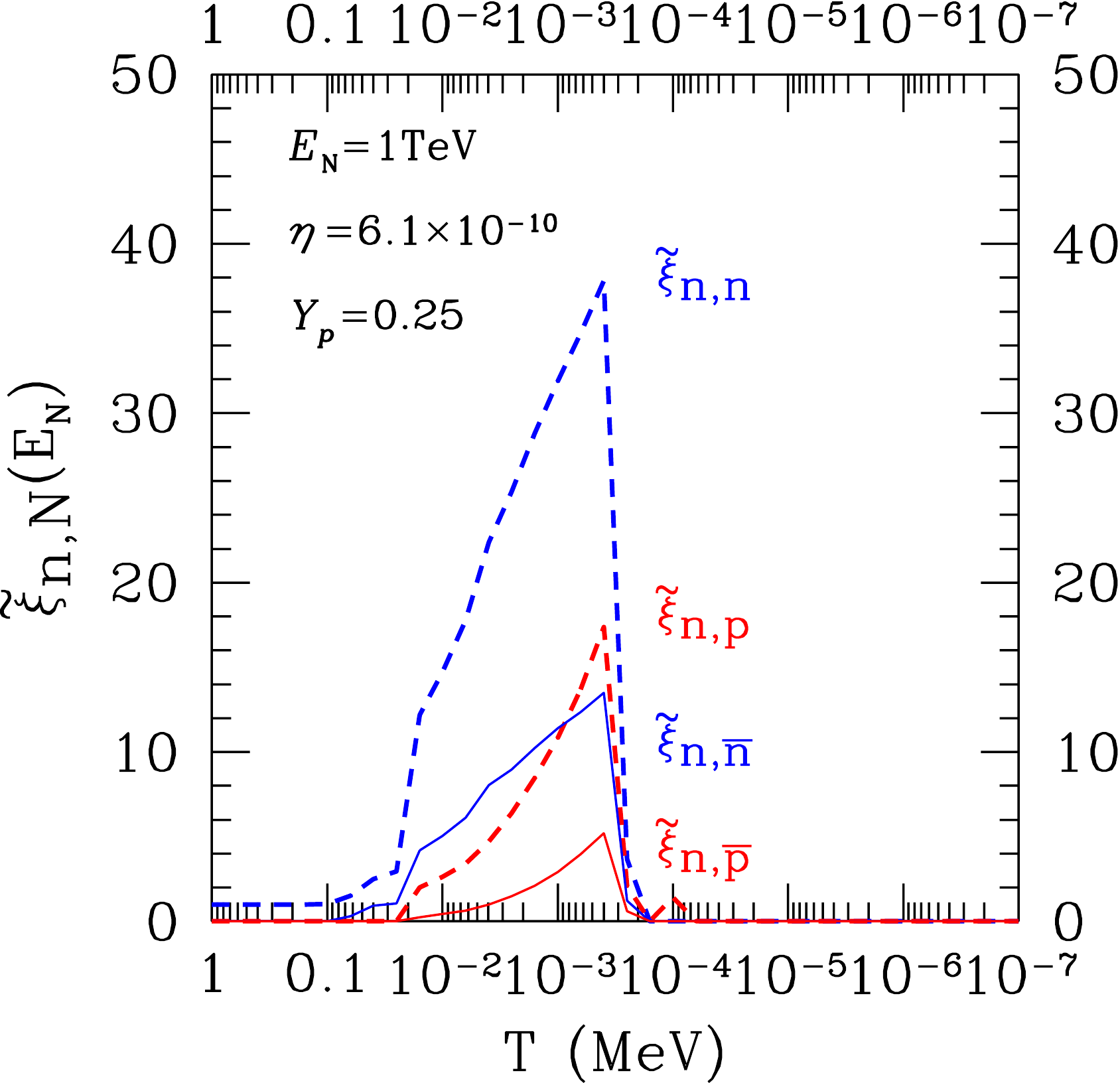}
  \end{center}
  \caption{$\tilde{\xi}_{n,N}$ for $T= 4$~keV as functions of the
    kinetic energy (left) and those for $E_N=1\ {\rm TeV}$ as functions
    of the cosmic temperature (right). Here, $N = n$, $\bar{n}$, $p$
    and $\bar{p}$, and we take $\eta = 6.1 \times 10^{-10}$ and $Y_p =
    0.25$.}
  \label{fig:xi_n}
  ~~\vspace{5mm}
  \begin{center}
    \includegraphics[height=0.3\textheight]{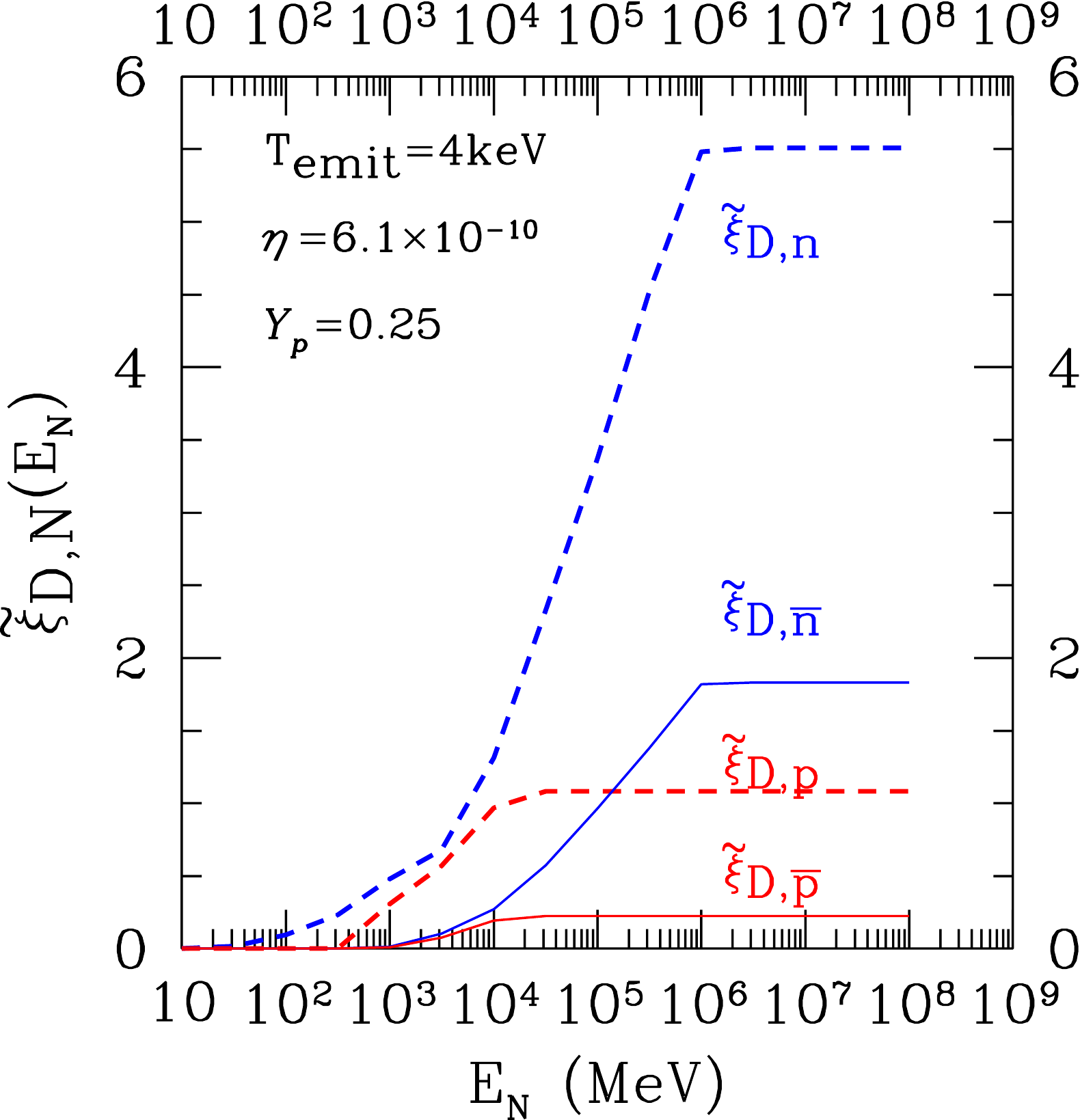}~~~
    \includegraphics[height=0.3\textheight]{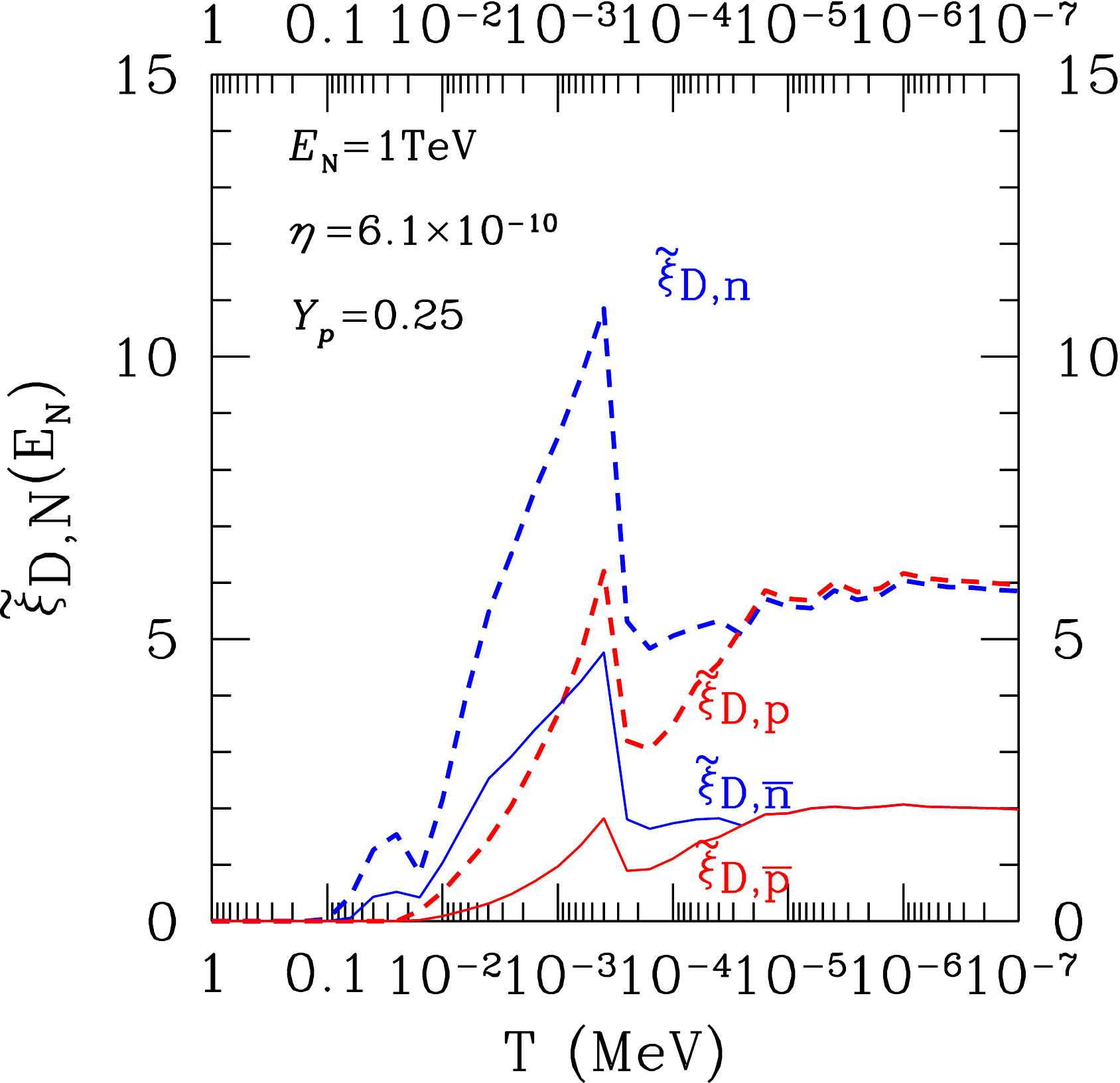}
  \end{center}
  \caption{$\tilde{\xi}_{{\rm D},N}$ for $T= 4$~keV as functions of the
    kinetic energy (left) and those for $E_N=1\ {\rm TeV}$ as functions
    of the cosmic temperature (right). Here, $N = n$, $\bar{n}$, $p$
    and $\bar{p}$, and we take $\eta = 6.1 \times 10^{-10}$ and $Y_p =
    0.25$.}
  \label{fig:xi_D}
\end{figure}

\begin{figure}
  \begin{center}
    \includegraphics[height=0.3\textheight]{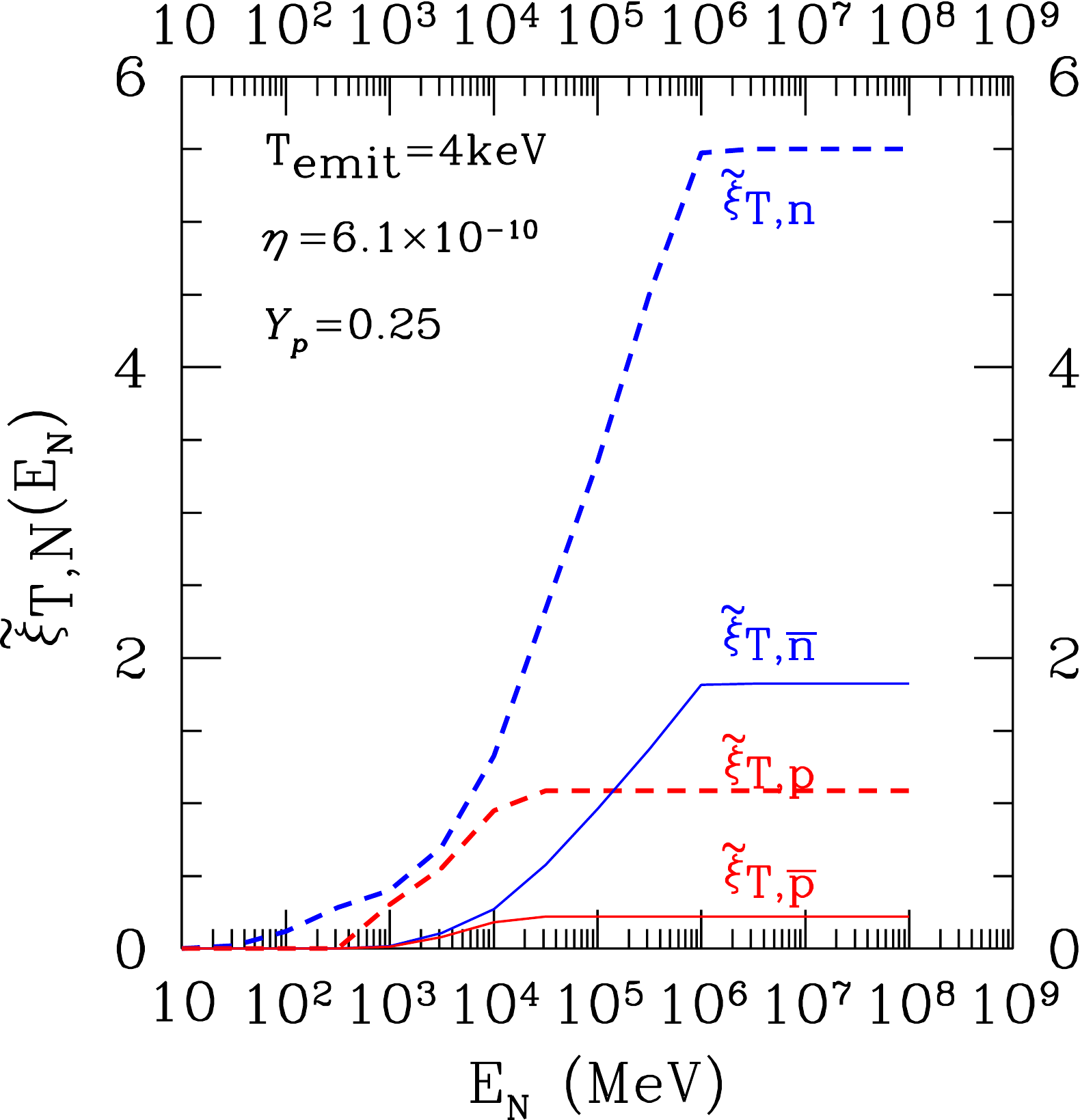}~~~
    \includegraphics[height=0.3\textheight]{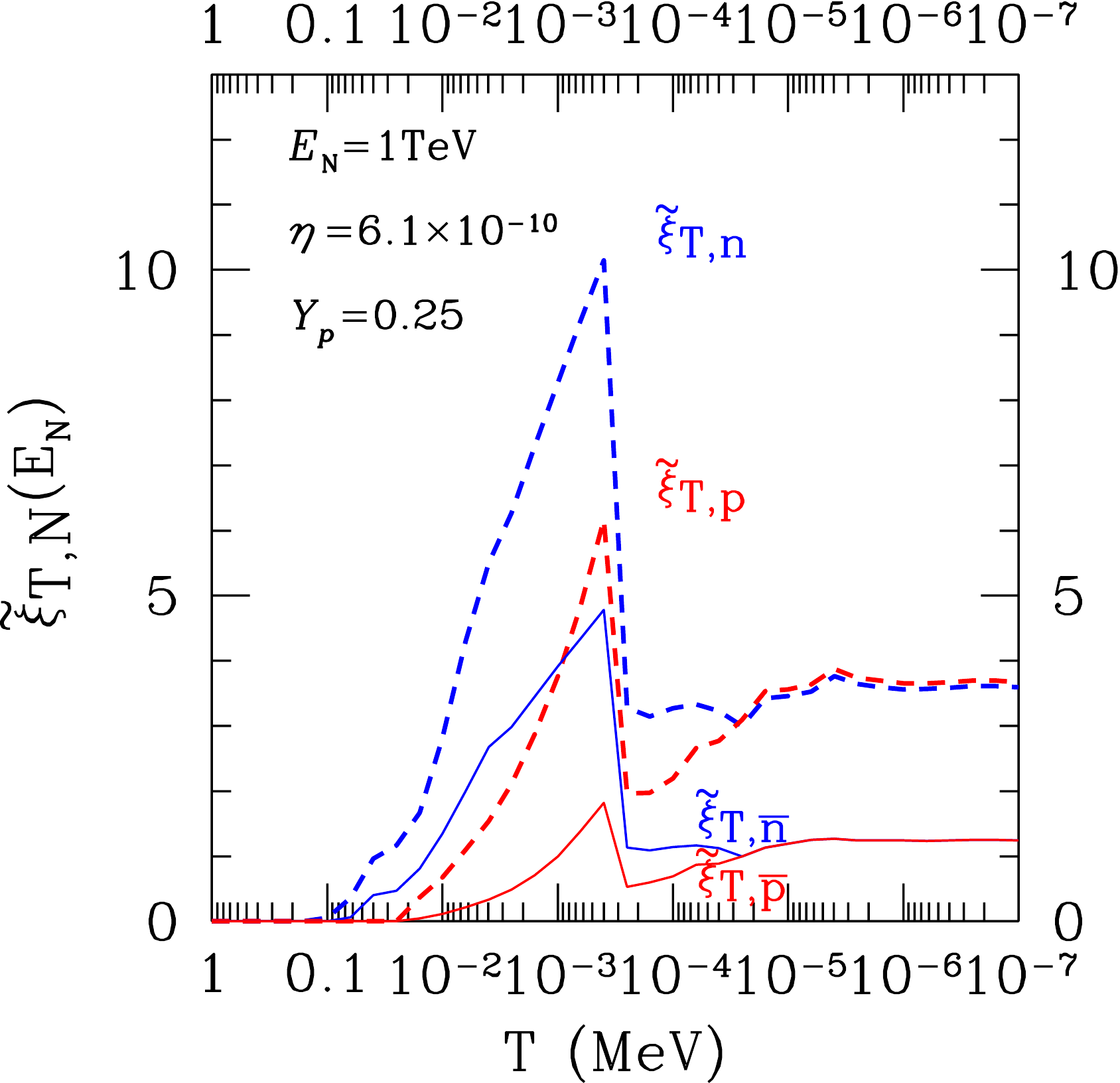}
  \end{center}
  \caption{$\tilde{\xi}_{{\rm T},N}$ for $T= 4$~keV as functions of
    the kinetic energy (left) and those for $E_N=1\ {\rm TeV}$ as
    functions of the cosmic temperature (right).  (In the subscript of
    $\tilde{\xi}_{{\rm T},N}$, T denotes tritium.)  Here, $N = n$,
    $\bar{n}$, $p$ and $\bar{p}$, and we take $\eta = 6.1 \times
    10^{-10}$ and $Y_p = 0.25$.}
  \label{fig:xi_T}
  ~~\vspace{5mm}
  \begin{center}
    \includegraphics[height=0.3\textheight]{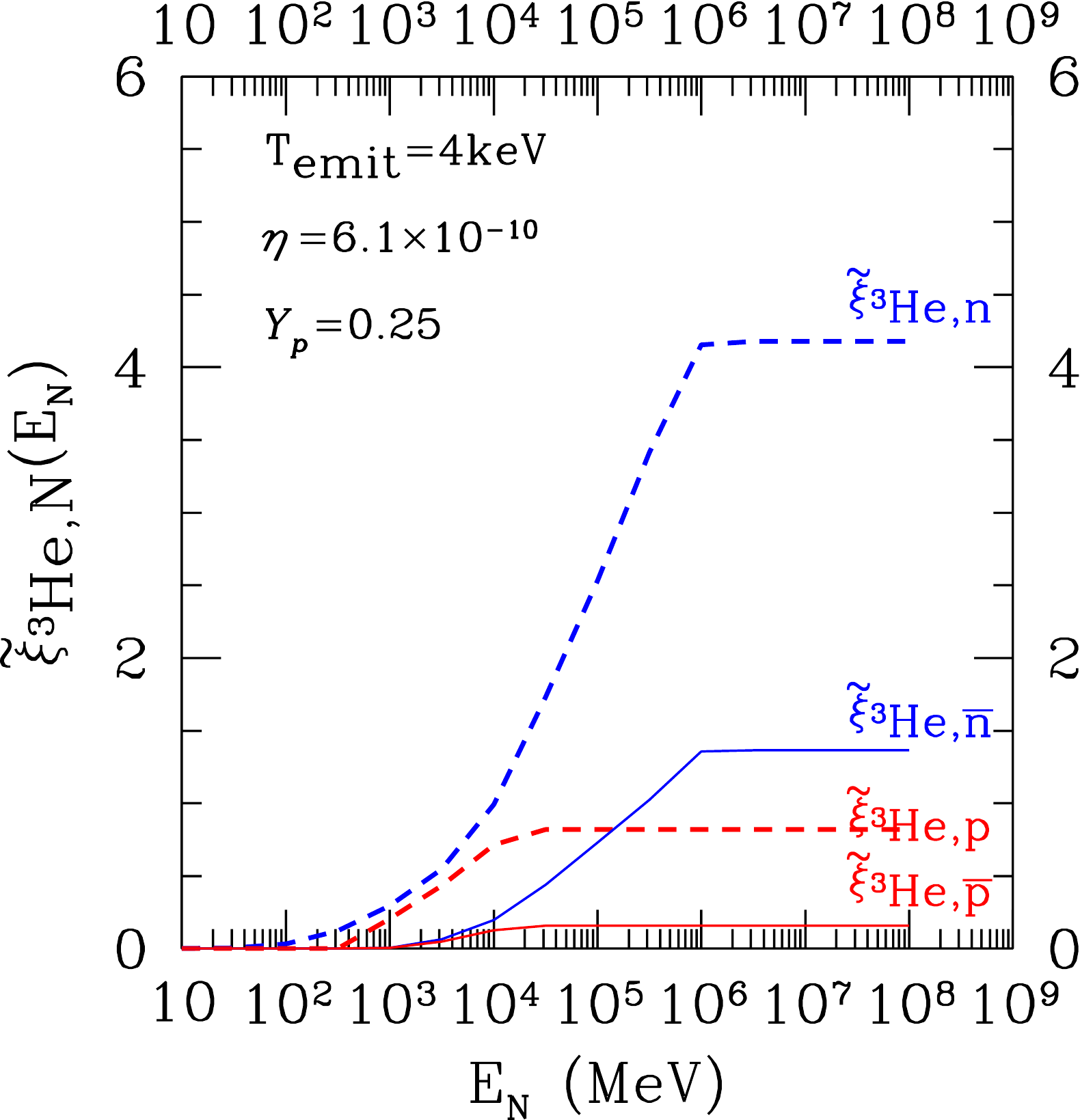}~~~
    \includegraphics[height=0.3\textheight]{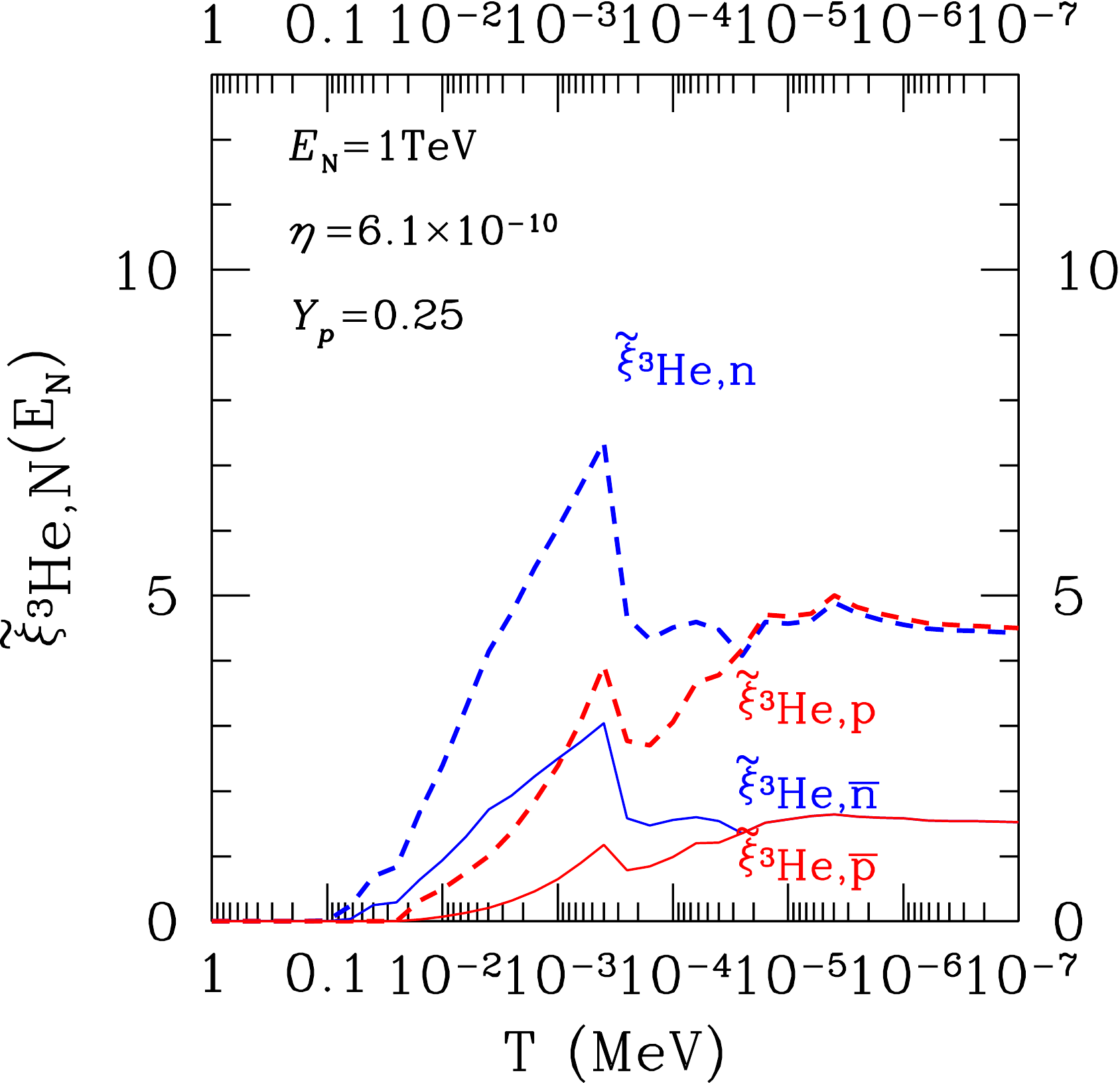}
  \end{center}
  \caption{$\tilde{\xi}_{{\rm ^3He},N}$ for $T= 4$~keV as functions of the
    kinetic energy (left) and those for $E_N=1\ {\rm TeV}$ as functions
    of the cosmic temperature (right). Here, $N = n$, $\bar{n}$, $p$
    and $\bar{p}$, and we take $\eta = 6.1 \times 10^{-10}$ and $Y_p =
    0.25$.}
  \label{fig:xi_He3}
\end{figure}

\begin{figure}
  \begin{center}
    \includegraphics[height=0.3\textheight]{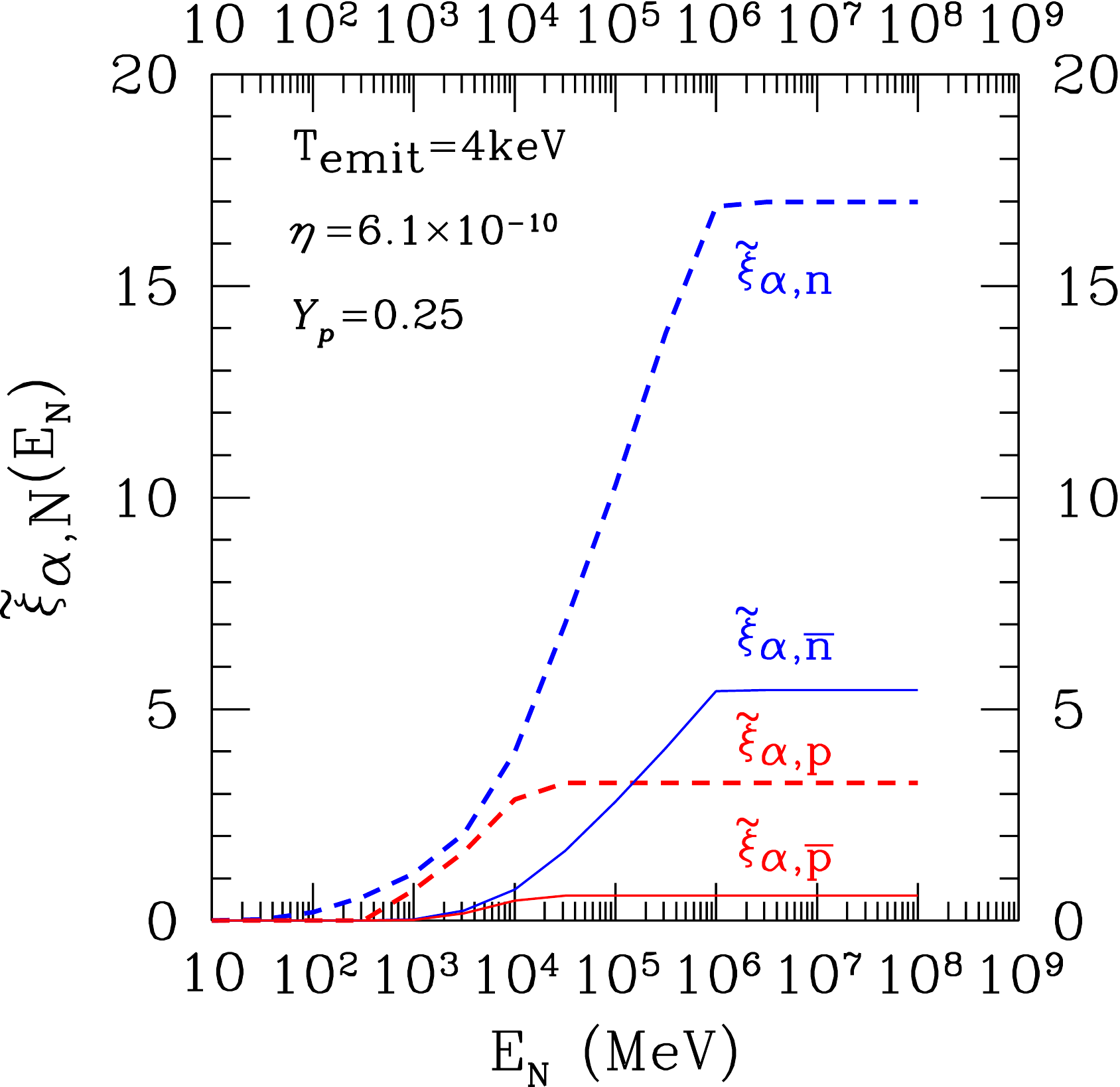}~~~
    \includegraphics[height=0.3\textheight]{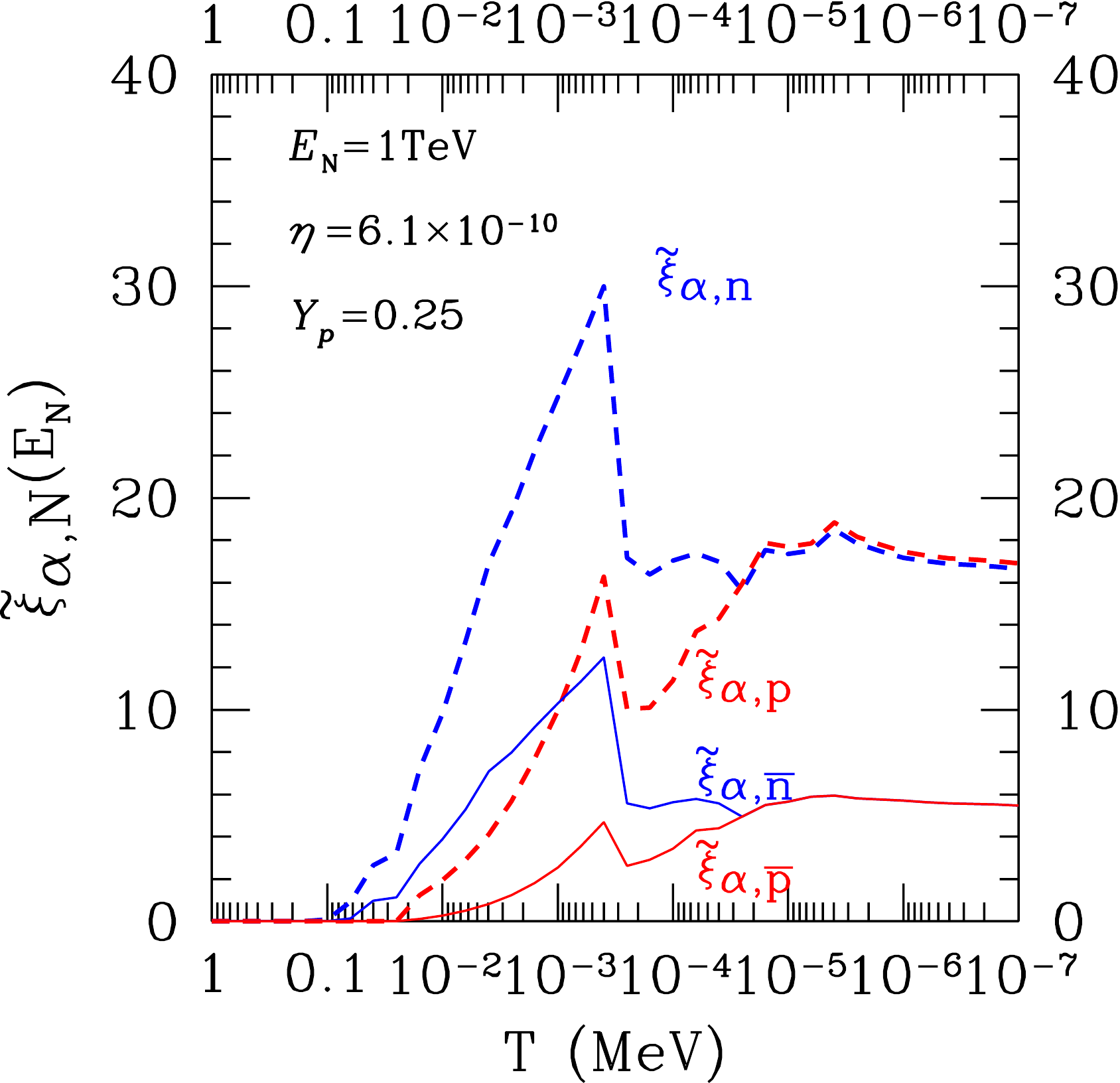}
  \end{center}
  \caption{$\tilde{\xi}_{\alpha,N}$ for $T= 4$~keV as functions of the
    kinetic energy (left) and those for $E_N=1\ {\rm TeV}$ as functions
    of the cosmic temperature (right). Here, $N = n$, $\bar{n}$, $p$
    and $\bar{p}$, and we take $\eta = 6.1 \times 10^{-10}$ and $Y_p =
    0.25$.}
  \label{fig:xi_Alpha}
\end{figure}


In Figs.~\ref{fig:xi_n}, \ref{fig:xi_D}, \ref{fig:xi_T},
\ref{fig:xi_He3} and \ref{fig:xi_Alpha}, we show $\tilde{\xi}_{n,N}$,
$\tilde{\xi}_{{\rm D},N}$, $\tilde{\xi}_{{\rm T},N}$,
$\tilde{\xi}_{{\rm ^3He},N}$, and $\tilde{\xi}_{\alpha,N}$,
respectively, as functions of the kinetic energy $E_N$ for $N = n$,
$\bar{n}$, $p$ and $\bar{p}$, taking $T= 4$~keV.  Notice that, at $T=
4$~keV, the energetic neutrons with $E_n > {\cal O}(1)$~TeV loose
their kinetic energy immediately down to $\sim 1$ TeV
\cite{Kawasaki:2008qe}.  Thus, $\tilde{\xi}_{\alpha,N}$ become
insensitive to $E_N$ for $E_N\gtrsim 1$ TeV.  On the same token,
the emitted high-energy protons with $E_n > {\cal O}(10)$~GeV loose their
kinetic energy down to $\sim 10$~GeV.  The figures also show how the
$\tilde{\xi}_{A_i,N}$ parameters depend on the cosmic temperature,
taking $E_N=1\ {\rm TeV}$.  As one can see, $\tilde{\xi}_{A_i,N}$
parameters increase as the temperature decreases until $T\sim 0.5\
{\rm keV}$; this is because the mean free paths of energetic nucleons
become longer as the temperature drops.  We can also see sharp
drop-offs of the $\tilde{\xi}$-parameters at $T\sim 0.5\ {\rm keV}$;
they come from the fact that, when the cosmic temperature is lower
than $\sim 0.5\ {\rm keV}$, the energetic neutrons decay before scattering
off the background nuclei and that the neutron contributions to the
hadrodissociations become negligible.

From the figures, we find that the injections of $\bar{n}$ and
$\bar{p}$ should change the total number of the destroyed $\alpha_{\rm
  BG}$ by $\sim 20$ -- $30\ \%$.  As will be shown later, the
constraints become stronger by 10 -- 30\ $\%$ when we include the
hadrodissociations by anti-nucleons. On the other hand, by the effects
of the interconversions at inelastic scatterings, the constraints
become weaker by 50 -- 80\ $\%$. That is because the high-energy protons
which can be produced by interconversion from a projectile neutron
tend to be stopped more easily than neutrons through electromagnetic
interactions inside the thermal plasma.

\begin{figure}
  \begin{center}
    \includegraphics[height=0.4\textheight]{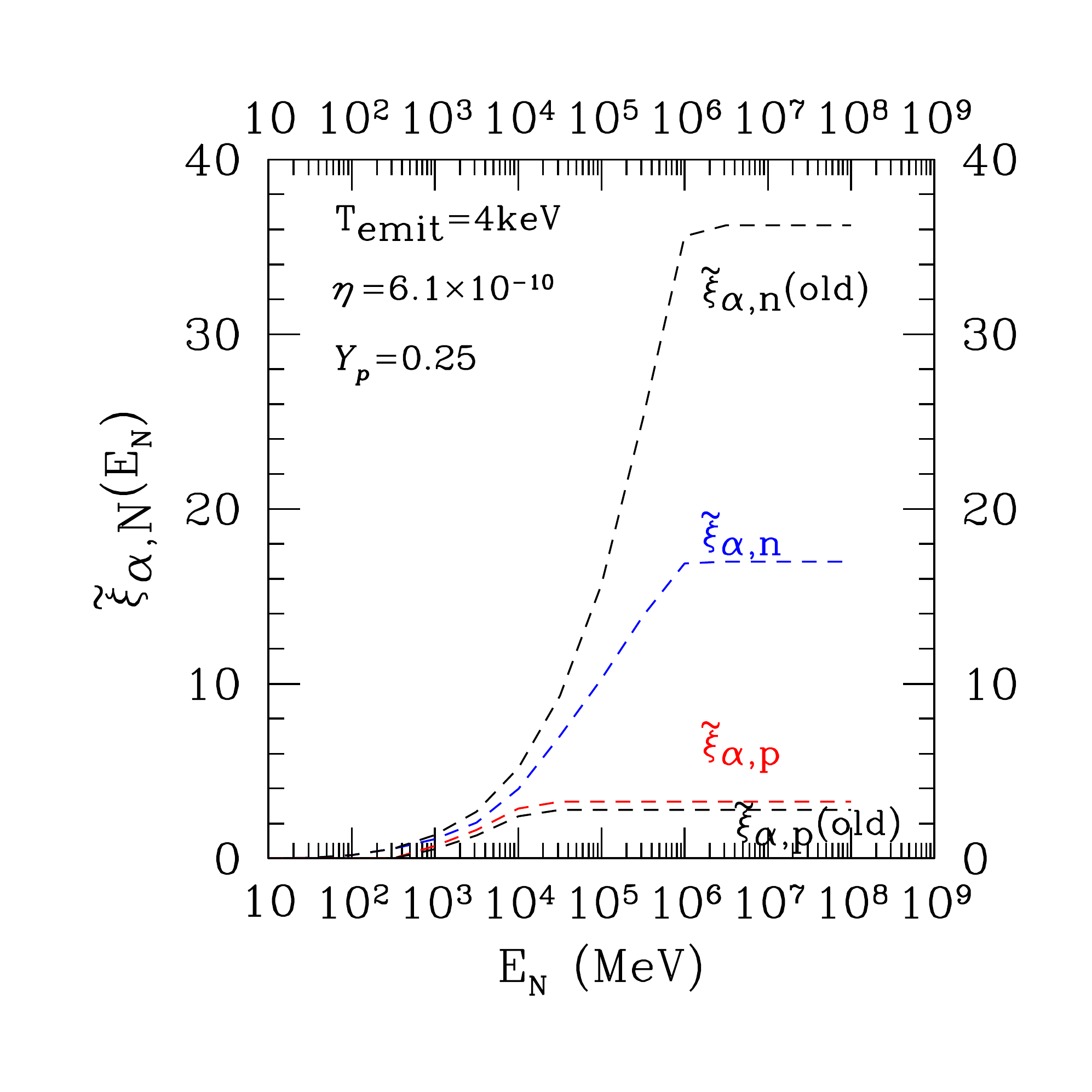}~~~
  \end{center}
  \vspace{-1cm}
  \caption{$\tilde{\xi}_{\alpha,N}$ for $T= 4$~keV as functions of the
    kinetic energy, based on the current method (red and blue) as well
    as those based on the old method which is without the
    interconversions between $n$ and $p$ at inelastic scattering
    (black).  Here, we take the same parameters as those used in the
    left panel of Fig.~\ref{fig:xi_Alpha}.}
  \label{fig:prodAlphaComp}
\end{figure}

For comparison, in Fig.~\ref{fig:prodAlphaComp}, we plot
$\tilde{\xi}_{\alpha,N}$ based on the current method as well as those
based on the old method \cite{Kawasaki:2004qu} which is without the
interconversions between $n$ and $p$ at inelastic scattering.  We can
see that the current $\tilde{\xi}_{\alpha,n}$ is reduced to about a
half of the old one because neutrons are converted to protons during
the hadronic shower evolutions. (Notice that protons are stopped more
easily than neutrons.)  On the other hand the current
$\tilde{\xi}_{\alpha,p}$ is larger than the old one because of the
conversion from protons to neutrons.

With $\xi_{A_i}(T,Y_p,\eta)$ being given, the effects of the
hadrodissociation are included into the Boltzmann equations which
govern the evolutions of the light element abundances; for $A_i= n$,
T, D, $^3$He, we use
\begin{align}
  \frac{dn_{A_i}}{dt} = 
  \left[ \frac{dn_{A_i}}{dt} \right]_{\rm SBBN}
  + \left[ \frac{dn_{A_i}}{dt} \right]_{\rm photodis}
  + n_X \Gamma_X \xi_{A_i},
\end{align}
while 
\begin{eqnarray}
    \frac{dn_{\rm ^4He}}{dt} = 
    \left[ \frac{dn_{\rm ^4He}}{dt} \right]_{\rm SBBN}
    + \left[ \frac{dn_{\rm ^4He}}{dt} \right]_{\rm photodis}
    - n_X \Gamma_X \xi_\alpha.
\end{eqnarray}
Here, the subscript ``SBBN'' and ``photodis'' imply the reaction rates
due to the SBBN and photodissociation processes.  Notice that the
effects of the $p\leftrightarrow n$ interconversions of the background
nucleons are included in the SBBN contributions by properly modifying
the number densities of the background $p$ and $n$.

For $^6$Li, $^7$Li and $^7$Be, we include non-thermal production
processes induced by secondary energetic T, $^3$He, $^4$He.  Such
energetic nuclei are produced by the hadronic scatterings of
energetic nucleons off the background $\alpha_\text{BG}$ as well as by the electromagnetic processes of
energetic photons.  For the non-thermal processes induced by hadrons,
the effects are parameterized by the following quantity (with
$A_f={\rm ^6Li}$, $^7$Li and $^7$Be) \cite{Dimopoulos:1987fz,
  Dimopoulos:1988zz, Dimopoulos:1988ue,Kawasaki:2004qu}
\begin{align}
  \xi_{A_f} =\, & 
  \sum_{A_i={\rm T}, {\rm ^3He}, {\rm ^4He}}
  \int d E_{A_i}^{\rm (in)}
  f_{A_i} (E_{A_i}^{\rm (in)})
  \int^{E_{A_i}^{\rm (in)}} 
  d E_{A_i} \left( \frac{dE_{A_i}}{dt} \right)^{-1}
  n_{\alpha} \sigma_{A_i+\alpha_{\rm BG}\rightarrow A_f+\cdots} (E_{A_i})
  \beta_{A_i} P_{A_f},
  \label{xi_LiBe}
\end{align}
where $f_{A_i}$ is the energy distribution of $A_i$ produced by the
scattering or hadrodissociation processes of $\alpha_{\rm BG}$,
$(dE_{A_i}/dt)^{-1}$ is the energy-loss rate, $\sigma_{A_i+\alpha_{\rm
    BG}\rightarrow A_f+\cdots}$ is the production cross section of
$A_f$, $\beta_{A_i}$ is the velocity, and $P_{A_f}$ is the survival
probability of $A_f$ after production.  (For $A_f={\rm ^6Li}$,
$A_i={\rm T}$, $^3$He, and $^4$He contributes, while only $A_i={\rm
  ^4He}$ is relevant for $A_f={\rm ^7Li}$ and $^7$Be.)  Then, for
$A_i=^6$Li, $^7$Li and $^7$Be, the Boltzmann equations are given in the
following form:
\begin{eqnarray}
  \frac{dn_{A_i}}{dt} = 
  \left[ \frac{dn_{A_i}}{dt} \right]_{\rm SBBN}
  + \left[ \frac{dn_{A_i}}{dt} \right]_{\rm photodis}
  + \left[ \frac{dn_{A_i}}{dt}
  \right]_{\gamma+\alpha_{\rm BG}\rightarrow\cdots}
  + n_X \Gamma_X \xi_{\rm A_i}.
\end{eqnarray}
Here, $[dn_{A_i}/dt]_{\gamma+\alpha_{\rm BG}\rightarrow\cdots}$
denotes the effects of the non-thermal production processes initiated
by the photodissociation of $\alpha_{\rm BG}$~\cite{Jedamzik:1999di},
which exists only for $A_i=^6$Li.

The expansion of the universe is determined by the Einstein equation;
the expansion rate of the universe is calculated with taking into
account the energy density of $X$.  In addition, the effect of the
entropy production due to the decay of $X$ is included.  Thus, the
value of $\eta$ is defined as the value after the entropy
production.\footnote
{We consider the case where the same amount of baryon and anti-baryon
  are produced in average by the decay.  For the case of an asymmetric
  decay, see, for example, Ref.\ \cite{Scherrer:1991yu}.}
The initial value of the baryon number density is set to realize the
required value of $\eta$.  These effects are relevant for the
parameter region in which $X$ once dominates the universe.  (As we
will see below, however, such a parameter region is mostly excluded by
the CMB constraints.)

\section{Constraints on Generic Decaying Particles}
\label{sec:const_generic}
\setcounter{equation}{0}

\subsection{Constraints from each light elements}
\label{subsec:const_each_element}

\begin{figure}
  \centering
  \includegraphics[width=0.49\textwidth]{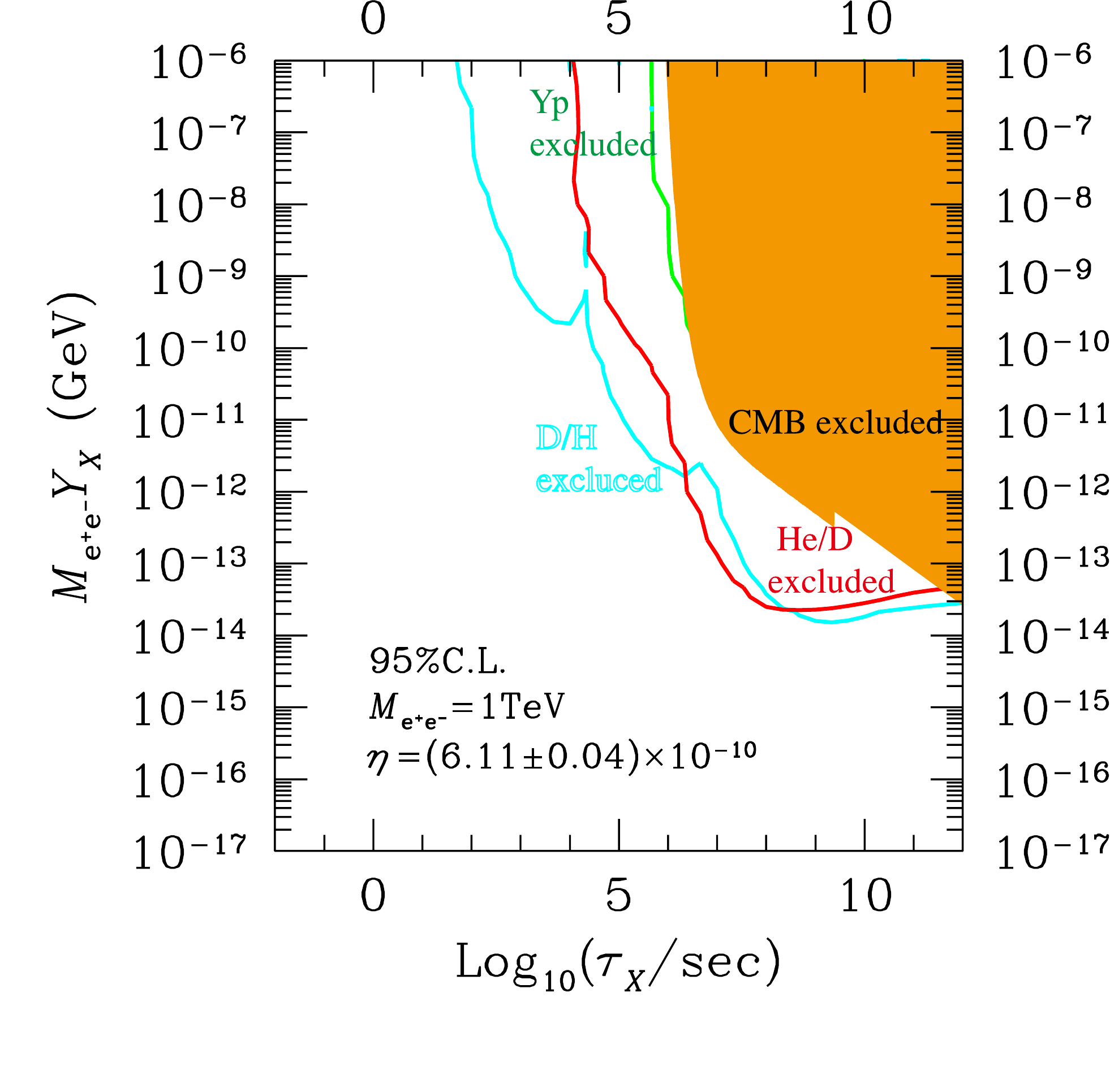} 
  \includegraphics[width=0.49\textwidth]{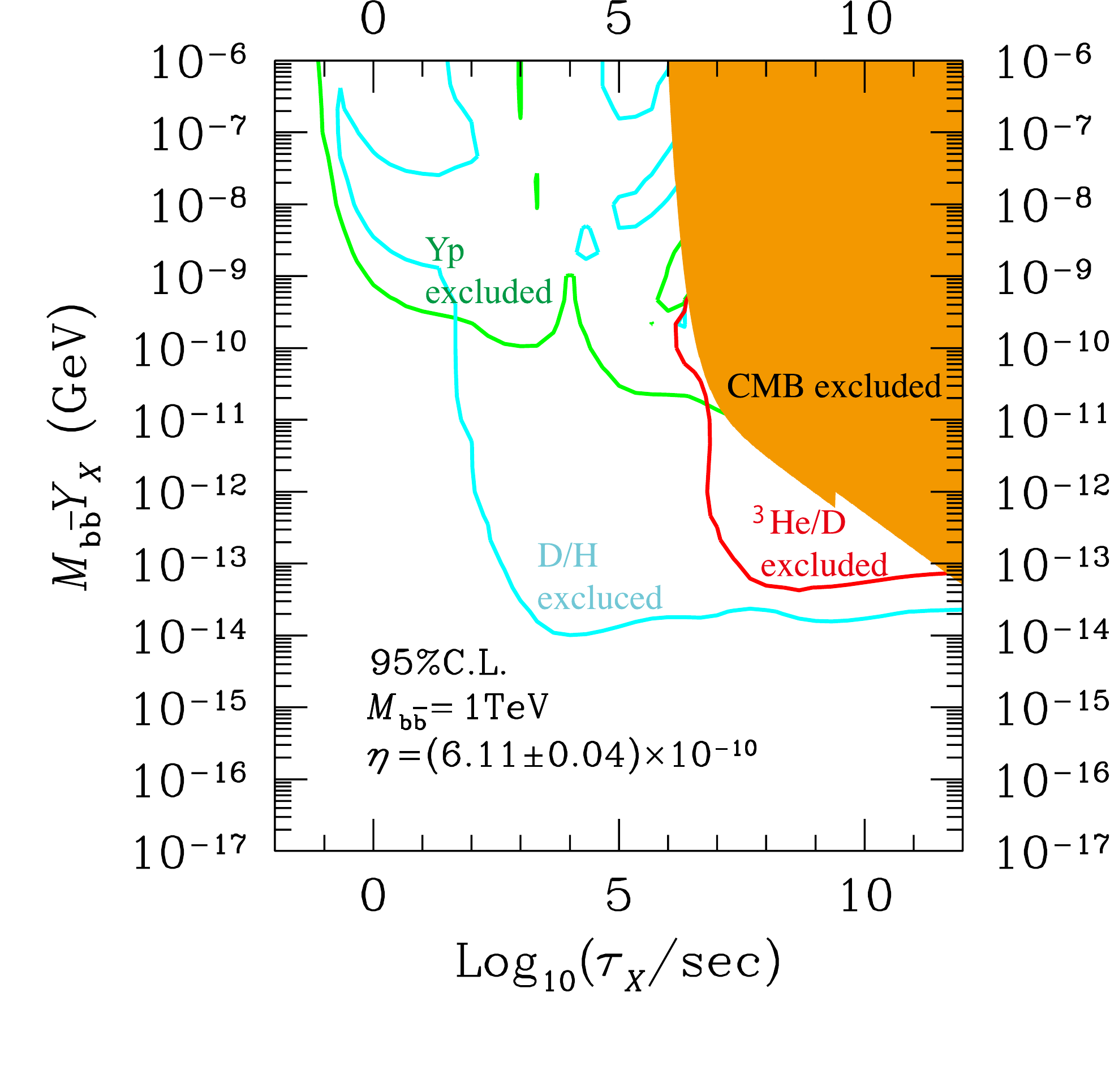} 
  \caption{\small%
    Constraints on $m_X Y_X$ vs.\ $\tau_X$ plane, assuming that the
    main decay modes are $e^{+}e^{-}$ (left) or $b\bar{b}$ (right)
    in case that the invariant mass into the daughter particles is
    $m_X=$~1TeV.
    The BBN constraints comes from $^4$He (green), D (cyan) and
    $^3$He/D (red).  The orange shaded region is excluded by the CMB
    spectral distortion.}
  \label{fig:generic_const_each_element}
\end{figure}


We first investigate the BBN constraints on generic decaying
particles.  In Fig.~\ref{fig:generic_const_each_element} we show the
constraints on the decaying particles which primarily decay into
$e^{+}e^{-}$ or $b\bar{b}$.  In the figure, we show the constraints on
the combination $m_X Y_X$ (with $m_X$ being the mass of $X$) as
functions of the lifetime $\tau_X$ for the final states $e^{+}e^{-}$
or $b\bar{b}$.\footnote
{In the figures, in order to indicate the final state, we use the
  notation $M_{FF'}$, which is equal to $m_X$, where $FF'$ corresponds
  to the final state particles of the $X$ decay.}

When the decaying particles mainly decay into $e^{+}e^{-}$, the
hadronic branching ratio is small.  In fact, hadrons are produced
through the decay process $X \rightarrow e^{+}+e^{-}+q+\bar{q}$, but
the branching ratio for a such process is suppressed by $\sim {\cal
  O}((\alpha/4\pi)^2)\sim 10^{-6}$.  Thus, most of the constraints are
due to radiative decay.  At the cosmic time $\tau_X \sim
10^4-10^6$~sec, only D is destroyed by energetic photons that are not
thermalized by photon-photon processes, so photodissociation of D
gives the stringent constraint for $\tau_X \sim 10^4-10^6$~sec.  At $t
\gtrsim 10^6$~sec, $^4$He is also destroyed by photodissociation
processes, which leads to non-thermal production of D and $^3$He.
Thus, the stringent constraints are imposed by overproduction of D and
$^3$He.\footnote
{Notice that the present constraints from $^3$He/D and D/H are almost
  the same while in the previous work~\cite{Kawasaki:2004qu} the
  constraint from $^3$He/D was severer than that from D/H.  This is
  because of the recent precise measurement of the abundance of D/H.}
It is seen that the abundance of the $X$ is also constrained for
$\tau_X\lesssim 10^{4}$~sec, which is due to hadrodissociation of
$^4$He.  Since the photons with energy larger than ${\cal O}(1)$~MeV
are quickly thermalized at $t\lesssim 10^{4}$~sec, they cannot destroy
the light elements.  On the other hand, hadrons like proton and
neutron can destroy $^4$He and produce D and $^3$He non-thermally.
The resultant constraint due to hadrons is weak because the hadronic
branching ratio is small in this case.

In the case where the $X$ mainly decays into $b \bar{b}$ (Fig.\
\ref{fig:generic_const_each_element} (right)), the stringent
constraints come from hadrodissociation of the light elements.  The
high energy quarks emitted in the decay induce hadronic showers in
which $^4$He nuclei are destroyed by energetic nucleons.  The
hadrodissociation of $^4$He leads to overproduction of D, which gives
a stringent constraint, in particular, for $\tau_X \sim
10^2-10^7$~sec.  The photodissociation of $^4$He also produces D and
gives a stringent constraint for $\tau_X \gtrsim 10^7$~sec where
effects of hadrodissociation and photodissociations are roughly
comparable.  In addition, non-thermal production of $^3$He by
photodissociation gives a significant constraint for $\tau_X \gtrsim
10^7$~sec.  For $\tau_X \lesssim 100$~sec the constraint coming from
$^4$He overproduction is most stringent.  At the early stage of BBN
($t\sim 1-100$~sec) interconversion of protons and neutrons is the
most important process which almost determines the final $^4$He
abundance.  The strongly interacting conversion increases $n/p$ from
its standard value.  As a result more $^4$He is produced, from which
we obtain the constraint for $\tau_X \lesssim 100$~sec.  From the
figure one notice that there appears the constraint from D/H for
$\tau_X \sim 0.1 - 100$~sec.  The constraint around $m_X Y_X \sim
10^{-9}$~GeV comes from overproduction of D due to larger $n/p$.
Since the change of $n/p$ affects the abundance of $^4$He more
significantly, the D/H constraint is weaker.  The D/H constraint
around $m_X Y_X \sim 10^{-7}$~GeV needs some caution.  In this
parameter region large $n/p$ already makes the abundance of $^4$He
much larger than the standard value.  The D abundance is sensitive to
the change of $^4$He abundance because D is residual after
synthesizing $^4$He.  The $^4$He abundance has ${\cal O}(10)$\%
theoretical uncertainty in this parameter region, which drastically
increases the uncertainty of the D abundance.  As a result, it makes
difficult to obtain a reliable constraint from D.  Since this region
is ruled out by overproduction of $^4$He, this is not an obstacle to
obtain constraints on the properties of $X$.

In Fig.\ \ref{fig:generic_const_each_element}, the constraints from
the CMB are also shown.  When the decay of $X$ injects electromagnetic
energy into the background plasma, the spectrum of the CMB is
distorted.  Since photon number changing processes, like the double
Compton scattering, are not efficient at $t \gtrsim 10^7$~sec, the
resultant CMB spectrum deviates from the Planck distribution if $X$
decays at such an epoch~\cite{Hu:1993gc,Chluba:2011hw,Chluba:2013wsa}.
The COBE set a stringent constraint on the spectrum distortion of the
CMB~\cite{Fixsen:1996nj}, from which we obtain the upper bound on the
abundance of $X$, as shown in
Fig~\ref{fig:generic_const_each_element}.  It is seen that the BBN
constraints are more stringent than those from the CMB for both decay
modes.\footnote { See also Ref.~\cite{Dimastrogiovanni:2015wvk}, for a
  possible mass dependence on this relation.  }  Notice that future
observations of the CMB spectrum such as PIXIE will improve the bounds
by a factor of 100 -- 500~\cite{Chluba:2013pya}.

\subsection{Constraints on various decay modes}

\begin{figure}
  \centering
  \includegraphics[width=0.48\textwidth]{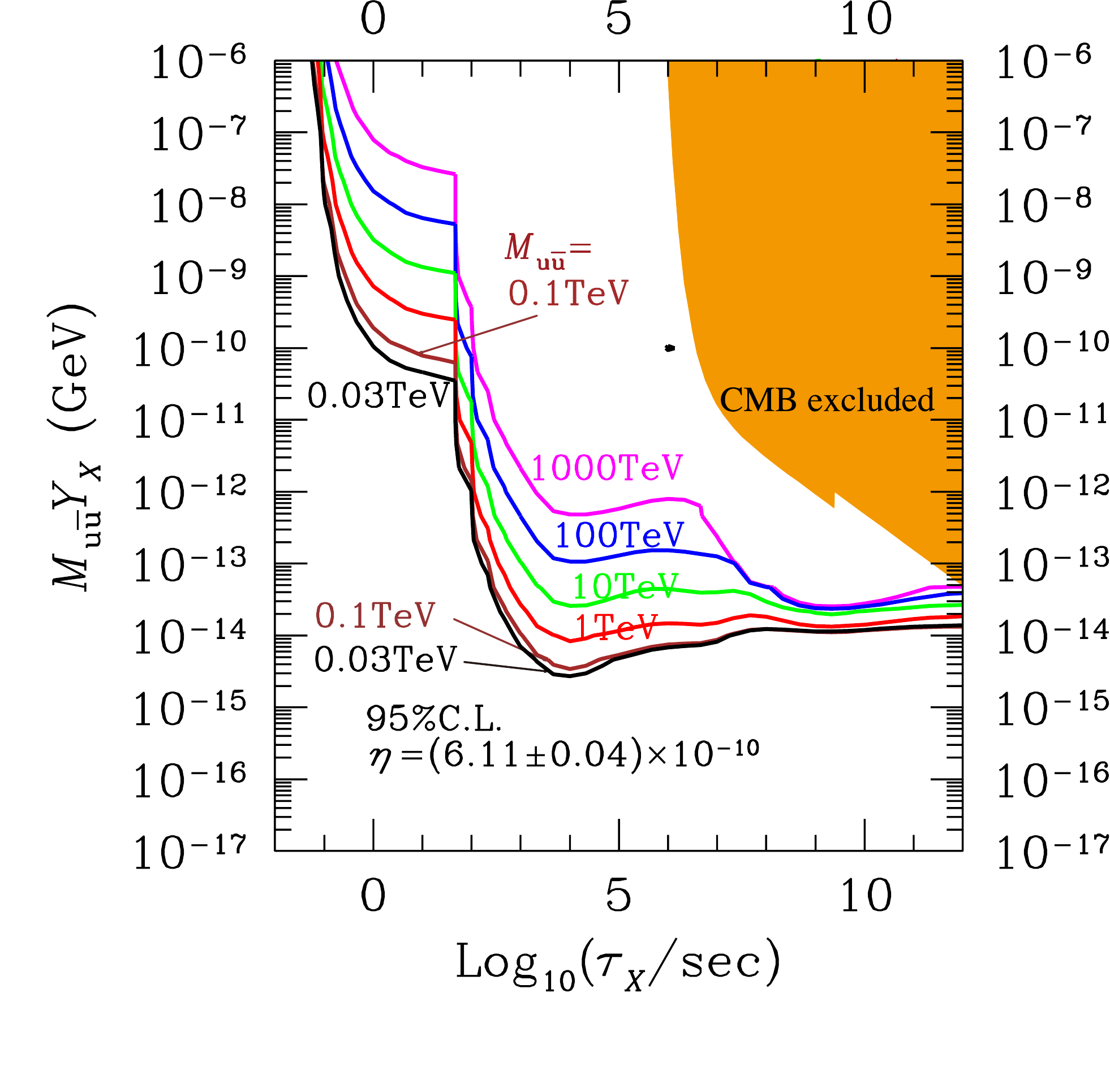} 
  \includegraphics[width=0.48\textwidth]{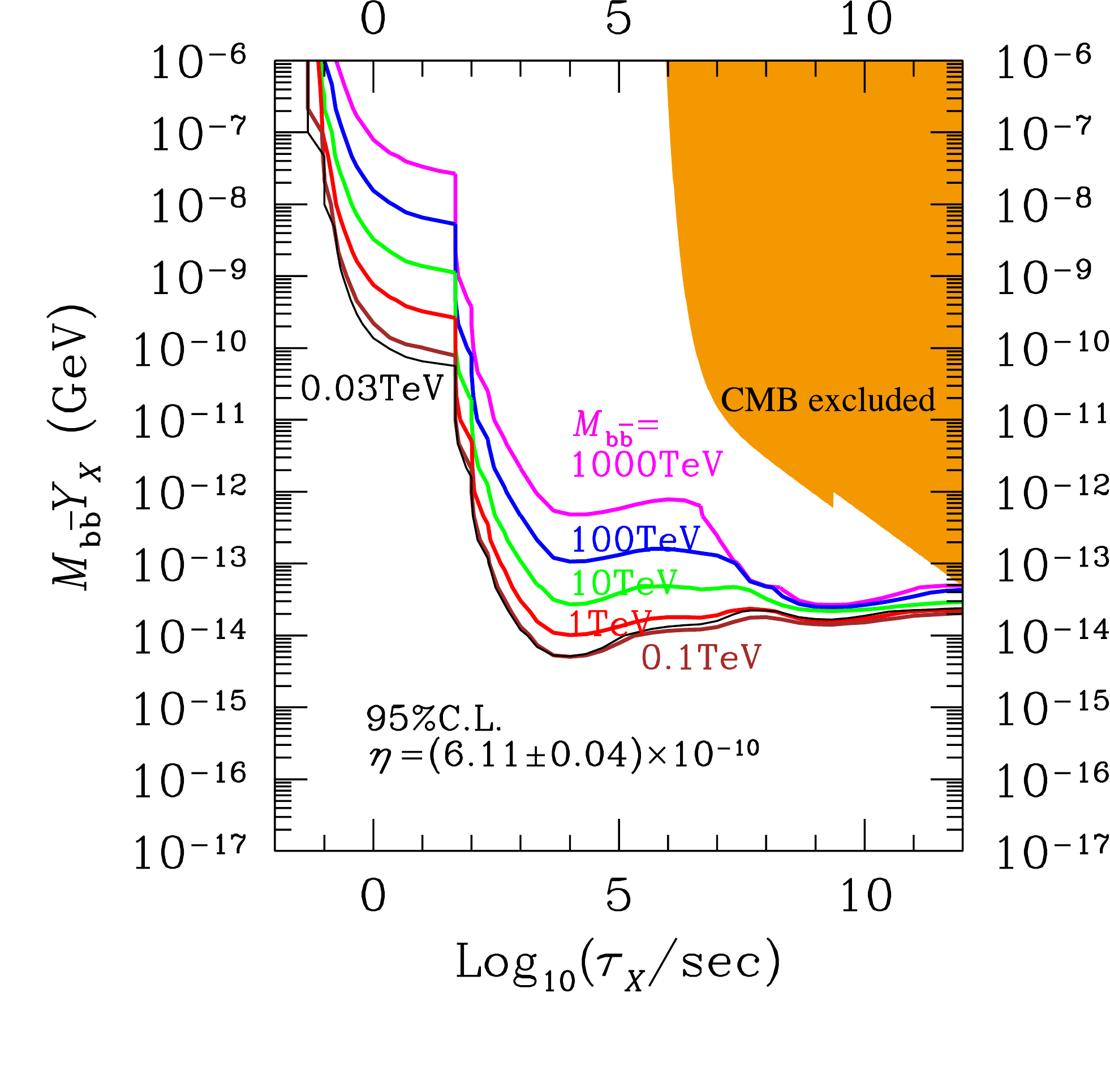} 
  \includegraphics[width=0.48\textwidth]{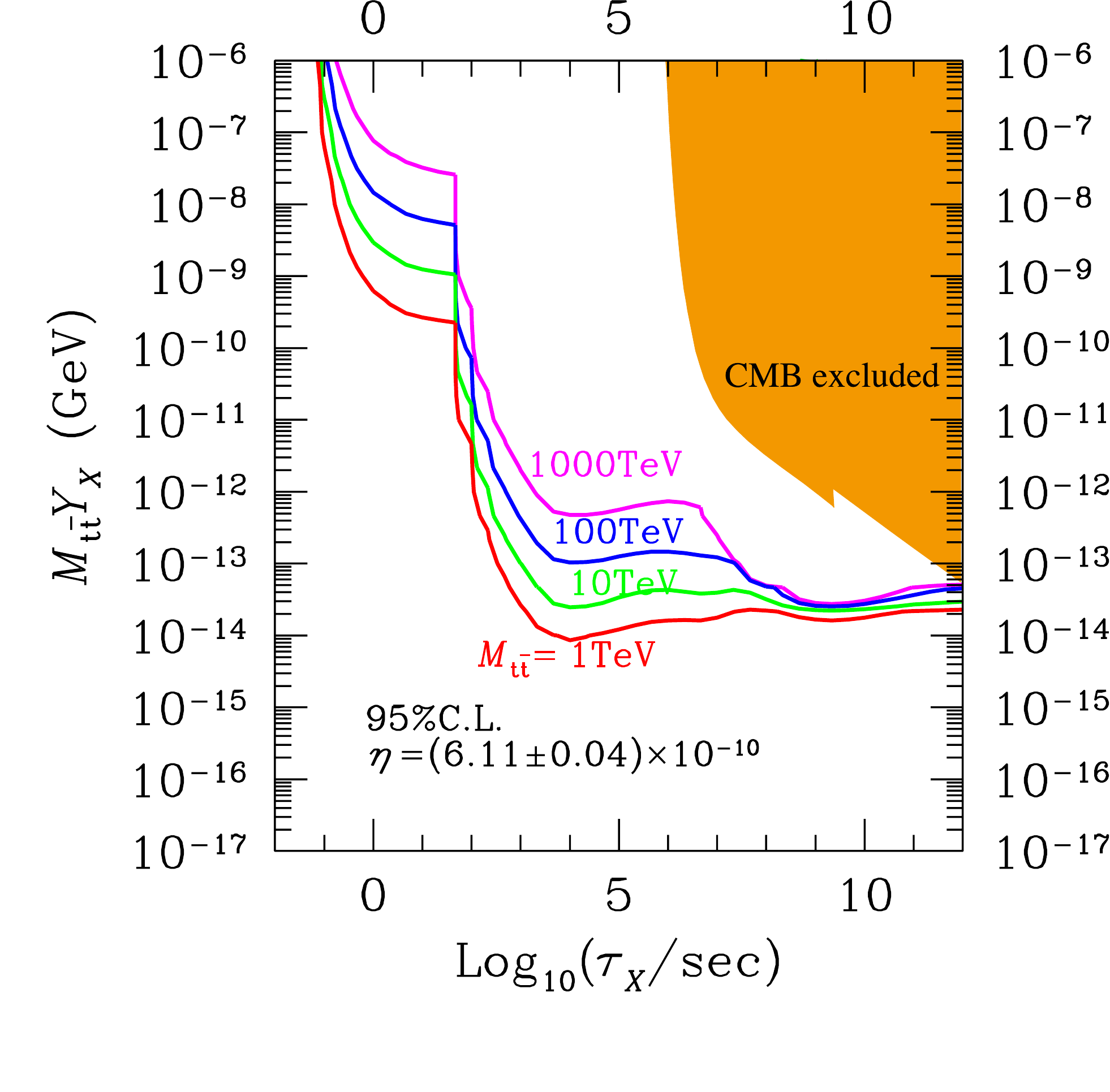} 
  \includegraphics[width=0.48\textwidth]{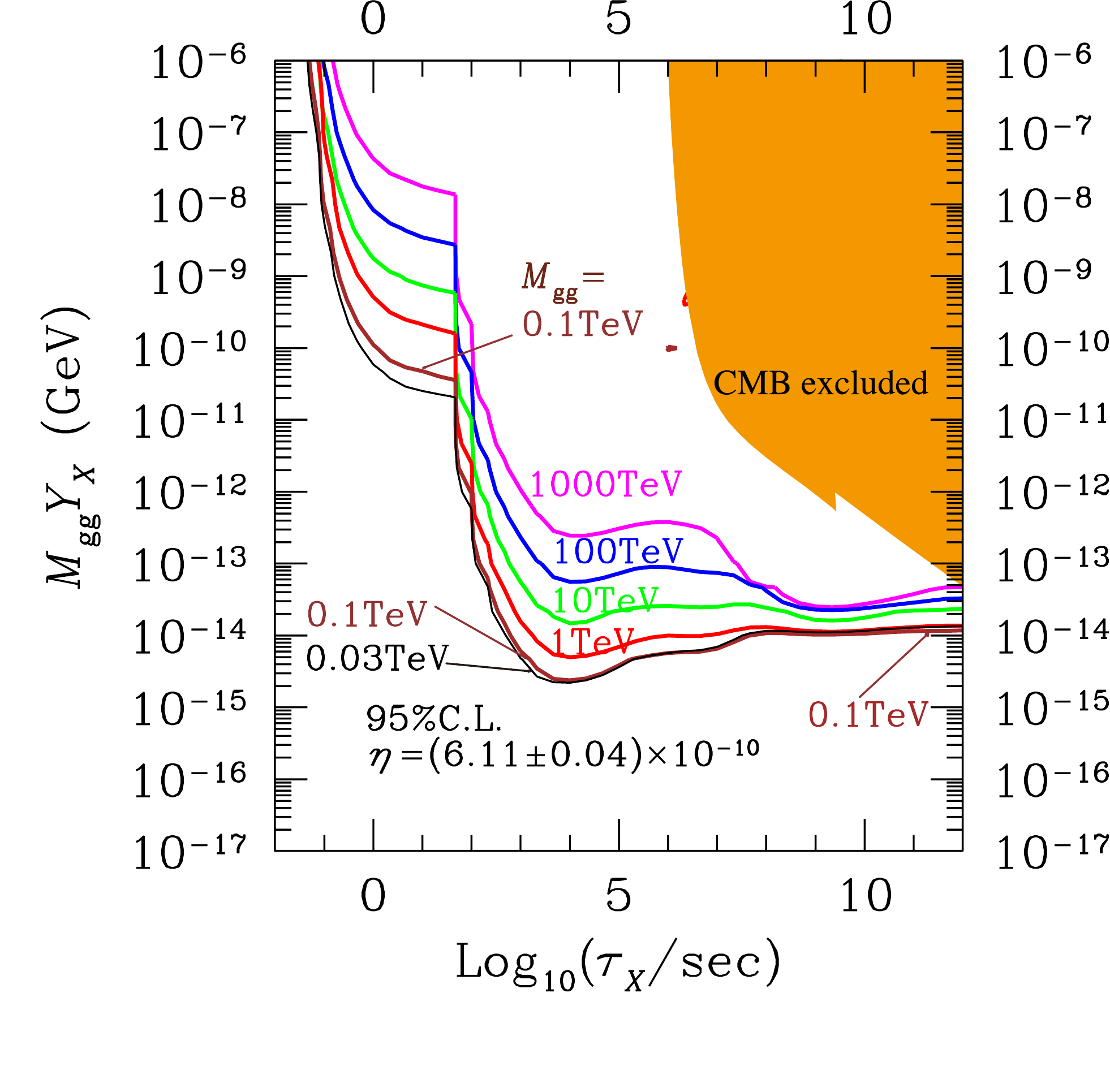} 
  \caption{\small%
    Constraints on $m_X Y_X$ vs.\ $\tau_X$ plane, assuming that the
    main decay modes are $u\bar{u}$ (upper left), $b\bar{b}$ (upper
    right), $t\bar{t}$ (lower left) and $gg$ (lower right).  The
    black, dark-red, red, green, blue and magenta solid lines denote
    the BBN constraints for $m_X=0.03$, $0.1$, $1$, $10$, $100$ and
    $1000$~TeV, respectively.  The orange shaded regions are excluded
    by the constraint from the CMB spectral distortion.}
	\label{fig:generic_const_1}
\end{figure}


Now, we show constraints for the cases with various main decay modes.

We first consider the cases where $X$ decays into colored particles.
In Fig.~\ref{fig:generic_const_1} we show the combined constraints on
the abundance and lifetime of $X$ whose main decay mode is $u\bar{u}$,
$b\bar{b}$, $t\bar{t}$ or $gg$; in Fig.~\ref{fig:generic_const_1},
only the most stringent constraint is shown combining the constraints
from D, $^4$He, and $^3$He/D.  For all the decay modes shown in this
figure, the decay products are colored particles and produce hadronic
showers whose total energy is roughly equal to $m_X$.  Therefore, the
resultant constraints are similar.  As described above, hadrons
produced by the decay affects the BBN for $\tau_X \lesssim 10^7$~sec.
As a result, the constraints from overproduction of $^4$He due to the
enhancement of $n/p$ is most stringent for $\tau_X \lesssim 10^2$~sec
while overproduction of D via hadrodissociation of $^4$He gives the
strongest constraint for $\tau_X \gtrsim 10^2-10^7$~sec.  We also show
how the constraints depend on $m_X$, taking $m_X = 0.03$, $0.1$, $1$,
$10$, $100$ and $1000$~TeV.  Since the number of hadrons produced
through the hadronization process depends on $m_X$ as $m_X^{\delta}$
with $\delta\sim 0.3$, the constraints on $m_X Y_X$ from
hadrodissociation become weaker as $m_X$ increases.  On the other
hand, the most stringent constraints for $\tau_X \gtrsim 10^7$~sec
come from the photodissociation of $^4$He, which leads to the
overproduction of D and $^3$He.  Since the effects of
photodissociations are determined by the total energy injection, the
constraint only depends on $m_X Y_X$ as confirmed in
Fig.~\ref{fig:generic_const_1}.

\begin{figure}
  \centering
  \includegraphics[width=0.48\textwidth]{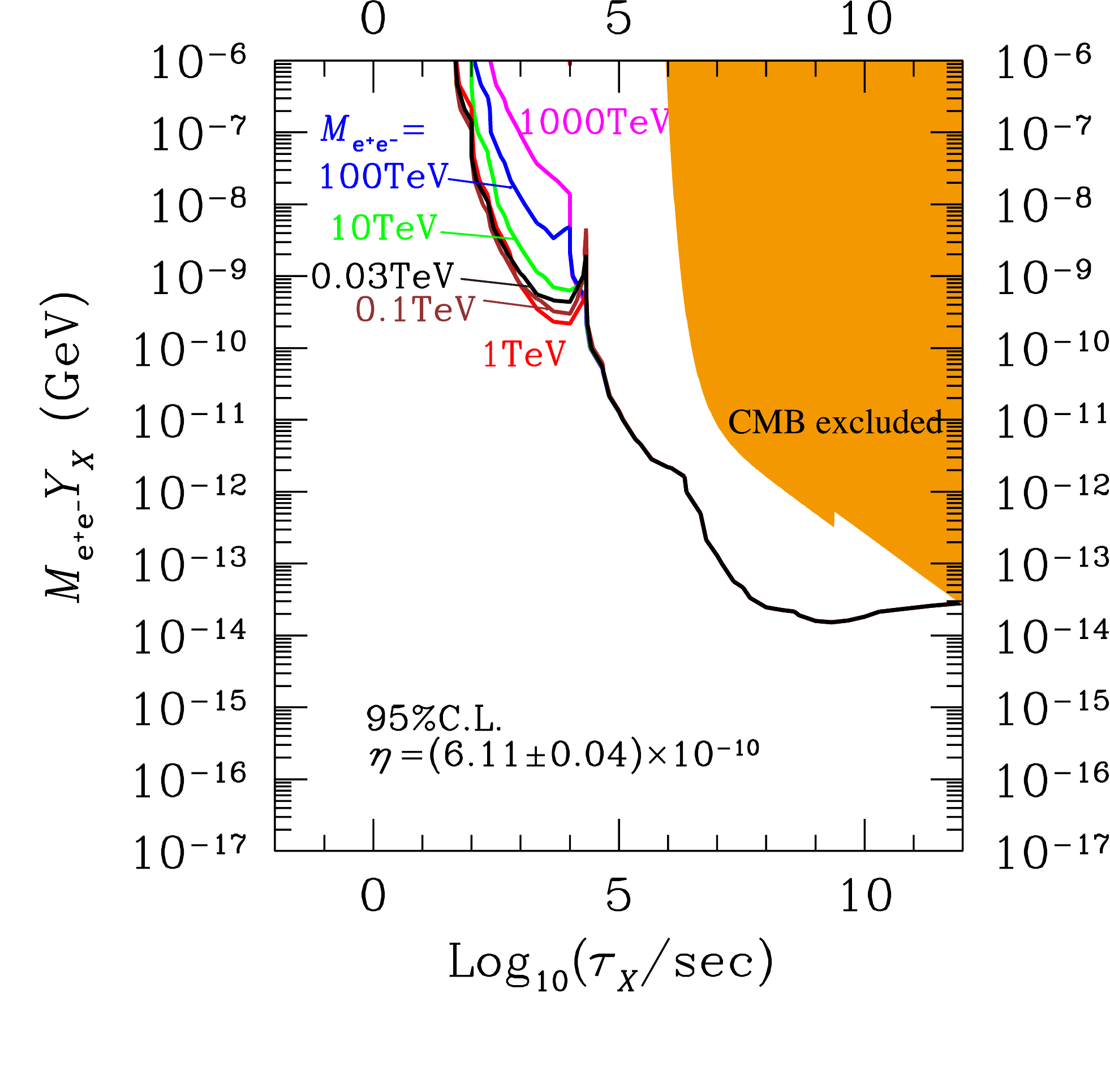} 
  \includegraphics[width=0.48\textwidth]{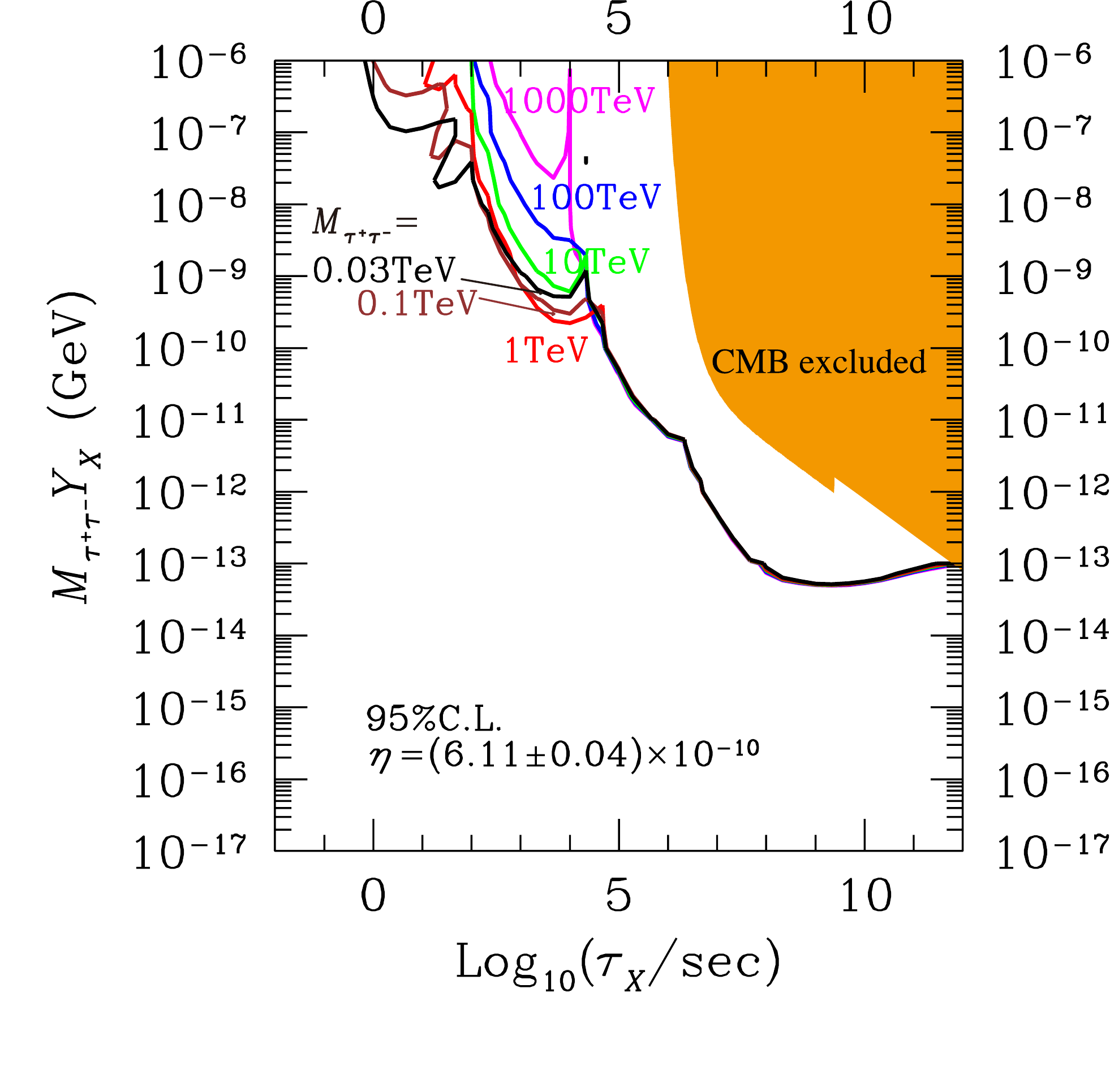} 
  \includegraphics[width=0.48\textwidth]{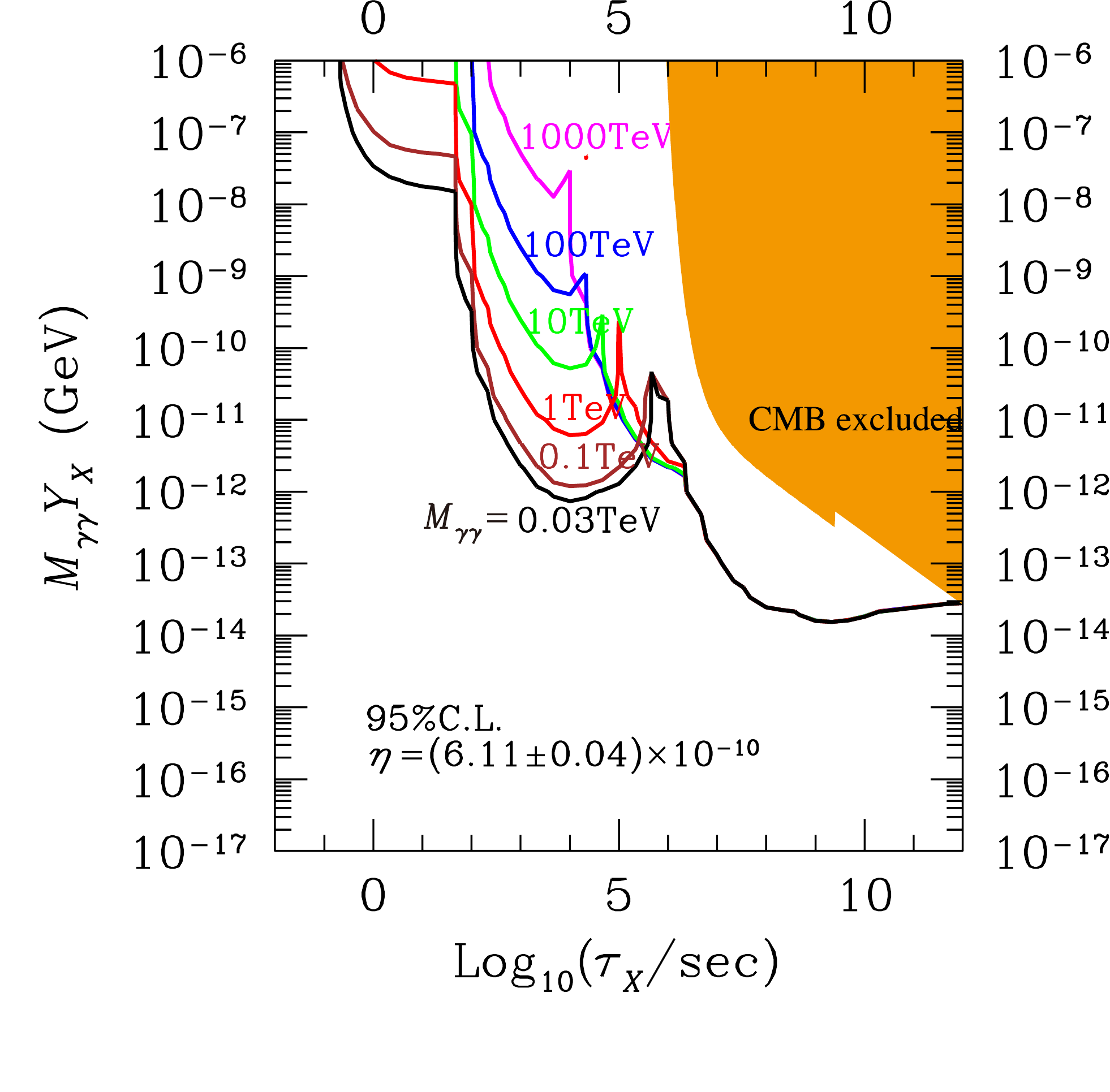}
  \includegraphics[width=0.48\textwidth]{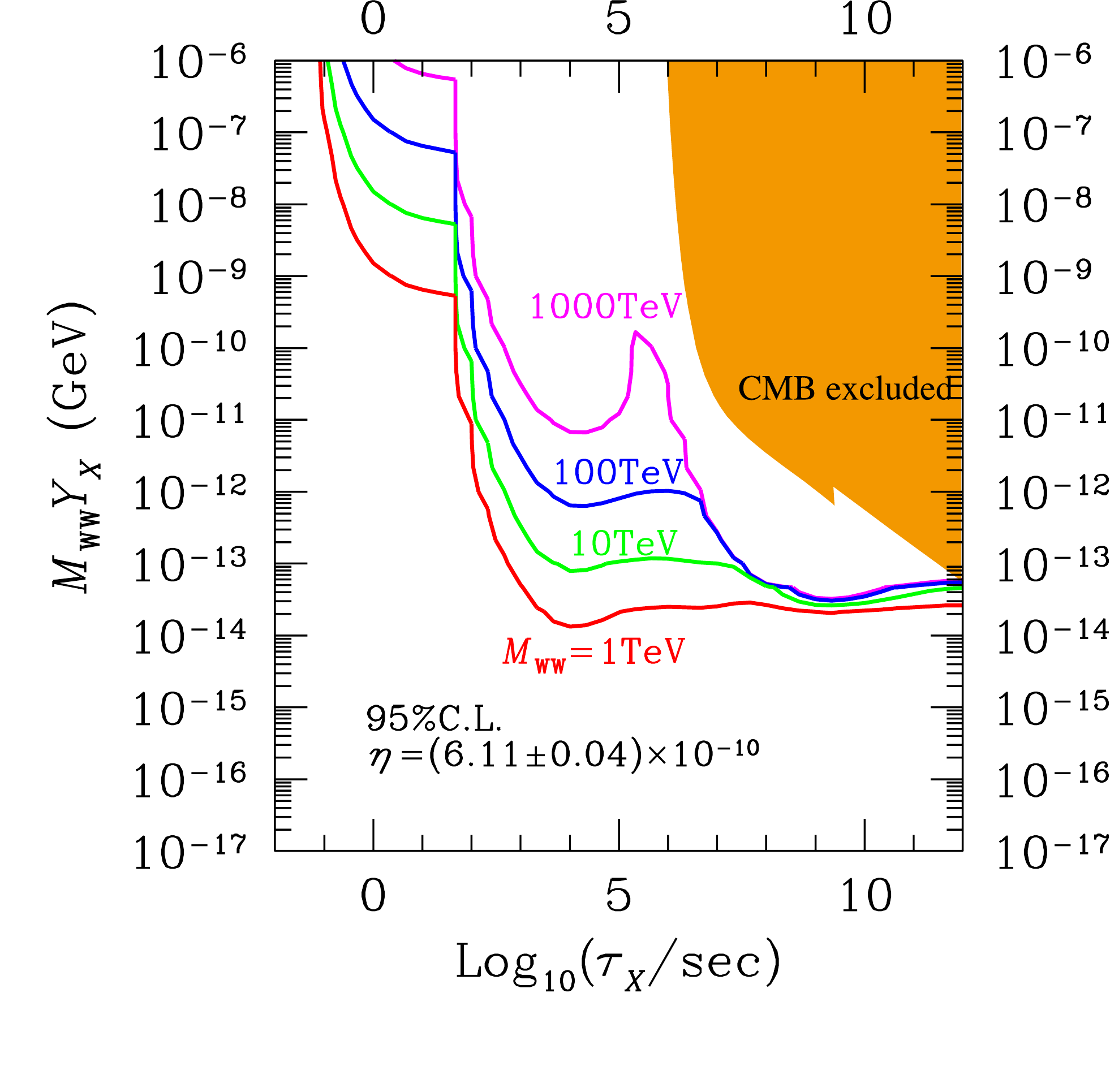} 
  \caption{\small%
    Constraints on $m_X Y_X$ vs.\ $\tau_X$ plane, assuming that the
    main decay modes are $e^{+}e^{-}$ (upper left), $\tau^{+}\tau^{-}$
    (upper right), $\gamma\gamma$ (lower left) and $W^{+}W^{-}$ (lower
    right).  The black, dark-red, red, green, blue and magenta solid
    lines denote the BBN constraints for $m_X=0.03$, $0.1$, $1$, $10$,
    $100$ and $1000$~TeV, respectively.  The orange shaded regions are
    excluded by the constraint from the CMB spectral distortion.}
  \label{fig:generic_const_2}
\end{figure}


In Fig.\ \ref{fig:generic_const_2}, we show the combined constraints
on $X$ which mainly decays into $e^{+}e^{-}$, $\tau^{+}\tau^{-}$,
$\gamma\gamma$ or $W^{+}W^{-}$, taking $m_X=0.03$, $0.1$, $1$, $10$,
$100$ and $1000$~TeV.  For the decay into $e^{+}e^{-}$, as described
in Section \ref{subsec:const_each_element}, the constraints are
determined by the photodissociation effect for $\tau_X \gtrsim 10^4$~sec
(destruction of D for $\tau_X \sim 10^4-10^6$~sec and overproduction
of $^3$He for $\tau_X \gtrsim 10^6$~sec).  Because the branching ratio
into $q\bar{q}$ is small, the effect of hadronic decay is significant
only for $\tau_X \lesssim 10^4$~sec where photodissociation does
not take place.  The constraints for the case that $X$ mainly decays
into $\tau^{+}\tau^{-}$ is also shown.  Since the branching ratio into
$q\bar{q}$ is as small as the $e^{+}e^{-}$ case, the constraints for
$\tau_X \gtrsim 10^4$~sec are almost same as
Fig.~\ref{fig:generic_const_2} (upper left); the constraints are
slightly weaker because some amount of the energy is carried away by
neutrinos.  The effect of hadrodissociation is seen for $\tau_X
\lesssim 10^4$~sec, which is similar to the $e^{+}e^{-}$ decay.  The
constraints from the change of $n/p$ are seen for $\tau_X \lesssim
10^2$~sec, which is due to the mesons produced by the $\tau$ decay.
The constraints on the decay mode $\gamma\gamma$ are shown in
Fig.~\ref{fig:generic_const_2} (lower left).  For $\tau_X \gtrsim
10^6$~sec, photodissociation is important and the constraint is
similar to that for $e^{+}e^{-}$ decay.  However, compared to the
$e^{+}e^{-}$ case, the $q\bar{q}$ production rate at the decay is
relatively large, $\sim {\cal O}(\alpha/4\pi)\sim 10^{-4}-10^{-3}$.
Therefore, the constraints from the hadronic processes is more
important than that from photodissociation for $\tau_X \lesssim
10^{6}$~sec.  In Fig.\ \ref{fig:generic_const_2} (lower right), we
show the constraints on the decay into $W^{+}W^{-}$.  Since $W$ bosons
further decay into hadrons with branching ratio of about $0.67$,
effects of the hadronic decay are important and hence the resultant
constraints looks similar to those on $q\bar{q}$ decay modes.

\begin{figure}[t]
\centering
    \includegraphics[width=0.48\textwidth]{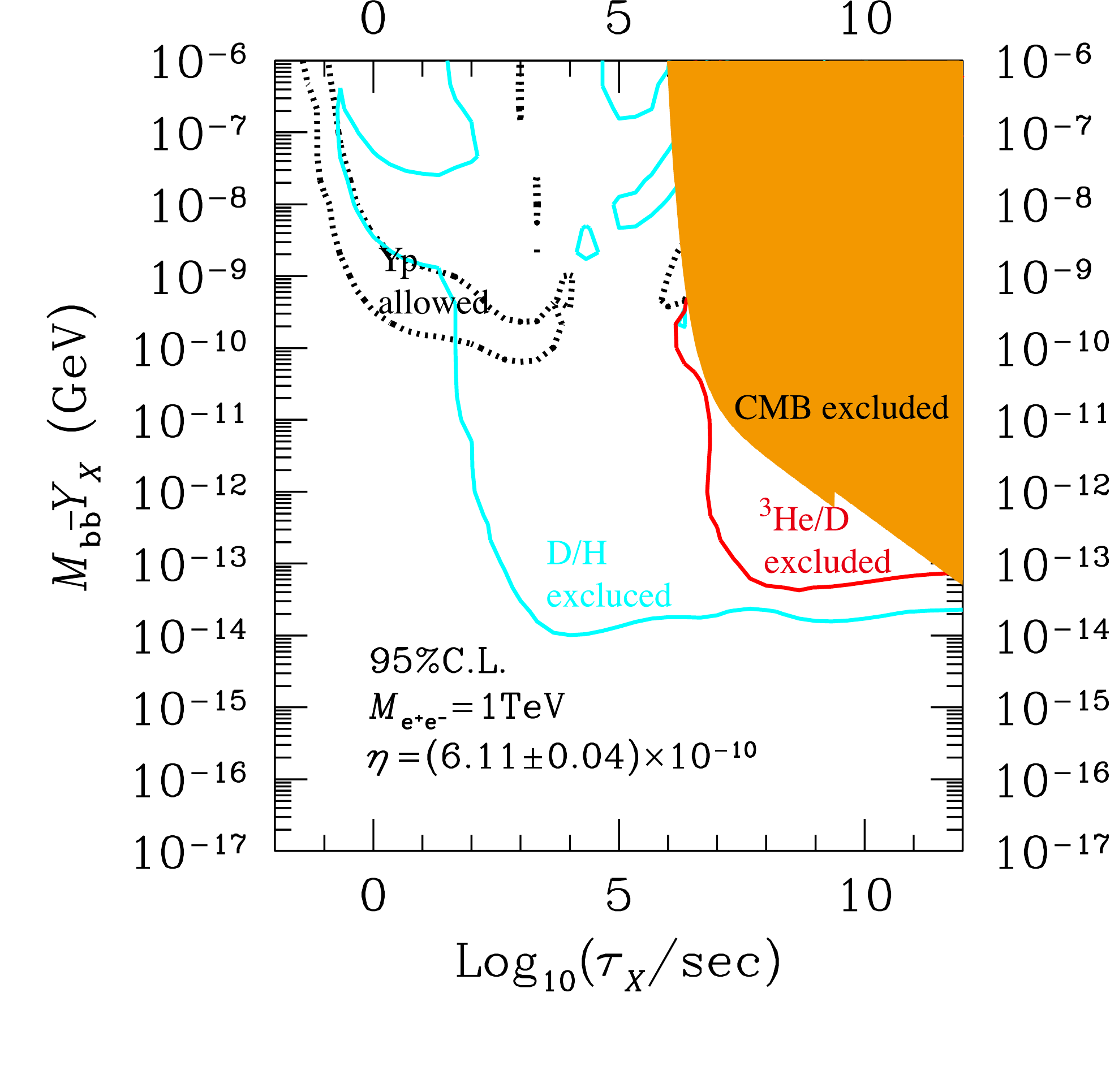} 
    \caption{\small%
      Constraints on $m_X Y_X$ vs.\ $\tau_X$ plane, assuming that
      $m_X=1\ {\rm TeV}$ and that the main decay mode is $b\bar{b}$.
      The solid cyan, solid red and dashed lines denote the BBN
      constraints from D, $^3$He/D and $^4$He, respectively.  Here, we
      use Eq.~(\ref{eq:obs_const_He4_ITG}) as the observational
      constraint on $Y_p$.  The orange shaded region is excluded by
      the CMB spectral distortion.}
  \label{fig:Izotov_const}
\end{figure}

So far we have adopted Eq.~(\ref{eq:obs_const_He4_AOS}) as the
observational constraint of $^4$He.  Now let us see how the
constraints change if we adopt the other constraint
(\ref{eq:obs_const_He4_ITG}).  In Fig.~\ref{fig:Izotov_const}, the BBN
constraints from D, $^3$He/D and $^4$He are shown, assuming that the
main decay mode is $b\bar{b}$.  As mentioned in Section \ref{sec:obs},
the $^4$He abundance~(\ref{eq:obs_const_He4_ITG}) estimated by Izotov,
Thuan and Guseva is significantly larger than that obtained by Aver,
Olive and Skillman given in Eq.~(\ref{eq:obs_const_He4_AOS}) and not
consistent with SBBN if we use the baryon-to-photon ratio
determined by Planck.  Their observation becomes consistent with BBN
if the abundance of $^4$He is increased by the decay of $X$.  When $X$
mainly decays into $b\bar{b}$, there appears a region consistent with
Eq.~(\ref{eq:obs_const_He4_ITG}) as well as with other light element
abundances, i.e., $\tau_X \sim 10^{-1}-10^4$~sec and $m_X Y_X \sim
10^{-10}-10^{-6}$~GeV.  This is due to the effect that the decay of
$X$ with such lifetime induces the $p\leftrightarrow n$ conversion,
resulting in the increase of the abundance of $^4$He.  If the $^4$He
abundance~(\ref{eq:obs_const_He4_ITG}) is confirmed, it may suggest
the existence of a long-lived hadronically decaying particle which
solves the discrepancy between SBBN and Eq.\
(\ref{eq:obs_const_He4_ITG}).  On the other hand, if the hadronic
branching ratio is much smaller, like the case that the main decay
mode is $e^{+}e^{-}$, there is no region consistent with
Eq.~(\ref{eq:obs_const_He4_ITG}) together with observational
constrains on D and $^3$He/D.

\begin{figure}
\centering
    \includegraphics[width=0.48\textwidth]{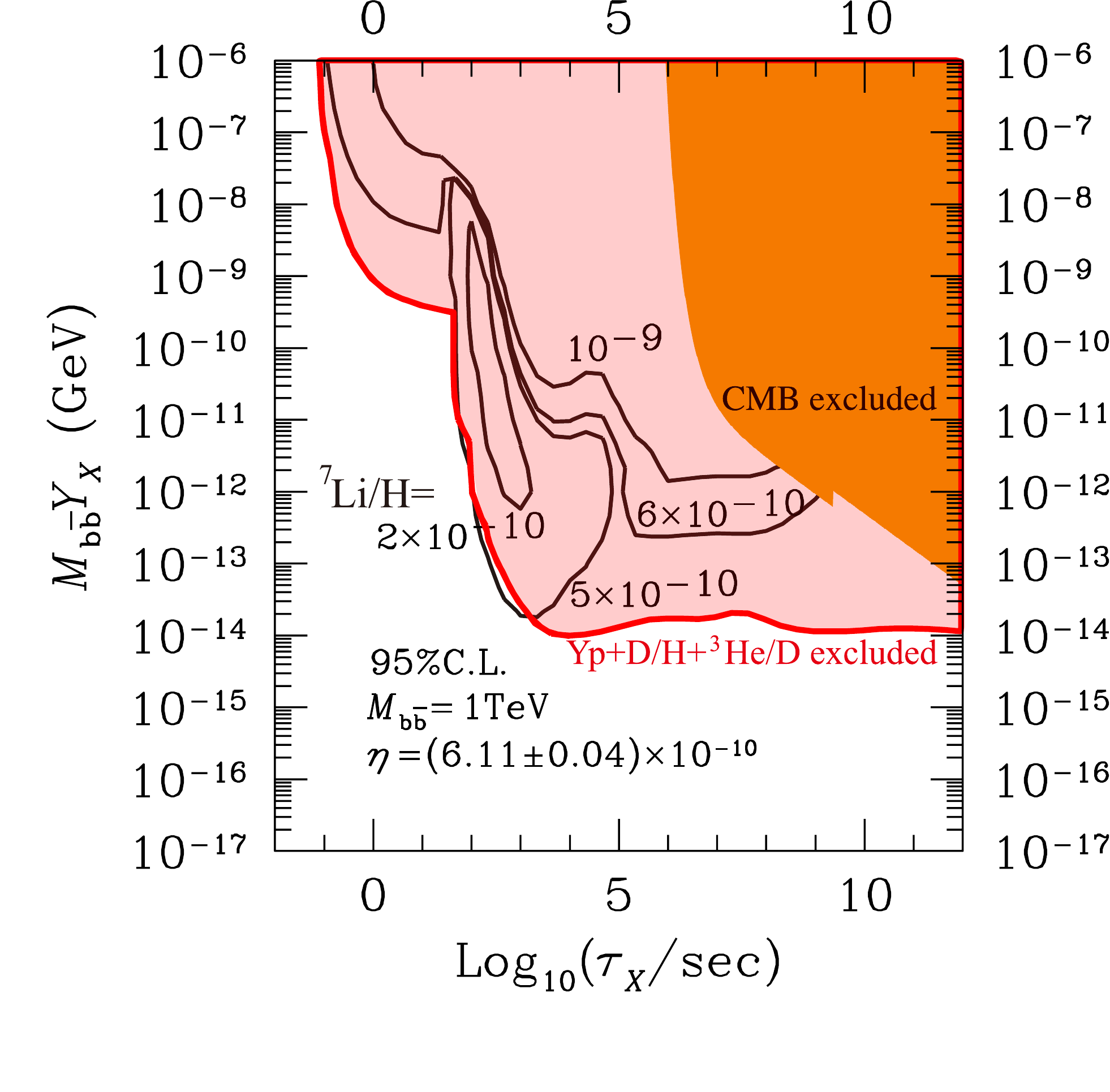} 
    \includegraphics[width=0.48\textwidth]{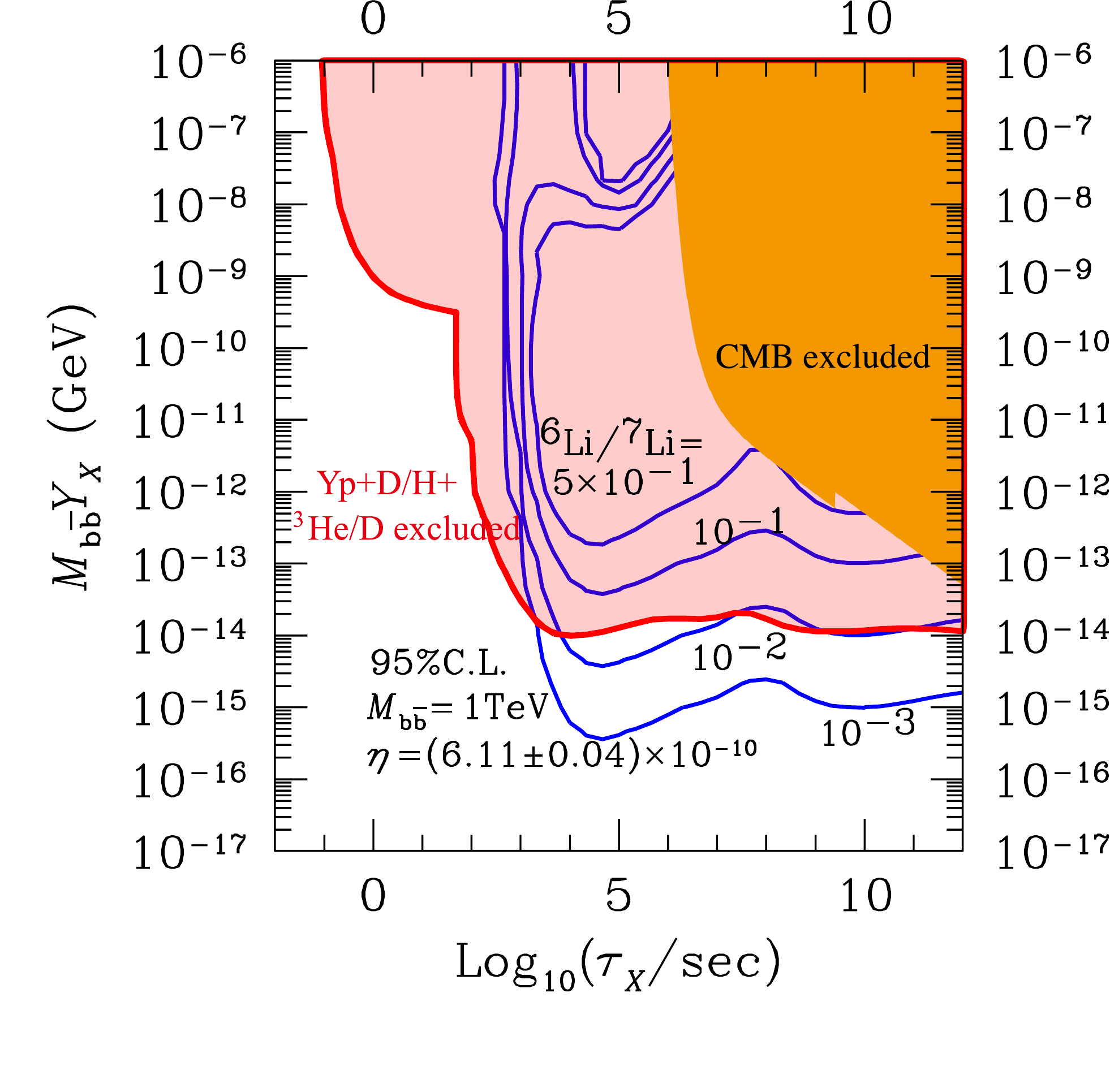} 
    \includegraphics[width=0.48\textwidth]{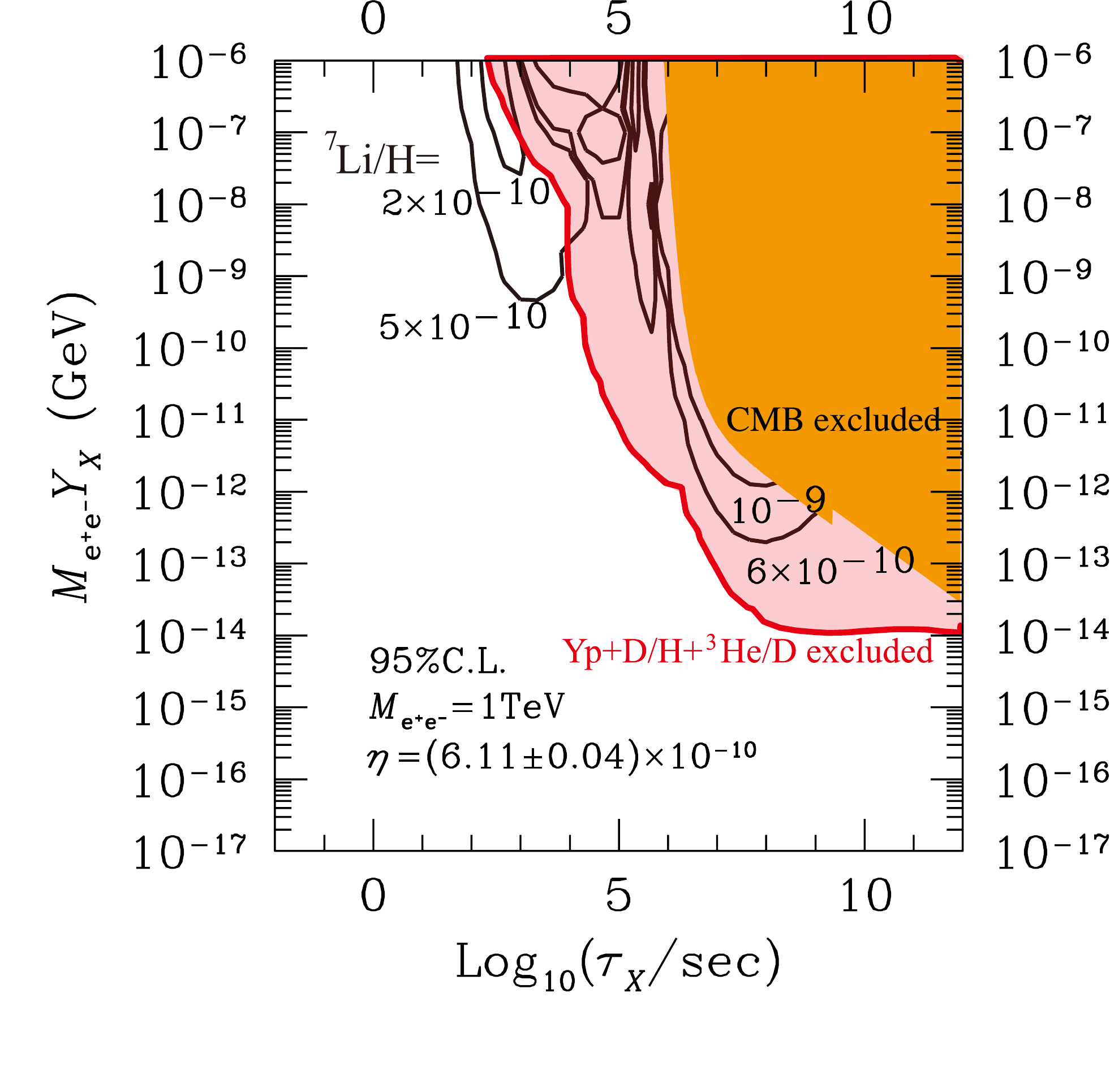} 
    \includegraphics[width=0.48\textwidth]{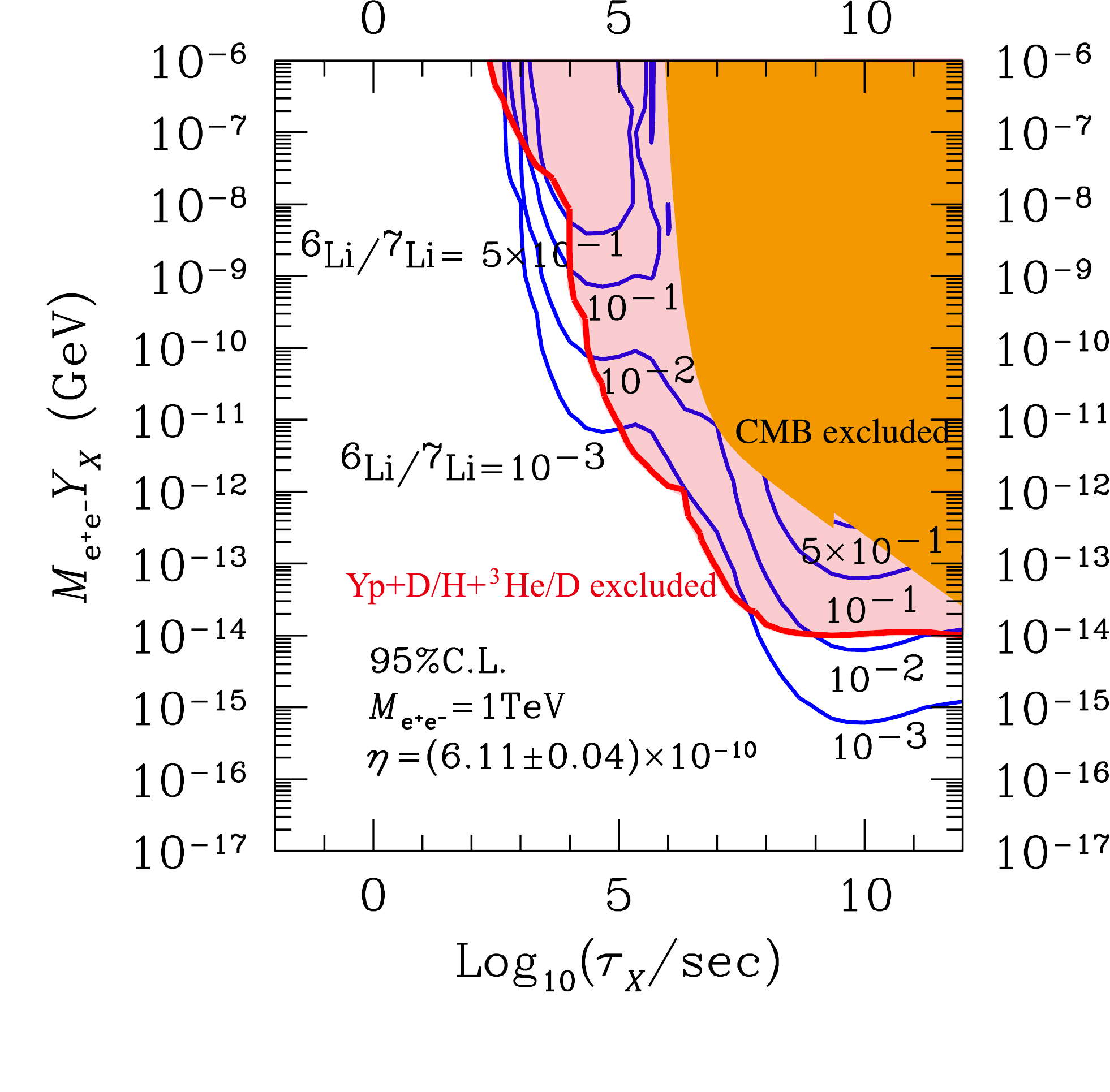} 
    \caption{\small%
      Abundances of $^7$Li/H (left) and $^6$Li/$^7$Li (right) for
      decay particles with mass $1$~TeV which decay mainly into
      $b\bar{b}$ (upper) and $e^{+}e^{-}$ (lower).  The constraints
      from other light elements and CMB are shown in pink shaded
      region (surrounded by red solid line) and orange shaded region,
      respectively.}
  \label{fig:Li_abundance}
\end{figure}

Before closing this section, we comment on the effects of the decaying
particles on abundances of $^7$Li and $^6$Li although we do not use
them to derive the constraints on the decaying particles.  In
Fig.~\ref{fig:Li_abundance}, abundances of $^7$Li and $^6$Li are
shown, assuming that the main decay mode is $b\bar{b}$ or
$e^{+}e^{-}$.  Notably, for $\tau_X \sim 10^2-10^3$~sec, primordial
$^7$Li abundance is reduced by non-thermal neutrons; $^7{\rm Be}$,
which is one of the origins of primordial $^7$Li, can be converted to
$^7$Li by non-thermal neutrons as $^7{\rm Be}+n\rightarrow ^7{\rm Li}
+p$, then $^7$Li is destroyed by the SBBN reaction $^7{\rm
  Li}+p\rightarrow ^4{\rm He} +^4{\rm He}$. Such effects may dominate
over the non-thermal production of $^7$Li due to energetic T, $^3$He,
and $^4$He from the photodissociation and scattering of $^4$He,
resulting in the net decrease of the $^7$Li abundance.  For the case
of $b\bar{b}$ mode, the constraints from D/H, $^3$He/D and $^4$He
exclude the parameter region where the $^7$Li abundance significantly
decreases.  On the other hand, $^7$Li abundance can be reduced for the
main decay mode of $e^{+}e^{-}$ without conflicting with other
constraints, if $\tau_X\sim{\cal O}(100)$~sec and $m_X Y_X \sim
10^{-7}$~GeV.  This might provide a solution to the $^7$Li problem in
SBBN.

In Fig.~\ref{fig:Li_abundance} it is seen that non-thermal production
of $^6$Li due to the decaying particles is significant for $\tau_X
\gtrsim 10^3$~sec.  Taking into account the constraints from the other
light elements, $^6$Li/$^7$Li can be as large as $10^{-2}$ for the
$b\bar{b}$ decay mode, while it can be ${\cal O}(0.1)$ for the
$e^{+}e^{-}$ mode.  On the contrary, $^6$Li is hardly produced in SBBN
(i.e., $^6{\rm Li}/^7{\rm Li}\lesssim 10^{-4}$).  Thus, if a
significant amount of $^6$Li/$^7$Li is observed in low-metal stars, it
would give an evidence for the existence of decaying particles in the
early universe.  In particular, for the $e^{+}e^{-}$ mode, it is
remarkable that $^6$Li/$^7$Li is significantly large with
simultaneously solving the $^7$Li problem at around $\tau_X \sim
10^3$~sec and $m_X Y_X\sim 10^{-7.5}$~GeV.

\section{Gravitino}
\label{sec:gravitino}
\setcounter{equation}{0}

One of the important candidates of long-lived particles is gravitino
in SUSY models.  Gravitino is the superpartner of graviton, and
interacts very weakly because its interaction is suppressed by inverse
powers of the (reduced) Planck scale $M_{\rm Pl}\simeq 2.4\times
10^{18}\ {\rm GeV}$.  Because of the weakness of its interaction, the
lifetime of gravitino may become so long that its decay products
affect the light element abundances (if gravitino is unstable).
Assuming $R$-parity conservation, gravitino becomes unstable if it is
not the lightest superparticle (LSP).  Even when gravitino is the LSP,
the next-to-the lightest superparticle (NLSP) decays into gravitino
with very long lifetime, and hence the BBN constraints on the
properties of the NLSP are derived, e.g., see Refs.\
\cite{Kawasaki:2008qe,Ishiwata:2009gs}.  Here, we pay particular
attention to the former case where gravitino is unstable because, in
such a case, we can obtain an upper bound on the reheating temperature
after inflation in order not to overproduce gravitino.  Applying the
analysis of the non-standard BBN processes discussed in the previous
sections, we study the effects of the gravitino decay on the light
element abundances and derive the upper bound on the reheating
temperature.

The primordial abundance of gravitino is sensitive to the reheating
temperature after inflation,\footnote
{Gravitino may be also produced non-thermally by the decay of the
  moduli fields or inflaton field \cite{Endo:2006zj, Nakamura:2006uc,
    Kawasaki:2006gs}.  We do not consider such contributions in our
  study to derive a conservative bound on the reheating temperature
  after inflation.}
which we define
\begin{eqnarray}
  T_{\rm R} \equiv
  \left(
  \frac{10}{g_*(T_{\rm R}) \pi^2} M_{\rm Pl}^2 \Gamma_{\rm inf}^2
  \right)^{1/4},
  \label{T_R}
\end{eqnarray}
with $\Gamma_{\rm inf}$ being the decay rate of the inflaton, and
$g_*(T_{\rm R})$ being the effective number of the massless degrees of
freedom at the time of reheating.  In our study, we use the value
suggested by the minimal SUSY standard model (MSSM), $g_*(T_{\rm
  R})=228.75$.

Gravitino is produced via the scattering processes of MSSM particles
in the thermal bath.  The Boltzmann equation for the number density of
gravitino (denoted as $n_{3/2}$) is given by
\begin{align}
  \frac{d n_{3/2}}{dt} + 3 H n_{3/2} = C_{3/2},
  \label{dot-ngrav}
\end{align}
where $C_{3/2}$ is the thermally averaged collision term.  The most
precise calculation of the collision term includes hard thermal loop
resummation to avoid infrared singularity \cite{Bolz:2000fu,
  Pradler:2006qh, Rychkov:2007uq}, and $C_{3/2}$ is parametrized as
\begin{align}
  C_{3/2} = \sum_{i=1}^3 \frac{3 \zeta (3) T^6}{16 \pi^3 M_{\rm Pl}^2}
  \left( 1 + \frac{M_i^2 (T)}{3 m_{3/2}^2} \right)
  c_i g_i^2 \ln \left(\frac{k_i}{g_i} \right),
\end{align}
where $i=1$, $2$, and $3$ correspond to the gauge groups $U(1)_Y$,
$SU(2)_L$, and $SU(3)_C$, respectively.  Here, $M_i (T)$ are the
gaugino mass parameters at the renormalization scale $Q=T$, and $g_i$
are the gauge coupling constants.  In addition, $c_i$ and $k_i$ are
numerical constants.  Refs.\ \cite{Bolz:2000fu, Pradler:2006qh} give
$(c_1,c_2,c_3)=(11, 27, 72)$ and $(k_1,k_2,k_3)=(1.266, 1.312,
1.271)$, while the coefficients from Ref.\ \cite{Rychkov:2007uq} are
$(c_1,c_2,c_3)=(9.90, 20.77, 43.34)$ and $(k_1,k_2,k_3)=(1.469, 2.071,
3.041)$.\footnote
{For the latter, we use the coefficients given in Ref.\
  \cite{Ellis:2015jpg}, which is based on Ref.\
  \cite{Rychkov:2007uq}.}

As one can see, the collision term is typically $C_{3/2}\sim T^6/
M_{\rm Pl}^{2}$ (as far as $T$ is higher than the masses of MSSM
particles).  Then, the yield variable of gravitino, which is defined
as $Y_{3/2}$ hereafter, is approximately proportional to the reheating
temperature for the parameter region of our interest.  Consequently,
the BBN constraints on the gravitino abundance can be converted to the
upper bound on the reheating temperature $T_{\rm R}$.

We numerically solve the Boltzmann equation \eqref{dot-ngrav} (as well
as the evolution equation of the universe based on Einstein equation)
to accurately calculate the primordial abundance of gravitino.
Assuming the MSSM particle content up to the GUT scale, the gravitino
abundance can be well fit by the following formula:
\begin{align}
  Y_{3/2} \simeq &\
  y_0 T_{\rm R}^{(8)}
  \left[
    1 + \delta_0^{(1)} \ln T_{\rm R}^{(8)} + \delta_0^{(2)} \ln^2 T_{\rm R}^{(8)}
  \right]
  \nonumber \\ &\
  + \sum_{i=1}^3
  y_i T_{\rm R}^{(8)}
  \left[
    1 + \delta_i^{(1)} \ln T_{\rm R}^{(8)} + \delta_i^{(2)} \ln^2 T_{\rm R}^{(8)}
  \right]
  \left( \frac{M_i^{\rm (GUT)}}{m_{3/2}} \right)^2,
  \label{Y32}
\end{align}
where $T_{\rm R}^{(8)}\equiv T_{\rm R}/10^8\ {\rm GeV}$ and $M_i^{\rm
  (GUT)}$ are the gaugino masses at the GUT scale which is taken to be
$2\times 10^{16}\ {\rm GeV}$ in our analysis.  Numerical constants in
the above fitting formula based on Refs.\ \cite{Bolz:2000fu,
  Pradler:2006qh} and those for Ref.\ \cite{Rychkov:2007uq} are
summarized in Tables \ref{table:ygravfitPS} and
\ref{table:ygravfitBS}, respectively.  The primordial abundance of
gravitino based on Refs.\ \cite{Bolz:2000fu, Pradler:2006qh} is
smaller than that based on Ref.\ \cite{Rychkov:2007uq}.  Thus, we
perform our numerical analysis with the former set of the coefficients
in order to derive a conservative constraint.  The bound based on Ref.\
\cite{Rychkov:2007uq} can be obtained by translating the bound on
$T_{\rm R}$ to that on the primordial abundance $Y_{3/2}$ using, for
e.g., Eq.\ \eqref{Y32}; the bound on $T_{\rm R}$ based on Ref.\
\cite{Rychkov:2007uq} is at most $2-3$ times more stringent than that
based on Refs.\ \cite{Bolz:2000fu, Pradler:2006qh}.

\begin{table}[t]
  \begin{center}
    \begin{tabular}{c|cccc}
      \hline\hline
      & $i=0$ & $i=1$ & $i=2$ & $i=3$
      \\
      \hline
      $y_i$ 
      & $2.3\times 10^{-14}$
      & $1.3\times 10^{-16}$ & $1.6\times 10^{-15}$ & $1.3\times 10^{-14}$
      \\
      $\delta_i^{(1)}$ & $0.015$ & $0.055$ & $0.013$ & $-0.042$
      \\
      $\delta_i^{(2)}$ & $-0.0005$ & $0.0081$ & $0.0001$ & $0.0006$
      \\
      \hline\hline
    \end{tabular}
    \caption{Numerical constants for the formula \eqref{Y32}, which
      gives a fitting formula for the primordial gravitino abundance,
      based on Refs.\ \cite{Bolz:2000fu, Pradler:2006qh}. }
    \label{table:ygravfitPS}
    \vspace{5mm}
    \begin{tabular}{c|cccc}
      \hline\hline
      & $i=0$ & $i=1$ & $i=2$ & $i=3$
      \\
      \hline
      $y_i$ 
      & $4.4\times 10^{-14}$
      & $1.5\times 10^{-16}$ & $2.1\times 10^{-15}$ & $3.2\times 10^{-14}$
      \\
      $\delta_i^{(1)}$ & $-0.014$ & $0.066$ & $0.015$ & $-0.089$
      \\
      $\delta_i^{(2)}$ & $0.0054$ & $0.0042$ & $0.0002$ & $0.0031$
      \\
      \hline\hline
    \end{tabular}
    \caption{Numerical constants for the formula \eqref{Y32} based on
      Ref.\ \cite{Rychkov:2007uq}.}
    \label{table:ygravfitBS}
  \end{center}
\end{table}

Because the interaction of gravitino is governed by the SUSY, the
partial decay rates of gravitino are determined once the mass spectrum
of the MSSM particles are known.  For a precise calculation of the
upper bound on the reheating temperature, we fix the mass spectrum of
the MSSM particles and calculate the decay widths of gravitino.  We
consider several sample points; the mass spectrum of each sample point
is summarized in Table \ref{table:MSSMmasses}.  The sample point 1 is
based on so-called CMSSM \cite{Martin:1997ns} parametrized by the
universal scalar mass $m_0$, unified gaugino mass $M_{1/2}$, universal
tri-linear coupling constants $A_0$ (with respect to the corresponding
Yukawa coupling constants), $\tan\beta$ (which is the ratio of the
vacuum expectation values of up- and down-type Higgs bosons), and the
sign of the SUSY invariant Higgs mass parameter $\mu$.  For the sample
point 1, we take $m_0 = 433\ {\rm GeV}$, $M_{1/2} = 970\ {\rm GeV}$,
$A_0 = -3020\ {\rm GeV}$, $\tan\beta=14$, and $\mu>0$.  As one can
see, in the sample point 1, all the MSSM particles are lighter than
$\sim 2.5\ {\rm TeV}$, and the discovery of some of the MSSM particles
are expected in future LHC experiment.  The sample point 2 is also
based on the CMSSM, with the underlying parameters of $m_0=5000\ {\rm
  GeV}$, $M_{1/2} =700\ {\rm GeV}$, $A_0 =-8000\ {\rm GeV}$,
$\tan\beta=10$ and $\mu > 0$.  At this point, the $\mu$-parameter is
relatively small, and the Higgsino-like neutralino becomes the LSP.
In the sample points 3 and 4, we consider the case where the sfermion
masses are above $100\ {\rm TeV}$ while gaugino masses are around the
TeV scale.\footnote
{Eq.\ \eqref{Y32} is applicable only when the mass scale of the MSSM
  particles is lower than the reheating temperature.  Although this
  condition is not satisfied when $T_R\lesssim 10^5\ {\rm GeV}$ for
  sample points 3 and 4, we use Eq.\ \eqref{Y32} throughout our
  analysis.  As we will see below, the upper bound on $T_R$ is found
  to be at least $\sim 10^5\ {\rm GeV}$, and our approximation is
  marginally acceptable for the reheating temperature of our
  interest.}
Such a mass spectrum is motivated in the so-called pure gravity
mediation model \cite{Ibe:2006de, Ibe:2011aa, ArkaniHamed:2012gw}, in
which scalar masses originate from a direct K\"ahler interaction
between the SUSY breaking field and the MSSM chiral multiplet, while
the gaugino masses are from the effect of anomaly mediation
\cite{Randall:1998uk, Giudice:1998xp}.  Then, the gaugino masses are
given in the following form:
\begin{eqnarray}
 M_1 &=& \frac{g_1^{2}}{16 \pi^{2}} \left( 11 F_\Phi + L \right),
  \label{M1}
  \\
 M_2 &=& \frac{g_2^{2}}{16 \pi^{2}} \left( F_\Phi + L \right),
  \\
 M_3 &=& \frac{g_3^{2}}{16 \pi^{2}} \left( -3 F_\Phi \right),
  \label{M3}
\end{eqnarray}
where $F_\Phi$ is the expectation value of the compensator
multiplet,\footnote
{In the model where the vacuum expectation value of the SUSY breaking
  field is much smaller than the Planck scale, $F_\Phi=m_{3/2}$.  In
  this analysis, however, we consider more general framework, and
  treat $F_\Phi$ as a free parameter.}
while $L$ parametrizes the effect of Higgs-Higgsino loop on the gaugino
masses.  In a large class of models, $L$ is of order $F_\Phi$ for the
scale below the masses of heavy Higgses and Higgsinos. The gaugino
masses for the sample points 3 and 4 are obtained by adopting
$(F_\Phi, L)=(131\ {\rm TeV}, 218\ {\rm TeV})$, and $(82\ {\rm TeV},
87\ {\rm TeV})$, respectively.  Notice that, above the mass scale of
the heavy Higgs and Higgsino, the gaugino mass parameters are obtained
by Eqs.\ \eqref{M1} $-$ \eqref{M3}, with taking $L\rightarrow 0$.  For
the calculation of the primordial abundance of gravitino, we take
account of this effect, and the thermally averaged gravitino
production cross sections for the points 3 and 4 are evaluated by
taking vanishing $L$.

\begin{table}
  \begin{center}
    \begin{tabular}{c|cccc}
      \hline\hline
      {} & {Point 1} & {Point 2} & {Point 3}  & {Point 4}\\
      \hline
      $m_{\tilde{u}_{R,1,2}}$ & $1907$ & $5242$ 
      & $\sim 1\times$ $10^5$ & $\sim 1\times$ $10^5$\\
      $m_{\tilde{d}_{R,1,2}}$ & $1898$ & $5054$ 
      & $\sim 1\times$ $10^5$ & $\sim 1\times$ $10^5$\\
      $m_{\tilde{u}_{L,1,2}} (m_{\tilde{d}_{L,1,2}})$ & $1980$ $(1982)$ &
		     $5082$ $(5083)$ 
      & $\sim 1\times$ $10^5$ & $\sim 1\times$ $10^5$\\
      $m_{\tilde{e}_{R,1,2}}$ & $562$ & $4801$ 
      & $\sim 1\times$ $10^5$ & $\sim 1\times$ $10^5$\\
      $m_{\tilde{e}_{L,1,2}} (m_{\tilde{\nu}_{L,1,2}})$ & $771$ $(767)$ &
	     $5093$ $(5092)$ 
      & $\sim 1\times$ $10^5$ & $\sim 1\times$ $10^5$\\
      $m_{\tilde{t}_1}$ & $977$ & $1455$ 
      & $\sim 1\times$ $10^5$ & $\sim 1\times$ $10^5$\\
      $m_{\tilde{t}_2}$ & $1635$ & $3603$ 
      & $\sim 1\times$ $10^5$ & $\sim 1\times$ $10^5$\\
      $m_{\tilde{b}_1}$ & $1608$ & $3607$ 
      & $\sim 1\times$ $10^5$ & $\sim 1\times$ $10^5$\\
      $m_{\tilde{b}_2}$ & $1843$ & $4990$ 
      & $\sim 1\times$ $10^5$ & $\sim 1\times$ $10^5$\\
      $m_{\tilde{\tau}_1}$ & $417$ & $4735$ 
      & $\sim 1\times$ $10^5$ & $\sim 1\times$ $10^5$\\
      $m_{\tilde{\tau}_2}$ & $732$ & $5062$ 
      & $\sim 1\times$ $10^5$ & $\sim 1\times$ $10^5$\\
      $m_{\tilde{\nu}_{\tau_L}}$ & $723$ & $5061$ 
      & $\sim 1\times$ $10^5$ & $\sim 1\times$ $10^5$\\
      $m_{\chi^0_1}$ & $417$ & $187$ & $1000$ & $500$ \\
      $m_{\chi^0_2}$ & $791$ & $-209$ & $1470$ & $880$ \\
      $m_{\chi^0_3}$ & $-1836$ & $321$
      & $\sim 1\times$ $10^5$ & $\sim 1\times$ $10^5$\\
      $m_{\chi^0_4}$ & $1838$ & $618$
      & $\sim 1\times$ $10^5$ & $\sim 1\times$ $10^5$\\
      $m_{\chi^\pm_1}$ & $791$ & $199$ & $1000$ & $500$ \\
      $m_{\chi^\pm_2}$ & $1839$ & $618$
      & $\sim 1\times 10^5$ & $\sim 1\times 10^5$\\
      $m_{\tilde{g}}$ & $2131$ & $1817$ & $3000$ & $2000$\\
      $m_h$ & $124$ & $126$ 
      & $125$ & $125$\\      
      $m_A$ & $1906$ & $1000$ 
      & $\sim 1\times$ $10^5$ & $\sim 1\times$ $10^5$\\      
      $M_{1,2,3}^{\rm (GUT)}$ & $970$ & $700$ & $(2583, 376, -1137)$ &
      $(1619, 236, -716)$ \\
      \hline\hline
    \end{tabular}
    \caption{The mass spectrum of the MSSM particles as well as the
      gaugino masses at the GUT scale for the sample points adopted in
      our analysis.  Here, $m_{\tilde{u}_{R,i}}$ and
      $m_{\tilde{u}_{L,i}}$ ($m_{\tilde{d}_{R,i}}$ and
      $m_{\tilde{d}_{L,i}}$) are masses of right- and left-handed sups
      (sdowns) in $i$-th generation, respectively, while
      $m_{\tilde{e}_{R,i}}$ and $m_{\tilde{e}_{L,i}}$
      ($m_{\tilde{\nu}_{L,i}}$) are masses of right- and left-handed
      charged sleptons (sneutrinos) in $i$-th generation,
      respectively.  In addition, $m_{\tilde{t}_1}$,
      $m_{\tilde{b}_1}$, and $m_{\tilde{\tau}_1}$ ($m_{\tilde{t}_2}$,
      $m_{\tilde{b}_2}$, and $m_{\tilde{\tau}_2}$) are lighter
      (heavier) stop, sbottom, and stau masses, respectively, while
      $m_{\tilde{\nu}_{\tau_L}}$ is the tau-sneutrino mass.
      Furthermore, $m_{\chi^0_i}$, $m_{\chi^\pm_i}$, and
      $m_{\tilde{g}}$ are neutralino, chargino, and gluino masses,
      respectively, while $m_h$ and $m_A$ are masses of the lightest
      Higgs boson and CP-odd Higgs boson, respectively.  All the mass
      parameters are given in units of GeV. }
    \label{table:MSSMmasses}
  \end{center}
\end{table}

With fixed mass spectrum of the MSSM particles, we vary the gravitino
mass and derive the upper bound on the reheating temperature as a
function of the gravitino mass.  In our calculation, all the two-body
tree-level decay processes of gravitino are taken into account.  We
also include three-body decay processes $\tilde{G} \rightarrow
\tilde{\chi}_0 q\bar{q}$ (with $\tilde{G}$, $\tilde{\chi}_0$ and $q$
denoting gravitino, the lightest neutralino and $u$, $d$, $s$, $c$ or
$b$ quark, respectively), if the mass splitting between gravitino and
the lightest neutralino is less than the $Z$-boson mass.  This is
because those three-body decay processes can have substantial
contribution to hadronic emissions from gravitino decays to the
lightest neutralino when the two-body decay process $\tilde{G}
\rightarrow \tilde{\chi}^0_1 Z$ is kinematically forbidden.
For the sample points 3 and 4, the lightest chargino
($\tilde{\chi}_1^\pm$) has degenerate
mass with the lightest neutralino and behaves as the LSP. We therefore
include the off-shell $W^\pm$ induced three-body decay processes,
$\tilde{G} \rightarrow \tilde{\chi}^\pm_1 q \bar{q}^{'}$, as well in our
calculation if $m_{3/2} -m_{\tilde{\chi}^\pm_1} < m_W$. 

The partial decay rates of gravitino are calculated by using
MadGraph5\_aMC@NLO v2.1 package \cite{Alwall:2014hca} with gravitino
interactions being implemented via FeynRules v2.3 \cite{Duhr:2011se,
  Degrande:2011ua, Christensen:2013aua, Alloul:2013bka}.  Subsequent
decay and the hadronization processes of the decay products of
gravitino are simulated by using PYTHIA 8.2 package
\cite{Sjostrand:2014zea}.  For given reheating temperature, the
primordial abundance of gravitino is calculated by numerically solving
Eq.\ \eqref{dot-ngrav}.  For the thermally averaged gravitino
production cross section, we adopt the results of Refs.\
\cite{Bolz:2000fu, Pradler:2006qh}.

\begin{figure}[t]
  \begin{center}
  \includegraphics[width=0.45\textwidth]{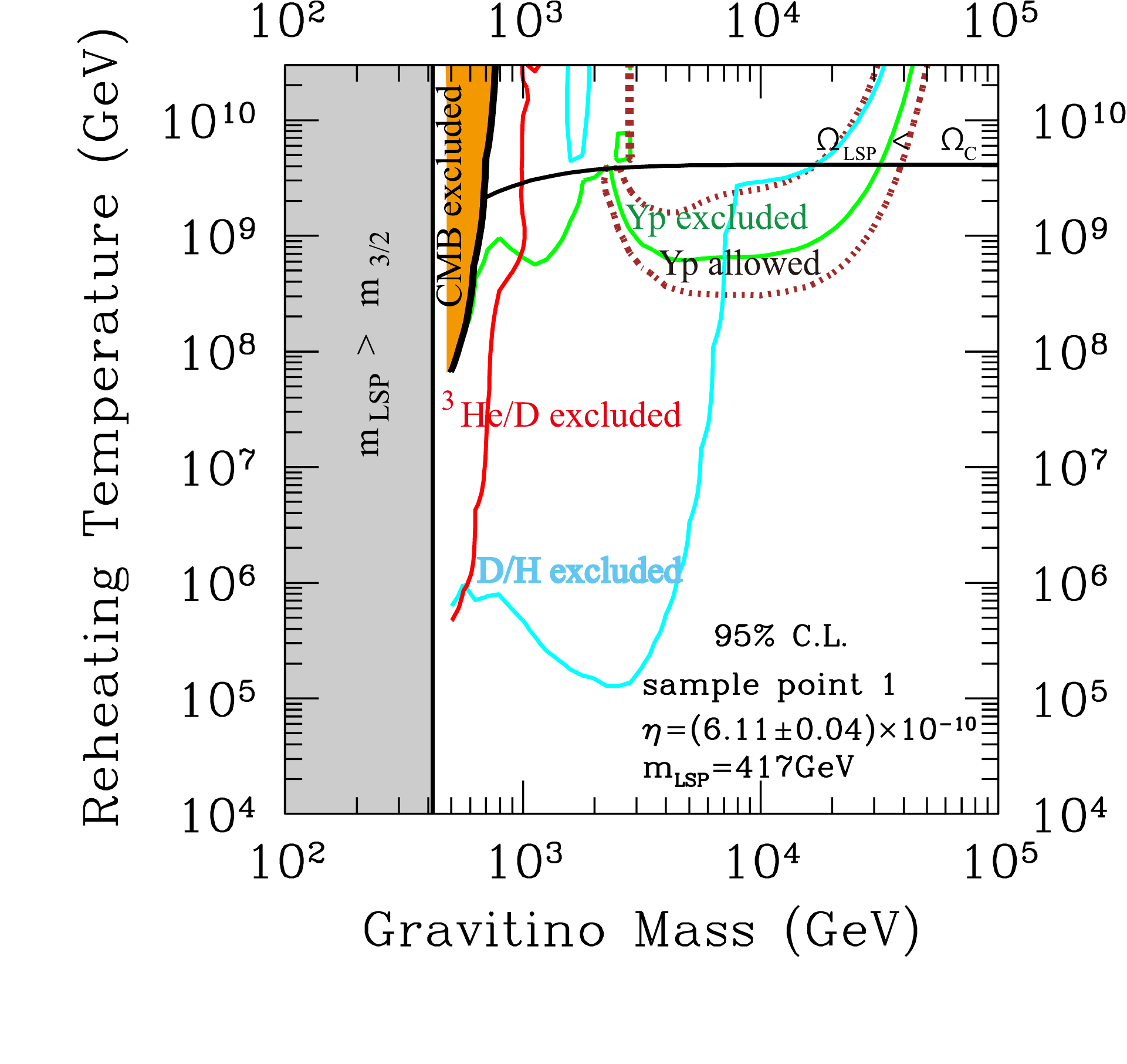} 
  \includegraphics[width=0.45\textwidth]{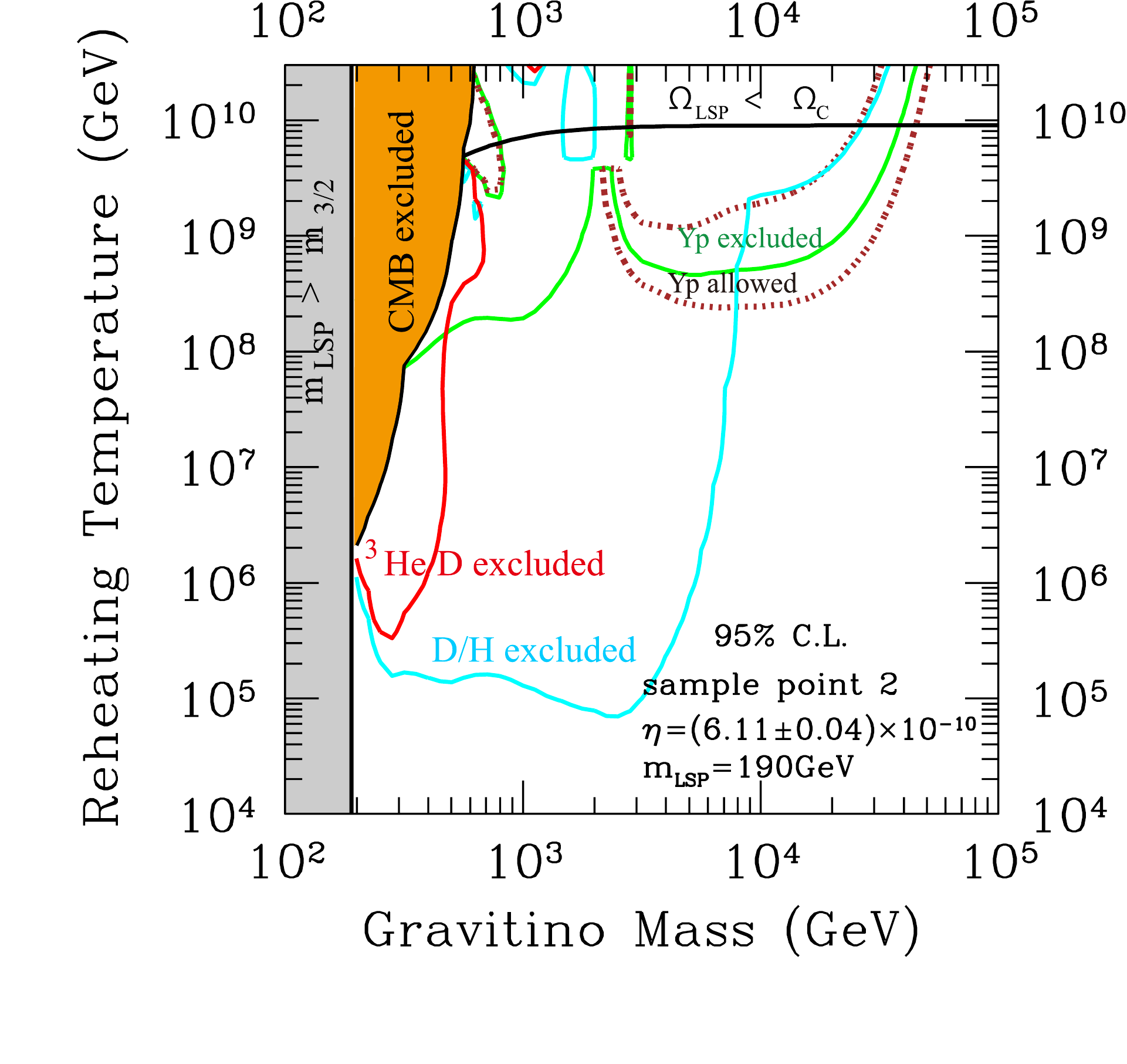} 
  \includegraphics[width=0.45\textwidth]{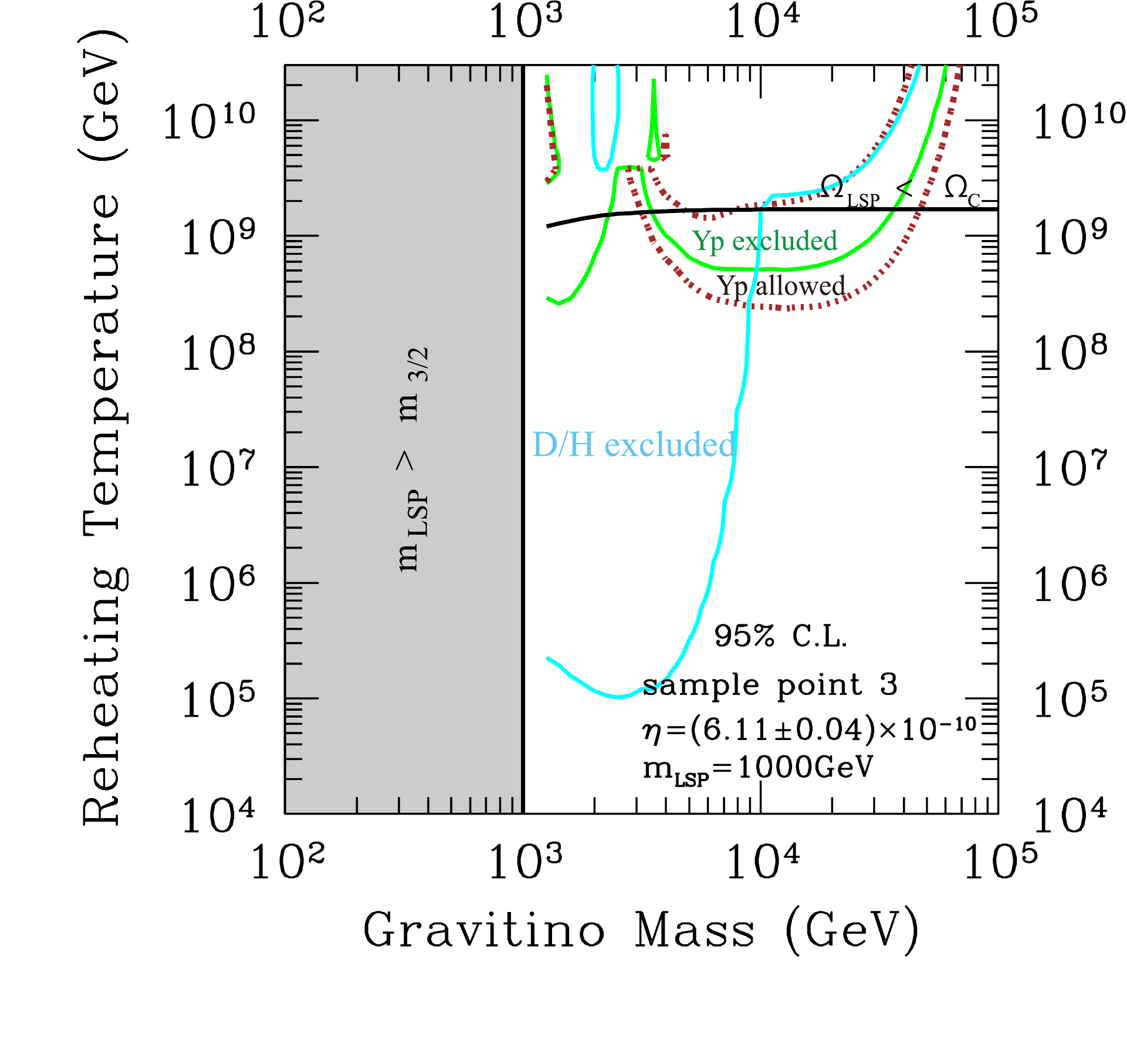} 
  \includegraphics[width=0.45\textwidth]{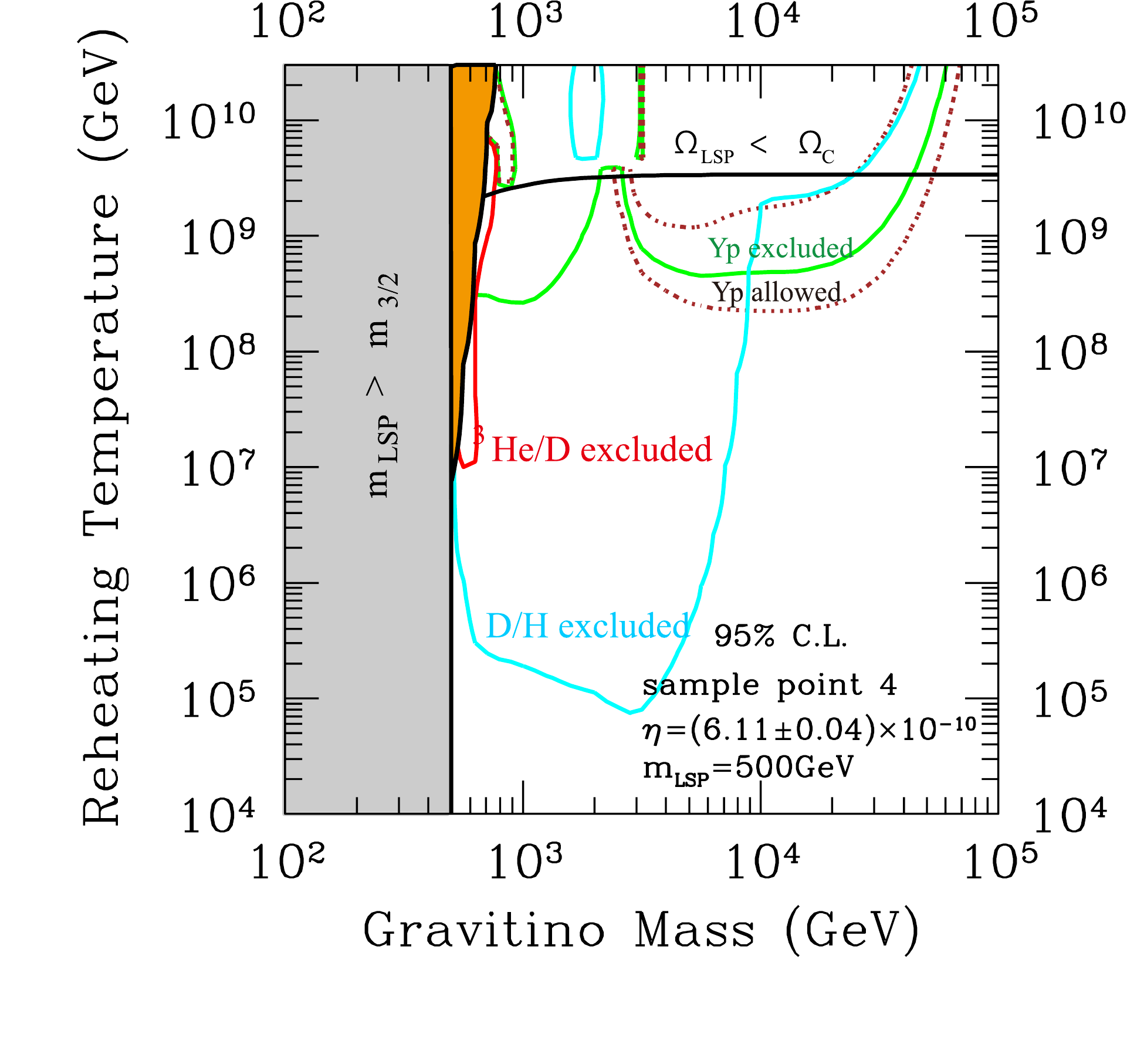} 
  \end{center}
 \vspace{-1.5cm}
 \caption{Upper bound on the reheating temperature $T_{\rm R}$ as a
   function of $m_{3/2}$ at 95$\%$ C.L. for models of Point 1 (upper
   left), Point 2 (upper right), Point 3 (lower left), and Point 4
   (lower right), respectively.  The regions surrounded by the
   brown-dotted line indicate the region consistent with Eq.\
   \eqref{eq:obs_const_He4_ITG}.}
\label{fig:mgravtrmax}
\end{figure}

Following the procedure discussed in the previous sections, we
calculate the light element abundances, taking into account the
effects of the decay products of gravitino.  The constraints on the
gravitino abundance is then translated to the upper bound on the
reheating temperature.  The upper bound on $T_{\rm R}$ for the sample
points 1, 2, 3, and 4 are shown in Fig.\ \ref{fig:mgravtrmax}.  As one
can see, a severe upper bound of $\sim 10^5-10^6\ {\rm GeV}$ is
obtained from the overproduction of D for relatively light gravitino
mass (i.e., for $m_{3/2}\lesssim$ a few TeV).  This is because, when
gravitino is lighter than a few TeV, the lifetime of gravitino is
longer than $\sim 10^3\ {\rm sec}$ so that the constraint from D is
significant, as discussed in the previous section.  On the contrary,
for heavier gravitino mass (i.e., for $m_{3/2}\gtrsim 10\ {\rm TeV}$),
for which the lifetime of  gravitino becomes shorter than $\sim
10^2\ {\rm sec}$, the most stringent constraint comes from the
overproduction of $^4$He due to the $p\leftrightarrow n$ conversion.

We also comment on the constraints based on the $^4$He abundance given
in Eq.~(\ref{eq:obs_const_He4_ITG}), which is inconsistent with the
SBBN prediction.  As mentioned in the previous section, with a
hadronically decaying long-lived particle, the primordial $^4$He
abundance may become consistent with Eq.~(\ref{eq:obs_const_He4_ITG}).
In the present case, gravitino has sizable branching ratio for
hadronic decay modes.  In Fig.\ \ref{fig:mgravtrmax}, we also show a
region consistent with the $^4$He abundance
(\ref{eq:obs_const_He4_ITG}) estimated by Izotov, Thuan and Guseva.
We can see that the allowed region exists for $m_{3/2}\sim {\cal
  O}(10)\ {\rm TeV}$ and $T_{\rm R}\sim {\cal O}(10^9)\ {\rm GeV}$.

Notice that the reheating temperature is bounded from above in order
not to overclose the universe by the LSP produced from the decay of
gravitino.  For the parameter region of our interest, gravitino decays
at the cosmic temperature lower than the freeze-out temperature of the
LSP.  Thus, the density parameter of the LSP from the decay of
gravitino is evaluated as
\begin{align}
  \Omega_{\rm LSP}^{\rm (decay)} = 
  \frac{m_{\rm LSP} Y_{3/2} s_{\rm now}}{\rho_{\rm crit}},  
\end{align}
where $s_{\rm now}$ is the entropy density of the present universe,
and $\rho_{\rm crit}$ is the critical density.  We show the contour of
$\Omega_{\rm LSP}^{\rm (decay)}=\Omega_c$ (with $\Omega_c\simeq 0.26$
\cite{Ade:2015xua} being the density parameter of the cold dark
matter).\footnote
{If the thermal relic abundance of the LSP is sizable, the bound
  should be imposed on the total mass density of the LSP.  However,
  because the thermal relic abundance is strongly dependent on the
  MSSM parameters, we show the bound based on $\Omega_{\rm LSP}^{\rm
    (decay)}$.}
We can see that, with the present choice of the MSSM mass spectrum,
the upper bound from the overclosure of the universe is $10^9-10^{10}\
{\rm GeV}$, which is less stringent when the gravitino mass is smaller
than $\sim 40-50\ {\rm TeV}$.

Before closing this section, we comment on the implication of our
result on the leptogenesis \cite{Fukugita:1986hr}, in which the baryon
asymmetry of the universe originates from the lepton asymmetry
generated by the decay of right-handed neutrinos.  In order to
generate enough amount of baryon asymmetry via thermal leptogenesis,
the reheating temperature is required to be higher than $\sim 10^9\
{\rm GeV}$ \cite{Giudice:2003jh, Buchmuller:2004nz}.  Thus, for a
viable scenario of thermal leptogenesis, scenarios realizing the
gravitino mass of ${\cal O}(10)\ {\rm TeV}$ is suggested, like the
pure gravity mediation scenario \cite{Ibe:2006de, Ibe:2011aa,
  ArkaniHamed:2012gw}.  Notice that such a scenario works irrespective
of the observational constraint on the $^4$He abundance,
Eq.~(\ref{eq:obs_const_He4_ITG}) or Eq.~(\ref{eq:obs_const_He4_AOS}).

\section{Conclusions and Discussion}
\label{sec:conclusions}
\setcounter{equation}{0}

We have revisited and updated the BBN constraints on long-lived
particles.  Compared with the previous analysis we have improved the
following points.  First, the SBBN reactions and their uncertainties
are updated.  Second, we have revised the hadronic shower calculation
taking into account $p\leftrightarrow n$ conversion in inelastic
scatterings of energetic nucleons off the background $p$ or $^4$He.
Third, we have included the effects of the hadronic showers induced by
the injections of energetic anti-nucleons ($\bar{p}$ and $\bar{n}$).
Finally we have used the most recent observational data for the
abundances of $^4$He and D and the cosmological parameters.

We have obtained the constraints on the abundance and lifetime of
long-lived particles with various decay modes.  They are shown in
Figs.~\ref{fig:generic_const_1} and \ref{fig:generic_const_2}.  The
constraints become weaker when we include the $p\leftrightarrow n$
conversion effects in inelastic scatterings because energetic neutrons
change into protons and stop without causing hadrodissociations.  On
the other hand, inclusion of the energetic anti-nucleons makes the
constraints more stringent.  In addition, the recent precise
measurement of the D abundance leads to stronger constraints.  Thus, in
total, the resultant constraints become more stringent than those
obtained in the previous studies.

We have also applied our analysis to unstable gravitino.  We have
adopted several patterns of the mass spectrum of superparticles and
derived constraints on the reheating temperature after inflation as
shown in Fig.~\ref{fig:mgravtrmax}.  The upper bound on the reheating
temperature is $\sim 10^5-10^6$~GeV for gravitino mass $m_{3/2}$ less
than a several TeV and $\sim 10^9$~GeV for $m_{3/2} \sim
\mathcal{O}(10)$~TeV.  This implies that the gravitino mass should be
$\sim \mathcal{O}(10)$~TeV for successful thermal leptogenesis.

In obtaining the constraints, we have adopted the observed $^4$He
abundance given by Eq.\ (\ref{eq:obs_const_He4_AOS}) which is
consistent with SBBN.  On the other hand, if we adopt the other
estimation (\ref{eq:obs_const_He4_ITG}), $^4$He abundance is
inconsistent with SBBN.
However, when long-lived particles with large hadronic branch have
lifetime $\tau_X \sim 0.1-100$~sec and abundance $m_X Y_X \sim
10^{-9}$, Eq.\ (\ref{eq:obs_const_He4_ITG}) becomes consistent with
BBN.

In this work, we did not use $^7$Li in deriving the constraints since
the plateau value in $^7$Li abundances observed in metal-poor stars
(which had been considered as a primordial value) is smaller than the
SBBN prediction by a factor $2$--$3$ (lithium problem) and furthermore
the recent discovery of much smaller $^7$Li abundances in very
metal-poor stars cannot be explained by any known mechanism.  However,
the effects of the decaying particles on the $^7$Li and $^6$Li
abundances are estimated in our numerical calculation.  Interestingly,
if we assume that the plateau value represents the primordial
abundance, the decaying particles which mainly decays into
$e^{+}e^{-}$ can solve the lithium problem for $\tau_X\sim
10^2-10^3$~sec and $m_X Y_X \sim 10^{-7}$.

\section*{Acknowledgments}
We thank Benjamin Fuks and Olivier Mattelaer for their supports in
simulating gravitino decays with MadGraph5\_aMC@NLO.  We thank Jens
Chluba for useful discussions. This work was partially supported by
JSPS KAKENHI Grant Nos.~JP17H01131 (M.K. and K.K), 17K05434 (M.K.),
26247042 (K.K.), 16H06490 (T.M.), and 26400239 (T.M.), and MEXT
KAKENHI Grant Nos.~JP15H05889 (M.K. and K.K.), and JP16H0877 (K.K.).


\end{document}